\documentclass[11pt]{article}
\pdfoutput=1

\usepackage[utf8]{inputenc}
\usepackage{multirow}
\usepackage{amsmath, amsfonts, amssymb}
\usepackage{comment}
\usepackage{graphicx}
\usepackage{float}
\usepackage{psfrag}
\usepackage{amsthm}
\usepackage{import}
\usepackage{bm}
\usepackage[usenames,dvipsnames,svgnames,table]{xcolor}
\usepackage{enumerate}
\usepackage{arydshln}
\usepackage{soul}
 \usepackage{slashed}
\usepackage{subfigure}
\usepackage{mathrsfs}
\usepackage{mathalfa}
\usepackage{longtable}
\allowdisplaybreaks
 \usepackage{a4wide}
  \usepackage{tikz}
  \usepackage{tikz-cd}
  \usetikzlibrary{shapes.geometric,arrows}
  \usepackage{tcolorbox}
\tikzstyle{startstop} = [rectangle, rounded corners, minimum width=3cm, minimum height=1cm,text centered, draw=black, fill=red!30]
\tikzstyle{io} = [trapezium, trapezium left angle=70, trapezium right angle=110, minimum width=3cm, minimum height=1cm, text centered, draw=black, fill=blue!30]
\tikzstyle{process} = [rectangle, minimum width=3cm, minimum height=1cm, text centered, draw=black, fill=orange!30]
\tikzstyle{decision} = [diamond, minimum width=3cm, minimum height=1cm, text centered, draw=black, fill=green!30]
\tikzstyle{arrow} = [thick,->,>=stealth]
\tikzcdset{row sep/normal=0.75cm}

  \usepackage{color}
  \definecolor{dark-gray}{gray}{0.20}
  \definecolor{gray}{gray}{0.30}
  \definecolor{light-gray}{gray}{0.80}
  \definecolor{dark-red}{rgb}{0.7,0,0}
  \definecolor{dark-green}{rgb}{0.1,0.4,0}
  \definecolor{dark-blue}{rgb}{0.3,0.3,0.7}
  \definecolor{light-blue}{rgb}{0.8,0.8,1}
      \definecolor{swamp}{RGB}{240, 199, 197}

  \usepackage{pifont}

\usepackage{newunicodechar} 
\newunicodechar{ỳ}{\`y}

\usepackage{setspace}

\newcommand{\be}{\begin{equation}}
\newcommand{\ee}{\end{equation}}

\def\be{\begin{equation}}
\def\ee{\end{equation}}
\def\bea{\begin{eqnarray}}
\def\eea{\end{eqnarray}}

\newcommand{\beq}{\begin{equation}}  \newcommand{\eeq}{\end{equation}}
\newcommand{\bal}{\begin{aligned}}   \newcommand{\eal}{\end{aligned}}
\def\beqa{\begin{eqnarray}}
\def\eeqa{\end{eqnarray}}

\newcommand{\cT}{\mathcal{T}}

\newcommand{\I}{\text{Im}\,}

\DeclareMathOperator{\diag}{Diag}

\DeclareMathOperator{\diff}{d}
\DeclareMathOperator{\rank}{rank}

\renewcommand{\Re}{\text{Re}\,}
\usepackage{stackengine}
\usepackage{calc}
\newlength\shlength
\newcommand\vv[2][0]{\setlength\shlength{#1pt}%
  \stackengine{-5.6pt}{$#2$}{\smash{$\kern\shlength%
    \stackengine{7.55pt}{$\mathchar"017E$}%
      {\rule{\widthof{$#2$}}{.57pt}\kern.4pt}{O}{r}{F}{F}{L}\kern-\shlength$}}%
      {O}{c}{F}{T}{S}}

\newcommand{\Mpl}{M_{\textrm{Pl}}}
\newcommand{\C}{\mathbb{C}}
\newcommand{\R}{\mathbb{R}}

\newcommand{\Z}{\mathbb{Z}}

\def\simleq{\; \raise0.3ex\hbox{$<$\kern-0.75em
      \raise-1.1ex\hbox{$\sim$}}\; }
   \def\simgeq{\; \raise0.3ex\hbox{$>$\kern-0.75em
      \raise-1.1ex\hbox{$\sim$}}\; }

\numberwithin{equation}{section}

\usepackage{jheppub}
\usepackage{hyperref}
\usepackage{cleveref}

\hypersetup{
	colorlinks=true,
	linkcolor=dark-blue,
	citecolor=dark-red,
	urlcolor=dark-green,
	linktoc=page
}

\theoremstyle{remark}

\crefname{appendix}{Appendix}{Appendices}

\title{\centering Asymptotic Accelerated Expansion in String Theory and the Swampland}

\author{Jos\'{e} Calder\'{o}n-Infante$^1$,} \author{Ignacio Ruiz$^1$,} 
\author{Irene Valenzuela$^{1,2}$} 
\affiliation{$^1$Instituto de F\'{i}sica Te\'{o}rica UAM-CSIC and Departamento de F\'{i}sica Te\'{o}rica, Universidad Aut\'{o}noma de Madrid, Cantoblanco, 28049 Madrid, Spain}
\affiliation{$^2$CERN, Theoretical Physics Department, 1211 Meyrin, Switzerland}

\preprint{CERN-TH-2022-153\\ \vspace*{-0.8cm} \hfill IFT-UAM/CSIC-22-110}
\emailAdd{j.calderon.infante@csic.es}
\emailAdd{ignacio.ruiz@uam.es}
\emailAdd{irene.valenzuela@cern.ch}

\abstract{
We study whether the universal runaway behaviour of stringy scalar potentials towards infinite field distance limits can produce an accelerated expanding cosmology \`{a} la quintessence. We identify a loophole to some proposed bounds that forbid such asymptotic (at parametric control) accelerated expansion in 4d $\mathcal{N}=1$ supergravities, by considering several terms of the potential competing asymptotically. We then analyse concrete string theory examples coming from F-theory flux compactifications on Calabi-Yau fourfolds, extending previous results by going beyond weak string coupling to different infinite distance limits in the complex structure moduli space. We find some potential candidates to yield asymptotic accelerated expansion with a flux potential satisfying $\gamma=\frac{\|\nabla V\|}{V}<\sqrt{2}$ along its gradient flow. However, whether this truly describes an accelerated expanding cosmology remains as an open question until full moduli stabilization including the Kahler moduli is studied. Finally, we also reformulate the condition for forbidding asymptotic accelerated expansion as a convex hull de Sitter conjecture which resembles a convex hull scalar WGC for the membranes generating the flux potential. This provides a pictorial way to quickly determine the asymptotic gradient flow trajectory in multi-moduli setups and the value of $\gamma$ along it.

}

\begin{document}
\hypersetup{pageanchor=false}
\makeatletter
\let\old@fpheader\@fpheader

\makeatother
\maketitle

\hypersetup{
    pdftitle={A cool title},
    pdfauthor={Jos\'{e} Calder\'{o}n-Infante, Ignacio Ruiz, Irene Valenzuela},
    pdfsubject={}
}

\newcommand{\remove}[1]{\textcolor{red}{\sout{#1}}}

\newpage

\section{Introduction}
\label{sec:intro}

String Theory seems to contain all necessary ingredients to describe our universe. However, it allows for a vast number of solutions and finding the concrete one that describes our universe remains a major challenge. In particular, each compactification has a moduli space given by the vacuum expectation values of scalar fields parameterizing the volumes and sizes of the compactification manifold, that can receive a potential from different sources.
The question is, where do we live in the string landscape? Is there any particular region of these moduli spaces that better resembles our universe?

We know very little about the bulk of the moduli space, but we have much more information  about the possible effective field theories arising near the infinite field distance boundaries of the moduli space. These infinite distance limits (known as asymptotic regions) correspond, from the perspective of the low energy Effective Field Theory (EFT),  to different perturbative descriptions of the full theory. They are special in the sense that they allow for parametric computational control, as we get a perturbative expansion on some parameter that can get parametrically small since it is fixed by the vev of the field sent to the limit. Furthermore, even if there are many different types of asymptotic limits, they all share some universal properties imposed by quantum gravity. These quantum gravity constraints are often formulated in terms of Swampland conjectures \cite{Vafa:2005ui,Brennan:2017rbf,Palti:2019pca,vanBeest:2021lhn,Grana:2021zvf}, and provide criteria to distinguish those EFTs that can be valid low energy descriptions of string theory from those that cannot. Interestingly, the Swampland constrains become more constraining in the asymptotic regions\footnote{For instance, the Distance Conjecture \cite{Ooguri:2006in} implies the existence of an infinite tower of states becoming light at every infinite distance limit. The Weak Gravity conjecture \cite{Arkani-Hamed:2006emk,Harlow:2022gzl} implies the existence of some state with a mass smaller than its charge, meaning that it becomes light at weak coupling (which is also at infinite distance). }, as it is there where quantum gravitational effects become important at scales much below $\Mpl$ and can have a significant impact on the low energy EFT. Hence, they present an opportunity to signal universal quantum gravity predictions or even provide new insights to explain the naturalness issues of our universe.

So what universal features arise at the asymptotic limits and how well they resemble our universe? As mentioned, they correspond to weak coupling limits in some parameter\footnote{Clearly, not necessarily in all parameters. The theory could admit a perturbative expansion in some parameter while being strongly coupled in others.}, which can be identified, so far in all string theory examples, with a $p$-form gauge coupling vanishing at infinite distance as remarked in \cite{Gendler:2020dfp}. There are also new light states and approximate global symmetries emerging. This is indeed familiar to our Standard Model of Particle Physics, which contains annoyingly unnatural small numbers, approximate global symmetries (e.g. B-L) and it is weakly coupled to gravity. Furthermore, even if the electromagnetic gauge coupling is only two orders of magnitude below $\Mpl$, it is small enough to allow for light fermions; if it were of order $\Mpl$, chirality would not be enough to guarantee a light spectrum \cite{Razamat:2020kyf}. Moreover, asymptotic limits in string theory are always characterized by a runaway potential such that the potential energy goes to zero asymptotically. This is more concretely formulated as an asymptotic version of the de Sitter conjecture \cite{Obied:2018sgi,Ooguri:2018wrx}, which is a generalization of Dine-Seiberg \cite{Dine:1985he} for any direction in field space. Hence, if we are far into the asymptotic regime, the vacuum energy is naturally small. This is a very interesting mechanism to get a small (time-dependent) vacuum energy that could explain the smallness of our cosmological constant without appealing to any sort of fine-tuning. Even if surprising from an EFT perspective, it becomes natural when integrating out the infinite number of states emerging in the limit, rather than just a finite number of them \cite{Montero:2022prj}. 

Hence,  could it be that we live  in an asymptotic limit? The asymptotic runaway potentials seem natural candidates to construct quintessence \cite{Peebles:1987ek, Ratra:1987rm, Caldwell:1997ii} in which the accelerated expansion is described by a rolling scalar field. However, unfortunately, getting a small positive vacuum energy is not enough to get an accelerated expanding cosmology,  it is also necessary that the slope of the potential is not too steep in order to yield acceleration. More concretely, for exponentially decreasing potentials, one needs that the exponential rate satisfies $\gamma\leq \sqrt{2}$ in four dimensions. The difficulties to get such a small exponent have pushed people to look for models beyond the asymptotic limits. In fact, it has even been proposed that quantum gravity does not allow in general for asymptotic accelerated expansion implying the lower bound $\gamma\geq \sqrt{2}$, which is known as the Strong de Sitter conjecture \cite{Rudelius:2021azq}. However, this has only been checked in a few examples, corresponding to the canonical limits associated to weak coupling and large volume in Type IIA/B \cite{Hertzberg:2007wc,Garg:2018zdg,ValeixoBento:2020ujr,Andriot:2020lea,Andriot:2022xjh,Cicoli:2021fsd}. Despite this, \cite{Cicoli:2021fsd} also made the strong claim that accelerated expansion cannot be achieved at any asymptotic limit at parametric control in string theory, even if the evidence came from a single type of limit. In this work, we want to emphasize that there are many other types of asymptotic limits in string theory, and that much work is required before concluding with confidence that asymptotic accelerate expansion is not possible in string theory/quantum gravity. As a first stone in this direction, we will analyze different limits in the complex structure moduli space of F-theory flux Calabi-Yau compactifications and find some potential candidates that exhibit small enough values for $\gamma$. Whether these limits provide a trustworthy accelerated cosmology will depend on the stabilization mechanism for the Kähler moduli, which goes beyond the scope of this paper. Moreover, we will identify the properties that the flux scalar potential must satisfy to yield $\gamma\leq \sqrt{2}$ and show how in principle one can easily violate the supergravity no-go in \cite{Hellerman:2001yi,Rudelius:2021azq} by considering several terms of the potential that dominate asymptotically, a scenario that had not yet been considered in the string theory literature.

Getting an accelerated cosmology is not enough to have a realistic quintessence model of our universe, as there are more phenomenological aspects to consider (see e.g. \cite{Hebecker:2019csg}); but it is the first crucial step. In this work, we will only focus on whether an accelerated expansion is possible asymptotically in string theory, ignoring for the moment other phenomenological issues. This is an important question by itself that can also teach us about deep properties of quantum gravity. As remarked in \cite{Rudelius:2021azq}, it could shed light on whether quantum gravity allows for asymptotic observables in an expanding cosmology.

The outline of the paper is as follows. In Section \ref{sec:asymp st} we review the conditions for an accelerated cosmology and its interplay with Swampland bounds, as well as the asymptotic form of a flux scalar potential in 4d $\mathcal{N}=1$ EFTs and in F-theory on Calabi-Yau fourfolds. We also analyze the possible asymptotic trajectories in this setup and provide a general recipe to determine the gradient flow trajectory and the exponential rate of the potential (the de Sitter coefficient $\gamma$) along it. In Section \ref{sec:examples}, we explain how to evade the no-go for accelerated expansion in \cite{Hellerman:2001yi,Rudelius:2021azq} and identify the features that the flux potential should satisfy to yield a small $\gamma$. We then compute it explicitly in concrete F-theory examples of different limits in the complex structure moduli space and find potential candidates for accelerated expansion. In Section \ref{sec: convex hull and SDC}, we reformulate the condition for accelerated expansion as a convex hull condition that takes into account all terms of the potential and can be used to quickly determine the de Sitter coefficient along the gradient flow trajectory. Furthermore, we compare our results with Swampland bounds on the potential and check the exponential rate of the tower of states predicted by the Distance Conjecture. In Section \ref{sec:open avenues} we discuss generalizations of our results to more moduli, take into account corrections beyond the strict asymptotic limit and provide some comments on Kähler moduli stabilization. Since the paper is quite long, we provide in Section \ref{sec:summ and conc} a self-contained summary of our results and conclusions.

\section{Asymptotic structure of flux potential\label{sec:asymp st}}

\subsection{Conditions for an accelerated expanding cosmology\label{SEC: cond acc}}

Consider that the uncompactified part of our Universe is described by a $d$-dimensional FLRW spacetime, with metric $g_{\mu\nu}$ and Hubble parameter $H=\frac{a'}{a}$ (where the tilde denotes differentiation with respect to the proper cosmological time and $a$ is the scale factor), inhabited by $n$ scalar fields $\{\varphi^a\}_{a=1}^{n}$ with moduli space metric $G_{ab}$ and potential $V(\varphi)$. Then, the effective action
\begin{equation}
\label{action}
    S=\int\diff ^d x\sqrt{-g}\left\{\frac{R}{2}+\frac{1}{2}g^{\mu\nu}G_{ab}\partial_\mu \varphi^a\partial_{\nu}\varphi
    ^b- V(\varphi)\right\}
\end{equation}
yields the following Friedmann and motion equations:
\begin{subequations}
\begin{align}
    \frac{(d-1)(d-2)}{2}H^2-\frac{1}{2}G_{ab}{\varphi^a}'{\varphi^b}'-V(\varphi)&=0 \, , \label{fried}\\
    {\varphi^a}''+\tilde{\Gamma}_{bc}^a{\varphi^b}'{\varphi^c}'+(d-1)H{\varphi^a}'+\partial^{a}V(\varphi)&=0 \, , \label{eqmot}
\end{align}
\end{subequations}
in Planck units. Derivatives with respect to the field $\varphi^{a}$ are denoted as $\partial_{a}$, these indices are raised with the metric in field space $G_{ab}$, and $\tilde{\Gamma}_{bc}^a$ are the Christoffel symbols associated to that metric.

For simplicity, let us first analyse the dyanamics under the \emph{slow-roll} conditions: (i) in \eqref{fried} the potential term dominates over the kinetic one ($V(\varphi)\gtrsim \frac{1}{2}G_{ab}{\varphi^a}'{\varphi^b}'$) and (ii) in \eqref{eqmot} the friction term dominates over the second order ones ($(d-1)H{\varphi^a}'\gtrsim{\varphi^a}''+\tilde{\Gamma}_{bc}^a{\varphi^b}'{\varphi^c}'$). This then translates to 
\begin{subequations}
\begin{align}
    H^2&=\frac{2}{(d-1)(d-2)}V(\varphi)\, ,\label{fried2}\\
   {\varphi^a}'&=-\frac{1}{(d-1)H}\partial^{a}V(\varphi) \, .\label{eqmot2}
\end{align}
\end{subequations}
These will be the equations that, within this slow-roll approximation, describe the dynamics of our scalar fields. From \eqref{eqmot2}, one finds that
\begin{equation}
    \frac{1}{2}G_{ab}{\varphi^a}' {\varphi^b}'=\frac{G_{ab}}{2(d-1)^2 H^2}\partial^{a}V(\varphi)\partial^{b}V(\varphi)=\frac{d-2}{4(d-1)}\gamma^2V(\varphi) 
\end{equation}
where we have defined
\beq\label{def dS}
\gamma\equiv \frac{\|\nabla V(\varphi)\|}{V(\varphi)}
\eeq
For later convenience, we will denote $\gamma$ as the \emph{dS coefficient}, since it is the coefficient playing the leading role in the dS Swampland conjecture reviewed in Section \ref{Sec:SwamplandBounds}.
Now, to avoid kinetic energy domination, i.e. to ensure $V(\varphi) \simgeq \frac{1}{2}G_{\varphi^a \varphi^b}{\varphi^a}' {\varphi^b}'$, we need $\gamma <2\sqrt{\frac{d-1}{d-2}}$, which for $d=4$ reduces to $\sqrt{6}\approx 2.4494$. 

Notice that having a positive potential is not enough to get an accelerated cosmology. A homogeneous universe is in accelerated expansion if the deceleration parameter $q$ is negative:
\begin{equation}\label{hubble ds}
    q=-\frac{a''a}{{a' }^2}=-1-\frac{{H}'}{H^2}<0,
\end{equation}
or equivalently $\epsilon<1$, defining the slow-roll parameter $\epsilon\equiv-\frac{H'}{H^2}$. In a multi-field model the solutions to the equations of motion will result in a trajectory $\{\varphi^a(t)\}_{a=1}^n$ over the moduli space. For slow roll, this allows us to get \cite{Hetz:2016ics, Achucarro:2018vey}
\begin{equation}\label{eq:eV}
    \epsilon_V\equiv\frac{d-2}{4}\frac{\|\nabla V\|^2}{V^2}=\epsilon\left(1+\frac{\Omega^2}{(d-1)^2H^2}\right),
\end{equation}
where $\Omega=\|\nabla_t \hat{T}\|$ is the non-geodesity factor, with $\hat{T}^a=\frac{{\varphi^a}'}{\|\varphi'\|}$ being the unit tangent vector to the trajectory and $\nabla_t X^a=\partial_t X^a+\tilde{\Gamma}_{bc}^aX^b{\varphi^c}'$
. It is then clear that for accelerated expansion one needs
\begin{equation}\label{eq:full ds}
    \gamma <\frac{2}{\sqrt{d-2}}\left(1+\frac{\Omega^2}{(d-1)^2H^2}\right)^{\frac{1}{2}} \, .
\end{equation}
Note that for highly non-geodesic trajectories, i.e. $\Omega\gg 1$, it is possible to have a universe in accelerated expansion even if $\gamma$ takes high values. However, as we will later show, the kind of solutions we will obtain asymptotically correspond to geodesics, so that $\Omega= 0$ and thus $\epsilon=\epsilon_V$ and the condition for accelerated expansion becomes

\beq
\gamma<\frac{2}{\sqrt{d-2}} \, .
\label{accel}
\eeq
This reduces to $\sqrt{2}$ for $d=4$ spacetime dimensions. Interestingly, even though we have derived this using slow roll for simplicity, the same conclusion holds for the asymptotic cosmology of a single field rolling down an exponential potential (see e.g. \cite{Rudelius:2022gbz}). Since all the solutions we will discuss can be reduced to this case (see section \ref{SEC:gradflow}), this criterion for asymptotic accelerated expansion will be applicable, even if the kinetic energy and the potential energy are comparable.\footnote{Beyond the slow-roll approximation, equation \eqref{eq:eV} should be replaced by
\begin{equation}
    \epsilon_V=\epsilon\left(1+\frac{\Omega^2}{(d-1-\epsilon)^2 H^2}-\frac{\eta}{2(d-1-\epsilon)}\right) \, ,
\end{equation}
where $\eta$ is the second slow-roll parameter (see e.g. \cite{Shiu:2023fhb}). For our solutions, we will find $\Omega\to0$ and $\eta\to0$ asymptotically. Hence, at late times $\epsilon_V=\epsilon$ holds and we recover the criterion for asymptotic accelerated expansion in \eqref{accel} even beyond the slow-roll approximation.}

One could wonder whether accelerated expansion is possible far from the slow-roll regime, i.e. for $\gamma >2\sqrt{\frac{d-1}{d-2}}$, so that the kinetic energy clearly dominates over the potential. It can be shown that in general, for a universe dominated by homogeneous scalar fields and their potential, the cosmological equation of state is given by $\omega(\varphi)=\frac{p(\varphi)}{\rho(\varphi)}=\frac{\frac{1}{2}\|{\varphi}'\|^2-V(\varphi)}{\frac{1}{2}\|{\varphi}'\|^2+V(\varphi)}$. It is immediate that if the potential dominates (i.e. as in slow-roll) then $\omega (\phi)\approx -1$, while for the case when the kinetic term dominates over the potential $\omega(\phi)\approx 1$. As a universe modeled by an equation of state $\omega$ experiences accelerated expansion\footnote{From the Friedmann equations in $d$ dimensions it can be shown \cite{Chen:2014fqa} that $q=\frac{1}{2}\sum_i\Omega_i[(d-3)+(d-1)\omega_i]$, where $\Omega_i=\frac{2}{(d-1)(d-2)H^2}\rho_i$ are the density parameters.} for $\omega<-\frac{d-3}{d-1}$ ($\omega<-\frac{1}{3}$ for $d=4$), then no accelerated expansion seems possible when the kinetic term of the scalar fields dominate over the potential, specifically when $\frac{1}{2}\|\varphi'\|^2\geq \frac{1}{d-2} V(\varphi)$ (at least for the type of solutions we will study in this paper).

\subsection{Swampland bounds\label{Sec:SwamplandBounds}}

One of the most fruitful outcomes of the Swampland program would be to determine general constraints that the scalar potential of an EFT consistent with quantum gravity must satisfy. This is a hard question, but it might be feasible to provide a definite answer at least at special corners of the field space parameterized by the scalar fields $\varphi^a$. In particular, there is  increasing evidence that the potential $V(\varphi^a)$ exhibits a very universal behavior as we approach a point which is at  infinite geodesic distance measured by the field metric $G_{ab}$ in \eqref{action}. In these \emph{asymptotic limits}, the potential universally satisfies
\beq
\label{dSconjecture}
\gamma=\frac{\|\nabla V(\varphi)\|}{V(\varphi)}\geq c_{d}
\quad \text{as}\quad
D(\varphi_0,\varphi(t))=\int_{t_0}^{t}\sqrt{G_{ab}{\varphi^a}'{\varphi^b}'}\diff t\rightarrow \infty \, ,
\eeq
with $c_{d}$ some order one factor that could depend on the space-time dimension $d$. This is an asymptotic version of the dS conjecture in \cite{Obied:2018sgi} (supported in \cite{Ooguri:2018wrx}) and it is realized in all known string theory compactifications (see e.g. \cite{Grimm:2019ixq,Junghans:2018gdb,Banlaki:2018ayh,Andriot:2020lea,Andriot:2022xjh}). It is a sort of generalization of the Dine-Seiberg problem \cite{Dine:1985he} to any direction in field space, and not only the weak string coupling limit. In particular, it excludes the presence of dS vacua at parametric control, since it is precisely in these asymptotic limits where we can keep parametric control of the corrections to the EFT. Notice, though, that the condition \eqref{dSconjecture} only applies in the asymptotic limits, and does not exclude the existence of a dS vacuum in the middle of the field space, although in such case we loose partial or complete computational control.

This bound is satisfied in string theory compactifications because the potential always seems to exhibit a runaway behavior towards the infinite distance limits, such that the potential energy decreases exponentially with the geodesic field distance, with exponential rate $\gamma$. Hence, while it cannot generate a dS vacuum, it is actually a natural candidate for a quintessence scenario in which the vacuum energy slowly varies in time. Whether this is indeed feasible depends on the exact value of the dS coefficient $\gamma$ since, as we saw in the previous section, it cannot be too large if we want it to describe an accelerated expanding cosmology. 

It is imperative then to get a better estimation of the factor $c_{d}$ in \eqref{dSconjecture}, which acts as a lower bound for $\gamma$. There have been two proposals setting the value of this factor:

\begin{itemize}
\item  \emph{Trans-Planckian Censorship Conjecture (TCC)} \cite{Bedroya:2019snp}, stating that any consistent EFT should not lead to cosmological expansions where perturbations with length scale greater than the Hubble radius could be traced back to trans-Planckian scales at an earlier time.  
This implies that at the asymptotic limits, $\gamma$ should be lower bounded by
    \begin{equation}\label{TCC bound}
        c_d^{\rm TCC}=\frac{2}{\sqrt{(d-1)(d-2)}}.
    \end{equation}

\item  \emph{Strong dS Conjecture} \cite{Rudelius:2021azq}, stating that the strong energy condition should be satisfied at late times in asymptotic limits of scalar field
space, so that
\beq\label{strong bound}
c_d^{\rm strong}=\frac{2}{\sqrt{d-2}} \, ,
\eeq
which forbids asymptotic accelerated expansion of the universe (see \eqref{accel}). This is motivated by imposing consistency of the conjecture under dimensional reduction \cite{Rudelius:2021oaz}.

\end{itemize}

Clearly $c_d^{\rm TCC}<c_d^{\rm strong}$, so that the TCC still allows for an accelerated expansion cosmology, unlike the Strong dSC. Notice, though, that already in \cite{Bedroya:2019snp} it was pointed out that the TCC bound could become stronger in the case of exponentially decreasing asymptotic potentials, implying in fact \eqref{strong bound}. However, we will continue denoting \eqref{TCC bound} as the TCC value by consistency with standard notation in the literature.

In this work, we are interested in whether this second bound, i.e $\gamma\geq c_d^{\rm strong}$, is satisfied in string theory compactifications. If it universally holds, it would be an extremely strong outcome preventing any accelerated expansion (and in particular, a description of our universe) at parametric control in quantum gravity. Notice that the strength of the bound comes from requiring that it holds at \emph{any asymptotic regime} and not only at perturbative string coupling, so that it is supposed to apply beyond the lamppost of known string compactifications, including to non-supersymmetric ones that are yet to be discovered. Although there have been a few works confirming such strong claim \cite{Hertzberg:2007wc,Garg:2018zdg,ValeixoBento:2020ujr,Cicoli:2021fsd}, they are restricted to a particular asymptotic limit associated to weak string coupling and large volume in Type II string theory. This is just a small corner in a vast step of possible asymptotic limits. Hence, before claiming that asymptotic accelerated expansion is not possible in string theory, we must study other types of limits. This work is a step forward in this direction.

Of course, to get a feasible quintessence model describing our universe, one needs to satisfy further phenomenological constraints and match the value of $\gamma$ with experiments. Here, we will not delve into these phenomenological constraints, but only ask the most basic question: \emph{Does string theory allow for an asymptotic accelerated expansion of the universe?}

\subsection{Asymptotic potential in 4d $\mathcal{N}=1$ EFTs}\label{SEC as pot}

In this paper we will focus on 4d $\mathcal{N}=1$ EFTs arising from string theory, with supersymmetry broken by the positive potential energy. The scalar fields can be grouped into chiral multiplets, so that the bosonic action reads
\begin{equation}
\label{action2}
    S=\int\diff ^d x\sqrt{-g}\left\{\frac{R}{2}+\frac{1}{2}g^{\mu\nu}G_{I\bar J}\partial_\mu \Phi^I\partial_{\nu}\bar\Phi^{\bar J}- V(\phi)\right\}
\end{equation}
and the $n$ complex fields $\Phi^I$ can be split as
\beq
\label{complexfields}
\Phi^I=\phi^I+is^I \, .
\eeq
In all known 4d $\mathcal{N}=1$ string compactifications, infinite distance limits are in one-to-one correspondence with axionic shift symmetries, such that if the limit $\I\Phi^I\rightarrow \infty$ is at infinite field distance, the field metric develops an approximate continuous shift symmetry for the axion $\phi^I$ at large values of $s^I$ which becomes exact in the limit. The non-periodic scalar $s^I$ is typically denoted as the \emph{saxion}. One direction of the correspondence is clear, since global symmetries can only be restored at infinite field distance in quantum gravity, while the other direction is non-trivial and underlies the Distant Axionic String Conjecture in \cite{Lanza:2021udy}.

In this work, we will indeed consider infinite distance limits associated to axionic symmetries, so that the asymptotic limits correspond to perturbative regimes characterised by the saxionic variables as described in \cite{Lanza:2020qmt,Lanza:2021udy}. The Kähler potential can be written as
\beq\label{Kahler saxions}
K=-\log (P(s) +\dots) \, ,
\eeq
where $P(s)$ is a polynomial function of the saxions preserving the continous axionic symmetry at the perturbative level, while the dots denote non-perturbatively suppressed corrections by appropriate powers of $e^{2\pi i \Phi}$. The field metric in \eqref{action2} is given by $G_{I\bar J}=2K_{I\bar J}=2\partial_I\partial_{\bar J}K$.

The superpotential takes the form
\beq\label{eq:superpot}
W=f_A\Pi^A(\Phi)+ \mathcal{O}(e^{2\pi i \Phi})\ , \quad f_A\in \mathbb{Z} \, ,
\eeq
where the perturbative part can always be written in terms of some holomorphic functions $\Pi^A(\Phi)$ known as \emph{periods}. The integers $f_A$ have the microscopic interpretation of internal fluxes in string compactifications. Since the discrete axionic shift symmetry is a gauge symmetry, it must be preserved by the EFT. Hence, shifting the axion as $\phi^I\rightarrow \phi^I+1$ yields a monodromy transformation on the periods 
\beq\label{eq: period mono}
\Pi^A(...,\Phi^I+1,..)=(\mathcal{R}_i)_B{}^A\Pi^B(...,\Phi^I,...)
\eeq
that is compensated by a transformation of the fluxes $f_A$
\beq
f_A\rightarrow(\mathcal{R}_i^{-1}\cdot f)_A \, .
\eeq
The 4d $\mathcal{N}=1$ F-term potential 
\beq\label{cremmer}
    V=e^K\left(K^{I\bar{J}}D_I W\bar{D}_{\bar{J}}\bar{W}-3|W|^2\right),\qquad \text{with }D_IW\equiv \partial_I W+W\partial_I K
\eeq
 can be re-written as
\beq\label{V from T}
V=\frac12 f_AT^{AB}f_B=\frac12( \| \partial \cT\|^2-\frac32 \cT^2)
\eeq
with
\beq\label{T from Pi}
\cT= 2e^{K/2}|f_A\Pi^A|\ , \quad  \| \partial \cT\|^2= 2K^{I\bar J}\partial_I \cT\bar \partial_{\bar J} \cT
\eeq
and 
\beq\label{def TAB}
T^{AB}=2M_{\rm P}^4e^K\rm{Re}\left(K^{\alpha\bar{\beta}}D_\alpha\Pi^A\bar{D}_{\bar{\beta}}\bar{\Pi}^B-3\Pi^A\bar{\Pi}^B\right) \, .
\eeq
Here $\cT$ can be formally interpreted as the tension of a $\frac12$BPS membrane with quantized charges equal to the fluxes $f_A$, such that $T^{AB}$ is the inverse gauge kinetic function and  $Q^2 = 2 V$ corresponds to the physical charge \cite{Lanza:2019xxg,Lanza:2020qmt,Herraez:2020tih}.

Consider now an asymptotic limit given by $\I \Phi^i\rightarrow \infty$. By the Nilpotent Orbit Theorem \cite{schmid}, the periods can be written as
\beq\label{eq: Nilpotent Orbit Theorem}
\Pi(\Phi)=e^{\Phi^iN_i}\Pi_0(\chi)
\eeq
up to non-perturbative corrections, where $\chi$ generically denotes those scalars not sent to a limit. Here, $N_i$ are nilpotent operators defined in terms of the above monodromy transformations as $N_i=\log \mathcal{R}_i$ and satisfying
\beq
\label{nilpot}
N_{(i)}^{d_i}\Pi_0\neq 0\ , \quad N_{(i)}^{d_i+1}\Pi_0=0,\qquad \text{with }N_{(i)}=N_1+...+N_i \, .
\eeq
By replacing this into the tension $\mathcal{T}$ in \eqref{T from Pi} one gets \cite{Lanza:2020qmt}
\beq\label{expansion T}
\mathcal{T}=2e^{K/2}\left|\rho(f,\phi)\cdot\prod_{j=1}^{n}\left(\sum_{k=1}^{d_j}1\frac{1}{k!}[i(s^j-s^{j+1})]^kN_{(j)}^j\right)\Pi_0\right|
\eeq 

where $\rho=e^{-\phi^iN_i}\cdot f$ is a polynomial function of the axions and the fluxes. Note that if we want to have $\mathcal{T}\rightarrow 0$ as we send the saxions to infinity, then some fluxes from $\rho$ must be turned off, as $e^{K/2}$ is of order $\prod_{i=1}^{n}(s^i)^{d_i/2}$ while the sum goes up to $k=d_i$. While for $n=1$ this translates to summing up to $k=\lfloor\frac{d_1}{2}\rfloor$, for $n>1$ competition between terms allows for more possibilities.

Since $e^K=1/P(s)$ is also a polynomial function on the saxions at perturbative level, this implies that asymptotically the tension (and therefore, the scalar potential) can be written as a finite expansion on the saxions. It becomes convenient to keep the $\rho$ functions explicitly in the scalar potential, so that we can split the axionic and saxionic part as
\beq
\label{VZ}
V=\frac12 \rho_AZ^{AB}\rho_B \, ,
\eeq
where $Z^{AB}$ will purely depend on the saxions at the perturbative level, and can be related to $T^{AB}$ in \eqref{def TAB} by the change of basis $T(s,\phi)=e^{\phi N}Z(s)e^{-\phi N}$ \cite{Herraez:2018vae,Grimm:2019ixq}.
This bilineal structure of $V$ has been noted in a series of works analyzing the dual formulation of the flux potential in terms of the 3-form gauge fields dual to the fluxes \cite{Bielleman:2015ina,Carta:2016ynn,Herraez:2018vae,Farakos:2017jme}
and has proven to be very useful to study moduli stabilization and the structure of flux vacua \cite{Valenzuela:2016yny,Marchesano:2019hfb,Marchesano:2020uqz,Marchesano:2021gyv}.

So far, the discussion is completely general and not tied to a particular string compactification. The string theory input will come on the concrete values of $Z^{AB}$ and $N_i$, since not everything is allowed in quantum gravity. For instance, in Section \ref{sec:examples} we will consider F-theory flux compactifications where the different possibilities can be classified by using Asymptotic Hodge Theory, as we explain next.

\subsection{Scalar flux potentials on CY$_4$ through Asymptotic Hodge Theory\label{sec: AHT}}

Let us consider the 4d $\mathcal{N}=1$ low energy EFTs arising from F-theory compactified on a Calabi-Yau fourfold with $G_4$ flux. This will allow us to revisit the canonical perturbative limit in Type IIA/B on flux orientifolds, but also analyze new asymptotic limits beyond weak string coupling in a parametrically controlled way.

The flux potential can be obtained via M-theory duality \cite{Becker:1996gj,Dasgupta:1999ss,Grimm:2014xva} 
and it is given by

\begin{equation}
\label{fluxpot}
    V_M=\frac{1}{\mathcal{V}^3}\left(\int_{Y_4}G_4\wedge\star G_4-\int_{Y_4}G_4\wedge G_4\right)=\frac{1}{\mathcal{V}^3}(\|G_4\|^2-\underbrace{\langle G_4,G_4\rangle}_{A_ {\rm loc}}) \, ,
\end{equation}
where $\star$ is the Hodge star operator on $Y_4$, through which our scalar potential depends on the complex structure moduli. The $A_{\rm loc}=\langle G_4,G_4\rangle$ term, independent of the moduli, must be fixed through the tadpole cancellation condition:
\begin{equation}\label{eq:tadpole}
    \frac{A_{\rm loc}}{2}=\frac{1}{2}\langle G_4,G_4\rangle=\frac{1}{2}\int_{Y_4}G_4\wedge G_4=\frac{\chi(Y_4)}{4!} \, ,
\end{equation}
where $\chi(Y_4)=\int_{Y_4}c_4(Y_4)$ is the Euler characteristic of the Calabi-Yau $Y_4$. This in turn imposes conditions on the quantized fluxes. In particular, it implies that the fluxes entering in the tadpole cannot take parametrically large values, as we will impose in the rest of the paper. Beyond this, we will assume that there is always some choice of fluxes satisfying the above tadpole constraint.

The scalar manifold can be split between the Kähler and the complex structure moduli spaces. Given a Calabi-Yau 4-fold $Y_4$ \cite{Haack:2001jz}, there are $h^{1,1}(Y_4)$ Kähler moduli arising from dimensionally reducing the complexified K\"{a}hler form $J$ and  $n=h^{3,1}(Y_4)$ complex moduli $z^i$ associated to the unique holomorphic $(4,0)$-form $\Omega$ of $Y_4$. In this paper, we are interested in the flux potential on the complex structure moduli space, so we will restrict to primitive fluxes $G_4\in H^4_p(Y_4,\mathbb{R})$. The flux potential \eqref{fluxpot} can then be equivalently derived from the following Kähler potential and superpotential \cite{Haack:2001jz},
\beq
  K=-\ln\left[\int_{Y_4}\Omega\wedge\bar{\Omega}\right]-3\log\mathcal{V}\ ,
\quad 
    W_{\rm cs}=\int_{Y_4}G_4\wedge \Omega,  \label{W G OMEGA}
\eeq
through \eqref{cremmer}. Since, at this level, the superpotential does not depend on the Kähler moduli, the no-scale condition cancels the $-3|W|^2$ term resulting in a positive definite scalar potential $V_M=e^KG^{i\bar{\jmath}}D_iW\overline{D_{j}W}$.

We are going to consider asymptotic limits approaching an infinite distance singularity in the complex structure moduli space, while keeping the overall volume $\mathcal{V}$ fixed. Around a degeneration loci $\{z^1=...=z^{n}=0\}$ of the moduli space, we introduce new coordinates $\Phi^j=\frac{1}{2\pi i}\log z^j$, separating their real (axion) and imaginary parts (saxions) $\Phi^j=\phi^j+ is^j$ as in \eqref{complexfields}. This way the $z^j\rightarrow 0$ limit corresponds to $\Phi^j\rightarrow i\infty$, or equivalently $s^j\rightarrow \infty$, with $\phi^j$ taking arbitrary (finite) values.

It is convenient to divide the asymptotic regions of the complex structure moduli space $\mathcal{M}_{\rm{cs}}$ into \emph{growth sectors} \cite{Grimm:2019ixq} depending on the growth rate of the saxions:
\begin{equation}\label{growth def}
    \mathcal{R}_{i_1i_2...i_{n}}=\left\{t^j=\phi^j+is^j:\frac{s^{i_1}}{s^{i_2}}>\gamma,..., \frac{s^{i_{n-1}}}{s^{i_{n}}}>\gamma,s^{i_{n}}>\gamma,\phi^j<\delta\right\},
\end{equation}
for some arbitrary $\gamma,\delta>0$. The set of all growth sectors $\{\mathcal{R}_{\sigma(1)\sigma(2)...\sigma(n)}\}_{\sigma\in\mathcal{S}_{n}}$ covers the whole space of possible asymptotic trajectories sending the saxionic fields to infinity while leaving axions finite. We will later show that in general the axions can be either stabilized or not come into play in the scalar potential $V_M$, being prevented through Hubble damping of reaching arbitrary high values that could bring us away from these growth sectors.

In the same spirit as in Section \ref{SEC as pot}, we rewrite \eqref{W G OMEGA} as $W_{\rm{cs}}=f_A\Pi^A$, with the quantized fluxes $f_A$ coming from the 4-form flux $G_4$ and the period vectors $\Pi^A$ from integrating $\Omega$ along an integral basis of $H_4(Y_4,\mathbb{R})$ . These periods undergo the monodromy transformation in \eqref{eq: period mono} and can be well approximated by the nilpotent orbit as in \eqref{eq: Nilpotent Orbit Theorem}.
Given some specific growth sector \eqref{growth def}, the monodromy matrices $\{N_i\}_{i=1}^{n}$ and the $\Pi_0$ vector, one can always define a set of $n$ commuting $\mathfrak{sl}(2,\C)$ triples $\{(N_i^+,N_i^-,Y_i)\}_{i=1}^{n}$ \cite{cattani_kaplan_schmid_1986}, with the standard commutation relations\footnote{This is,
$
        [Y_j,N_k]=-2N_j\delta_{jk},\qquad
        [N_j^\pm,N_k]=[N_j^+,N_k^-]=Y_j\delta_{jk},\qquad
        [Y_j,N_k^\pm]=\pm2N_j^\pm\delta_{jk}
$}. Defining now the $Y_{(i)}=Y_1+...+Y_i$ matrices, one can split the primitive cohomology group into orthogonal subspaces as
\begin{equation}\label{lattice splitting}
    H^4_{\rm p}(Y_4,\R)=\bigoplus_{\bm{l}\in\mathcal{E}}V_{\bm{l}} \, ,
\end{equation}
where $\mathcal{E}\subseteq\{0,...,8\}^{n}$, with $V_{\bm{l}}$ defined through
\begin{equation}
    v_{\bm{l}}\in V_{\bm{l}}\Longleftrightarrow Y_{(i)}v_{\bm{l}}=(l_i-4)v_{\bm{l}},\qquad\text{ for }i=1,...,n \, .
\end{equation}
Given $\bm{8}=(8,...,8)$, one has that $\dim V_{\bm{l}}=\dim V_{\bm{8}-\bm{l}}$, so that $V_{\bm{l}}\simeq V_{\bm{8}-\bm{l}}$, and that $\langle V_{\bm{l}},V_{\bm{l}'}\rangle=0$ unless $\bm{l}+\bm{l}'=\bm{8}$. Hence, given a flux $G_4\in  H^4_{\rm{p}}(Y_4,\R)$ we can decompose it as
\begin{equation}
    G_4=\sum_{\bm{l}\in\mathcal{E}}G_4^{\bm{l}},\qquad\text{with }G_4^{\bm{l}}\in V_{\bm{l}} \, .
\end{equation}
The goal is to provide the asymptotic form of the Hodge norm $\|G_4\|^2$ (and thus of the flux potential through \eqref{fluxpot}) in the limit $s^I\rightarrow \infty$. 
As described in \cite{Grimm:2019ixq}, one can introduce a limiting Hodge norm    $ \|v\|^2_\infty$ which is defined using only the structure at the limiting locus and it is independent of the coordinates $s^I$. The vector spaces are orthogonal with respect to this inner product, so that the limiting flux norm satisfies the following direct sum decomposition:
\begin{equation}\label{eq:dir sum norm}
    \|G_4\|^2_\infty=\sum_{\bm{l}\in\mathcal{E}}\|G_4^{\bm{l}}\|_\infty^2 \, .
\end{equation}
When moving a bit away from the degeneration loci, we can use the sl(2)-orbit theorem to provide the leading dependence of the Hodge norm on the saxions,
\begin{equation}
    \|G_4\|^2\sim\sum_{\bm{l}\in\mathcal{E}}\left(\frac{s^1}{s^2}\right)^{l_1-4}...\left(\frac{s^{n-1}}{s^{n}}\right)^{l_{n-1}-4}(s^{n})^{l_{n}-4}\|\rho_{\bm{l}}(G_4,\phi)\|^2_{\infty}=\sum_{\bm{l}\in\mathcal{E}}\prod_{j=1}^{n}(s^j)^{\Delta l_i}\|\rho_{\bm{l}}(G_4,\phi)\|^2_{\infty},
\end{equation}
where we introduce $\Delta l_i=l_i-l_{i-1}$ (with $l_0=4$). The $\rho_{\bm{l}}$ functions are defined as in \eqref{expansion T} by
\begin{equation}
    \rho(G_4,\phi)=e^{-\phi^iN_i}G_4=\sum_{\bm{l}\in\mathcal{E}}\rho_{\bm{l}}(G_4,\phi) \, .
\end{equation}
It is then easy to see that the Hodge norm $ \|G_4\|^2$ takes the form 
\begin{equation}
     \|G_4\|^2=\sum_{\bm{l}\in\mathcal{E}}Z^{\bm{l}}(s)\|\rho_{\bm{l}}(G_4,\phi)\|^2_{\infty}=\sum_{\bm{l}\in\mathcal{E}}\varrho_{\bm{l}}(\phi)^\intercal Z^{\bm{l}}(s)\varrho_{\bm{l}}(\phi)
\end{equation}
where, thanks to asymptotic Hodge theory, we can now provide the leading behavior of the metric $Z(s)$ in \eqref{VZ}. Using \eqref{eq:dir sum norm}, this is given by\footnote{In general, the $V_l$ subspaces can be higher dimensional, so one should expand $\rho(\phi)=v^{\bm{l}}_{i}\varrho_{\bm{l}}^{i}(\phi)$ in a suitable basis $\{v^{\bm{l}}_{1},\dots,v^{\bm{l}}_{\text{dim}V_l}\}$ for $V_l$. In that case, one gets $Z^{\bm{l}}(s)\|\rho_{\bm{l}}(G_4,\phi)\|_\infty^2=\sum_{ij}Z_{ij}^{\bm{l}}(s)\varrho_{\bm{l}}^{i}(\phi)\varrho_{\bm{l}}^{j}(\phi)$ with $ Z_{ij}^{\bm{l}}(s)=\prod_{k=1}^{n}(s^{k})^{\Delta l_k}\mathcal{K}^{\bm{l}}_{ij}$, and $\mathcal{K}^{\bm{l}}$ symmetric, definite positive, constant Gram matrices given by the limiting Hodge product $\mathcal{K}^{\bm{l}}_{ij}= \langle \star_\infty v_i^l, v_j^l\rangle$.}
\beq
  Z^{\bm{l}}(s)=\prod_{k=1}^{n}(s^{k})^{\Delta l_k}\mathcal{K}^{\bm{l}},
\eeq
 
 which is a positive-definite matrix in the strict asymptotic regime, as $\mathcal{K}^{\bm{l}}$ are positive numbers. Notice that if we move further away from the asymptotic regime, the orthogonality that allows for a direct sum in \eqref{eq:dir sum norm} gets broken and one needs to consider non-diagonal terms that are suppressed by powers $\mathcal{O}\left(\frac{s^i}{s^j}\right)$ with $i>j$ in a given growth sector \eqref{growth def}. However, since we are interested in the asymptotic trajectories approaching the infinite distance limit, it will be enough for our purposes to stay in the strict asymptotic regime.

On the other hand $\varrho_{\bm{l}}^{i}(\phi)$ are polynomials of the axions and the internal fluxes. It can be shown that
    \begin{equation}\label{rho prop}
        \partial_{\phi^j}\varrho_{\bm{l}}^{i}(\phi)=-\varrho_{\bm{l}'}^{i}(\phi),\qquad\text{with }\bm{l}'=(l_1,...,l_{j-1},l_j+2,l_{j+1}+2,...,l_{n}+2),
    \end{equation}
    so that, as $l_j\in\{1,...,8\}$, we have that those $\rho_{\bm{l}}$ with higher $l_j$ will have smaller degree with respect to $\phi^j$. 
    This will be of great use when discussing axion stabilization in Section \ref{brief stab ax}.

Plugging everything together, we arrive to the following expression for the asymptotic scalar potential $V_M$ in a given growth sector:
\begin{align} \label{eq:asymptotic-potential}
    V_M
    &=\frac{1}{\mathcal{V}_0^3}\left(\sum_{\bm{l}\in\mathcal{E}}\prod_{j=1}^{n}(s^j)^{\Delta l_j}\|\rho_{\bm{l}}(G_4,\phi_0)\|^2_{\infty}-A_{\rm loc}\right)\equiv\frac{1}{\mathcal{V}_0^3}\left(\sum_{\bm{l}\in\mathcal{E}} A_{\bm{l}} \prod_{j=1}^{n}(s^j)^{\Delta l_j}\right) \, .
\end{align}
 We have denoted $A_{\bm{l}}=\|\rho_{\bm{l}}(G_4,\phi_0)\|^2_{\infty}$ and $A_{\bm{4}}-A_{\rm loc}\rightarrow A_{\bm{4}}$ to shorten notation. Note that apart from $A_{\bm{4}}$, all the other $A_{\bm{l}}$ axionic prefactors are non-negative.

Finally, the K\"{a}hler potential  in the strict asymptotic regime for a  given growth sector $\mathcal{R}_{1...n}$ reads \cite{Grimm:2019ixq}
\begin{equation}\label{asymp Kahler}
    K_{\rm cs}\sim -\log\left[\left(\frac{s^1}{s^2}\right)^{d_1}...\left(\frac{s^{n-1}}{s^{n}}\right)^{d_{n-1}}(s^{n})^{d_{n}}f(\zeta,\bar{\zeta})+\dots\right]=-\sum_{j=1}^{n}\Delta d_j\log s^j-\log f(\zeta,\bar{\zeta})+\dots,
\end{equation}
where $\Delta d_i=d_i-d_{i-1}$ (with $d_0=0$) and $d_i$ is the nilpotency order of the monodromies $N_{(i)}$ (see \eqref{nilpot}). Here, $f(\zeta,\bar{\zeta})$ involves the saxions remaining finite, so it  can be neglected asymptotically. From this, it is immediate to obtain the asymptotic field metric
\begin{equation} \label{eq:field-metric}
    G=2 \partial_{I}\partial_{\bar{J}} K_{\rm cs}=\diag\left(\frac{\Delta d_{j}}{2 (s^j)^2}\right)_{i=1}^{n}\, ,
\end{equation}
to leading order. It is important to notice that whenever $\Delta d_i=0$ (which e.g. will always occur  for $n\geq 4$ as $d_i\leq 4$), the leading term vanishes and one needs to consider subleading corrections as explained in detail in \cite{Bastian:2020egp,Bastian:2021eom}. This in particular spoils the diagonal character of \eqref{eq:field-metric}. A toy model example of working with this non-diagonal matrices featuring subleading terms appears in Appendix \ref{toy model}.\\

\subsection{Gradient flow trajectory \label{SEC:gradflow}}
We are interested in analyzing the physics along the asymptotic trajectories to which the potential sends the scalars to infinite distance in the field space. These asymptotic trajectories should solve the equations of motion including the potential, and should be able to approach an infinite distance boundary of the field space without being obstructed by the potential, meaning that the potential must go to zero asymptotically. This way, they correspond to the infinite distance limits that can be described by the EFT with a finite cut-off.

Intuitively, these trajectories will asymptotically approach a gradient flow trajectory where the shape of the potential (and the field metric) runs the show. Parameterizing the trajectory as $s^j(\lambda)$ such that $\lambda\rightarrow \infty$ is at infinite distance, the gradient flow condition is given by
\begin{equation}\label{grad flow}
    \dot{s}^{j} = - \mathcal{F}(\lambda) \, \partial^{j}V_{M}(\lambda) \, ,
\end{equation}
where\footnote{As differential operators, we will distinguish the covariant $\nabla=(\partial_1,...,\partial_n)$ from the contravariant $\vec{\partial}=(\partial^1,...,\partial^n)$. Note that $\|\nabla f\|=\|\vec{\partial}f\|$ for any function $f$.} $\partial^{j}V_{M}=G^{jk} \partial_{k}V_{M}$ is evaluated along the trajectory itself. $\mathcal{F}(\lambda)$ is some smooth, positive function and $\dot{s}^j$ represents differentiation of $s^j$ with respect to $\lambda$. This parameter does not necessarily have a physical meaning, with $\mathcal{F}$ accounting for this reparametrization freedom. One can check that this trajectory satisfies the equations of motion in the slow-roll approximation, i.e. when the friction term dominates over the second order ones, such that the equation of motion \ref{eqmot} reduces to
\beq
   {\varphi^a}'=-\frac{1}{(d-1)H}\partial^{a}V(\varphi) \, . 
   \eeq
This is equivalent to \eqref{grad flow} after a change of variables between $\lambda$ and the cosmological time $t$, so that
\begin{equation}\label{Fcal}
    \mathcal{F}(\lambda)\equiv\frac{1}{(d-1)H}\frac{\diff t}{\diff \lambda} \, .
\end{equation}
Knowing the expression of $\mathcal{F}$ (which is immediate once the solution $\vec{f}$ is obtained) and using \eqref{fried2}, then it is easy to see that if we remain in the slow-roll regime, the relation between $\lambda$ and the cosmological time is given by
\begin{equation}\label{time lambda}
    t-t_0=\int_{\lambda_0}^{\lambda}\sqrt{\frac{2(d-1)}{d-2}V(\vec{f}(\tau))}{\mathcal{F}(\tau)}{\diff\tau} \, .
\end{equation}
Now, when parameterizing in term of the traveled distance $D=\int_{0}^\lambda\sqrt{G_{ij}\dot{\Phi}^i(\tau)\dot{\Phi}^j(\tau)}\diff \tau$, from \eqref{def dS} and \eqref{grad flow} we have $V(D)\sim V_0e^{-\gamma D}$ and $\mathcal{F}(D)=\|\nabla V\|^{-1}$, so that\footnote{Notice that as 
$$
\frac{1}{2}G_{ab}\varphi'^a\varphi'^b\sim\left(\frac{\partial D}{\partial{t}}\right)^2\sim e^{-\gamma D},
$$
both the potential and the kinetic energy of the fields decay at the same rate, so that it is not possible to go from $V\gg \frac{1}{2}G_{ab}\varphi'^a\varphi'^b$ at initial times to parametrically $V\ll \frac{1}{2}G_{ab}\varphi'^a\varphi'^b$ at asymptotic limits.} $t\sim e^{\frac{\gamma}{2}D}$. This way the $D \rightarrow \infty$ limit is not reached in finite time for the studied type of potentials (that asymptotically go to zero). As we will see, for some of our goals, such as relating the obtained results with the Swampland Distance Conjecture, this parametrization with respect to the geodesic distance $D$ will be very convenient. We refer to Sections \ref{sec:distance conj} and \ref{sec: Convex hull} for more on this.\\

One could wonder if the above gradient flow trajectory still solves asymptotically the equations of motion  when we are no longer in the slow-roll regime. When $\frac{1}{2}G_{ab}{\phi^a}'{\phi^b}'\simgeq V(\phi)$ the kinetic terms might become important in our equations, so that $H\sim \|\phi'\|$ and ${\varphi^a}''+\tilde{\Gamma}_{bc}^a{\varphi^b}'{\varphi^c}'$ can now be of the order of the damping term $H{\phi^a}'$. If the potential gradient $\vec{\partial} V$ is not strong enough, then the scalar fields might no longer follow the gradient flow and rather be described by the geodesics equations with a damping external force. This makes the resolution of the equations of motion much more complicated in general. Hence, from now on, we restrict ourselves to gradient flow trajectories. In the following we show that these gradient flow trajectories are guaranteed to solve the full equations of motion (even beyond the slow roll approximation) if they are geodesics.

Following \cite{Achucarro:2018vey}, we define the unit tangent and normal vectors to the trajectory as $\hat{T}^{i} = \frac{\dot{s}^{i}}{\|\dot{s}\|}$ and $\hat{N}^i=-\frac{\nabla_t\hat{T}^i}{\|\nabla_t\hat{T}\|}$, respectively. This way the gradient of the potential can be split as
    \begin{equation}\label{grad decomposition}
        \partial^a V=\hat{T}^a V_D+\hat{N}^a V_N,\quad\text{with }\left\{\begin{array}{l}
             V_D=\hat{T}^a\partial_a V\\
             V_N=\hat{N}^a\partial_a V
        \end{array}
        \right.
    \end{equation}
    One can now project the equations of motion \eqref{eqmot} on the tangent and normal directions to show that in terms of the traveled distance $D$ \cite{Achucarro:2018vey}
    \begin{align}
        D''+3HD'+V_D&=0 \, , \label{eqMotD}\\
        \Omega&=\frac{V_N}{D'} \, . \label{eqMotN}
    \end{align}
    From \eqref{grad decomposition} one obtains that a set of fields following a gradient flow obeys $V_N=0$. This translates through \eqref{eqMotN} to $\Omega=\|\nabla_t\hat{T}\|=0$, in turn corresponding to a geodesic trajectory. This means that for solutions to the equations of motion, gradient flow trajectories must correspond to geodesics and vice versa. As it will be later shown for a diagonal metric and Appendix \ref{toy model} for a non-diagonal one, our gradient flow solutions indeed fulfill the geodesic equation.

As we will show now, for the type of scalar potentials we are interested in (this is, power-like in the saxions), these geodesic/gradient flow trajectories are asymptotic trajectories that solve the equations of motion at late times. Recall that geodesic trajectories for the field metric are purely saxionic, as this is obtained from the K\"{a}hler potential from \eqref{Kahler saxions}, asymptotically depending only on $\{s^i\}_{i=1}^{n}$. Hence, in order to have an asymptotic non-geodesic trajectory, one would need that the axions (and not only the saxions) take parametrically large field values. However, as we will see in Subsection \ref{brief stab ax} (and in more detail in Appendix \ref{stab ax}), the axions will always be either eventually stabilized or not come into play in the scalar potential, independently of the asymptotic gradient flow trajectory followed by the saxions\footnote{See however \cite{Buratti:2018xjt} for a non-gradient flow solution in String Theory in which an axion approaches a parametrically large distance in a space-like dependent configuration.}. Therefore, it will be enough for our purposes to study the saxionic trajectories that solve the gradient flow equation. We devote the rest of the section to explain, in general, how to find this solution explicitly. This reduces the setup to a single field cosmology running down an exponential potential, for which the bound \eqref{accel} on asymptotic accelerated expansion holds (see e.g. \cite{Rudelius:2022gbz}). 

From the discussion in the previous subsections, we can consider a scalar potential with a power-like dependence on the saxions, assuming that the axions have already been stabilized or do not enter the potential:
\begin{equation}\label{eq:pot}
    V=\sum_{\bm{l}\in\mathcal{E}}A_{\bm{l}}\prod_{k=1}^n(s^k)^{l_k},
\end{equation}
with $\mathcal{E}\subseteq\Z^n$ and $A_{\bm{l}}\in \R$. When particularizing to the potentials allowed by asymptotic Hodge Theory (see \eqref{eq:asymptotic-potential}), we must replace  $ l_i\rightarrow \Delta l_i$, which get very constrained and  characterize the type of asymptotic limit. In order for this potential to be admissible, we need that asymptotically $V\geq 0$ and that $V\rightarrow 0$ along at least one trajectory towards infinity. 

On power-like potentials such as \eqref{eq:pot} we can distinguish the following three kinds of
terms \cite{Grimm:2019ixq}:
\begin{itemize}
    \item $\mathcal{E}^{\rm heavy}$, consisting in those terms which blow up for any trajectory going to infinity, such as $A_{21}(s^1)^2(s^2)$. As we ask our potentials to asymptotically go to zero, these kind of terms need to be canceled, for which some of the quantized fluxes might need to be set to 0.
    \item $\mathcal{E}^{\rm light}$, which consists in those terms which go to zero for any trajectory going to infinity (some saxions can be left finite, but different from 0), such as $\frac{A_{-1,-4}}{s^1(s^2)^4}$. While these terms are allowed in our potentials, they will not give way to interesting results.
    \item $\mathcal{E}^{\rm rest}$, comprising those terms which blow up or go to zero depending on the trajectory we take. For example, take $A_{-1,2}\frac{(s^2)^2}{s^1}$, which goes to zero for $s^1\gg (s^2)^2$ but blows up for $s^1\ll (s^2)^2$. Note that in order for the saxions not to be sent to 0, we will need that several terms of $\mathcal{E}^{\rm rest}$ are present when dominant. It is through this competition between terms that we will obtain ``valleys'' or ``attractors'' along which the saxions go to infinity at a lower rate than if they were pushed by a single $\mathcal{E}^{\rm light}$ term, providing possible counterexamples for the dS conjecture bounds.
\end{itemize}

As we will be using power-like potentials $V$ and we expect the complex-structure moduli space metrig $G$ to be also power-like, solving \eqref{grad flow} will result in power-like trajectories, which asymptotically will be dominated by their leading term, so that asymptotically, the gradient flow will always take the form
\begin{equation} \label{eq:saxions-ansatz}
	s^{i}(\lambda) =f^i(\lambda)= \alpha^{i} \lambda^{\beta^{i}} \, ,
\end{equation}
for some constants $\alpha^{i}>0$ and $\beta^{i}\in\R$. One can check that these trajectories are geodesics of the metric in \eqref{eq:met}. 
In the following, we provide a general recipe to derive $\alpha^i$ and $\beta^i$, which will be translated into a graphical convex hull condition in Section \ref{sec: Convex hull}.

While \eqref{grad flow} can be solved for any kind of metric and all of the general results hold for them, for simplicity we will consider the following complex structure K\"{a}hler potentials and the associated diagonal field space metrics,
\begin{equation}\label{eq:met}
    K=-\log\prod_{i=1}^n(s^i)^{d_i},\qquad G=\diag\left(\frac{d_1}{2(s^1)^2},...,\frac{d_n}{2(s^n)^2}\right) \, ,
\end{equation}
as obtained in \eqref{asymp Kahler} in the strict asymptotic regime upon replacing $\Delta d_i\rightarrow d_i$ for simplicity in this section. All the specific examples analyzed in Section \ref{sec:examples} will exhibit this type of diagonal metrics. By taking derivatives with respect to the saxions in \eqref{eq:pot} and using the metric \eqref{eq:met} to raise the index we particularize the gradient flow equations \eqref{grad flow}. By dividing those for two different saxions $s^{i}$ and $s^{j}$ with $i\neq j$ we obtain:
\begin{equation} \label{eq:saxionic-flow}
	\frac{\partial_{\lambda}\log s^{i}}{\partial_{\lambda}\log s^{j}} = \frac{d_{j}}{  d_{i}} \frac{\sum_{\bm{l}\in\mathcal{E}}   l_{i} A_{\bm{l}} \prod_{k=1}^{n}(s^k)^{  l_k}}{\sum_{\bm{l}\in\mathcal{E}}   l_{j} A_{\bm{l}} \prod_{k=1}^{n}(s^k)^{  l_k}} \, .
\end{equation}
From this expression we can already realize something interesting: Both the numerator and denominator in the rhs are linear combinations of the same monomials in the saxions. As a consequence, this part of the equation goes to a constant for any limit at infinite distance that the gradient flow may explore. It is then evident that the solutions will take the form in \eqref{eq:saxions-ansatz}.

Up to normalization (which can be absorbed by $\lambda$) $\vec{\beta}$ will be fixed by the asymptotic behavior of the rhs in \eqref{eq:saxionic-flow}, which converges to different values depending on the monomials dominating the expression, in turn determined by $\vec{\beta}$ through the followed trajectory.
In practice,  taking $\beta^{i}$ as parameterizing all the admissible asymptotic trajectories, we should pick the one maximizing the gradient of the potential along the corresponding direction, since the gradient flow should be driven towards the steepest descendant trajectory. Although this seems intuitive, let us derive it more carefully to express the result in the form of a concrete optimization problem.

Taking into account that, since $\log$ is a monotonously growing function, $V$ and $\log V$ have the same gradient flow trajectories, we project $\vec{\partial} \log V$ on the unit tangent vector $\hat T$ to some certain trajectory of the  \eqref{eq:saxions-ansatz} kind, so that for a potential in \eqref{eq:asymptotic-potential} and metric in \eqref{eq:field-metric} we obtain
\begin{equation} \label{eq:dlogV}
	\hat{T}\cdot\vec{\partial} \log V=\hat{T}^{i}\partial_{i}\log V = \frac{\sum_{\bm{l}\in\mathcal{E}} A_{\bm{l}} \left( \prod_{k=1}^{n}(\alpha^k)^{  l_k}\right) (\hat{\beta}^{i}   l_{i}) \lambda^{\beta^{j}   l_{j}} }{\sum_{\bm{l}\in\mathcal{E}} A_{\bm{l}} \left( \prod_{k=1}^{n}(\alpha^k)^{  l_k}\right) \lambda^{\beta^{j}   l_{j}} } \, ,
\end{equation}
where we have defined $\hat{\beta}^{i} = \frac{\beta^{i}}{|\vec{\beta}|}$, with $|\vec{\beta}|$ being the norm of the vector $\vec\beta$ taken with the metric that is induced by the field space, such that $|\vec{\beta}|=\|\dot{\vec{f}}(\lambda=1)\|=\|\dot{\vec{f}}(\lambda)\|\lambda$ (constant), with $\vec{f}(\lambda)$ being the trajectory followed by the saxions, \eqref{eq:saxions-ansatz}. For instance, for the diagonal metric \eqref{eq:met}, it reads
\begin{equation}\label{norm beta}
	|\vec{\beta}| = \sqrt{\frac{1}{2}\sum_{i=1}^{n} d_i(\beta^i)^2} \, .
\end{equation}

Most importantly, notice that for $\lambda\to\infty$ the rhs of \eqref{eq:dlogV} drastically simplifies. The sum in both numerator and denominator will be dominated by the terms maximizing the exponent of $\lambda$. Since it is this exponent what appears in the numerator, and it has to be the same for all the terms that dominate, we can take it out of the sum. We then see that all the dependence on $A_{\bm{l}}$ and $\alpha^{i}$ goes away and we end up with the simple result
\begin{equation}
	\hat{T}^{i}\partial_{i}\log V = \max_{\bm{l}\in\mathcal{E}} \left\{ \hat{\beta}^{i} l_{i} \right\} \, .
\end{equation}

Therefore, as expected, the solution to the gradient flow equations will be obtained as the $\hat{\beta}$ giving the steepest descent direction, which is equivalent to maximizing $-\hat{T}^{i}\partial_{i}\log V$ (or minimizing $\hat{T}^{i}\partial_{i}\log V$). This is, the one corresponding to the following optimization problem:\footnote{Although in the way the optimization problem is written it could seem as a continuous problem, as it will be shown in Section \ref{sec: Convex hull}, it is actually a discrete problem. We will also argue in Appendix \ref{APP1} that the solution to this optimization problem is unique, thanks to the absence of tachyons in the asymptotic regime of flux compactifications.}
\begin{equation} \label{eq:min-method}
	\min_{\hat{\beta} \in \mathbb{S}^{n}} \left\{ \max_{\bm{l}\in\mathcal{E}} \left\{ \hat{\beta}^{i} l_{i} \right\} \right\} \quad \text{with} \quad \mathbb{S}^{n} = \left\{\hat{\beta}\in\mathbb{R}^{n}: |\hat{\beta}|^{2} = \frac{1}{2}\sum_{i=1}^{n} d_i(\hat{\beta}^i)^2=1 \right\} \, .
\end{equation} 
In this setup (diagonal hyperbolic metric), the $\mathbb{S}^{n}$ region consists in a ellipsoid. However, other metrics in field space can induce more complicated metrics on the space of $\vec{\beta}$ (again, see e.g. \ref{app met} in Appendix \ref{toy model}). There exists the possibility that the obtained trajectory sends some of the saxions to zero or to a finite minimum. In this case the obtained solution would not be valid, as we cannot trust the above form of the potential outside the asymptotic region. One must then check that for the obtained solution $\hat{\beta}^i\geq 0$ and  $\hat{\beta}^{i}  l_{i}^{\rm dom}=\max_{\bm{l}\in\mathcal{E}} \left\{ \hat{\beta}^{i} l_{i} \right\}\leq 0$, so that the potential sends the saxions to infinity and decays along said trajectory.

Analogously, one must check that the obtained solution stays in the growth sector \eqref{growth def}  in which our potential is defined, which in terms of $\hat{\beta}$ implies the additional restriction
\begin{equation}\label{check growth sector}
    \hat{\beta}^{i_1} \geq \hat{\beta}^{i_2} \geq \cdots \geq \hat{\beta}^{i_{n}}\geq 0 \, .
\end{equation}

On the other hand, although for the purposes of this paper we will only be interested in the value of $\vec{\beta}$, once this is known one can easily obtain the value of $\vec{\alpha}$, thus completely determining the saxion trajectory. Given some trajectory $\vec{\beta}$, one can identify the dominant terms $ \mathcal{E}^{(0)}\subseteq\mathcal{E}$ by being those such that $\hat{\beta}^i l_i=\max_{\bm{l}\in\mathcal{E}} \left\{ \hat{\beta}^{i} l_{i} \right\}$. Then, asymptotically \eqref{eq:saxionic-flow} reduces to
\begin{equation}\label{obt alpha}
    \frac{\beta^i}{\beta^j} = \frac{d_{j}}{  d_{i}} \frac{\sum_{\bm{l}\in \mathcal{E}^{(0)}}   l_{i} A_{\bm{l}} \prod_{k=1}^{{n}}(\alpha^k)^{  l_k}}{\sum_{\bm{l}\in \mathcal{E}^{(0)}}   l_{j} A_{\bm{l}} \prod_{k=1}^{{n}}(\alpha^k)^{  l_k}} \, ,
\end{equation}
where now $\{\alpha^k\}_{k=1}^n$ are the unknowns, with the rest of variables being known parameters. Notice that we have $n-1$ equations, but a constant overall factor can be absorbed into $\lambda$, resulting only in $n-1$ needed values for the $\alpha^k$. For an admissible solution and positive $A_{\bm{l}}$ for $\bm{l}\in \mathcal{E}^{(0)}$ there will always exist a solution for the above system of equations. This solution can be either unique (the potential selects a unique trajectory towards infinity) or infinite (there exists a continuous family of solutions, all with the same $\vec{\beta}$ behavior, going to the same singularity, as this is determined by $\hat{\beta}^i$).
These two possibilities give rise to the two types of gradient flows that will be discussed next and play a central role in the paper.

\subsection{de Sitter coefficient}

There are two types of trajectories that can solve the gradient flow condition asymptotically, depending on the specific flux potential under consideration. The distinction between these two possible scenarios is essential for our work, since only one of them can violate the strong dS bound \eqref{strong bound} and potentially lead to asymptotic accelerated expansion.

\begin{figure}[htbp]
\centering
\subfigure[Case (I): $V_M=\frac{1}{su}$]{\includegraphics[width=0.45\textwidth]{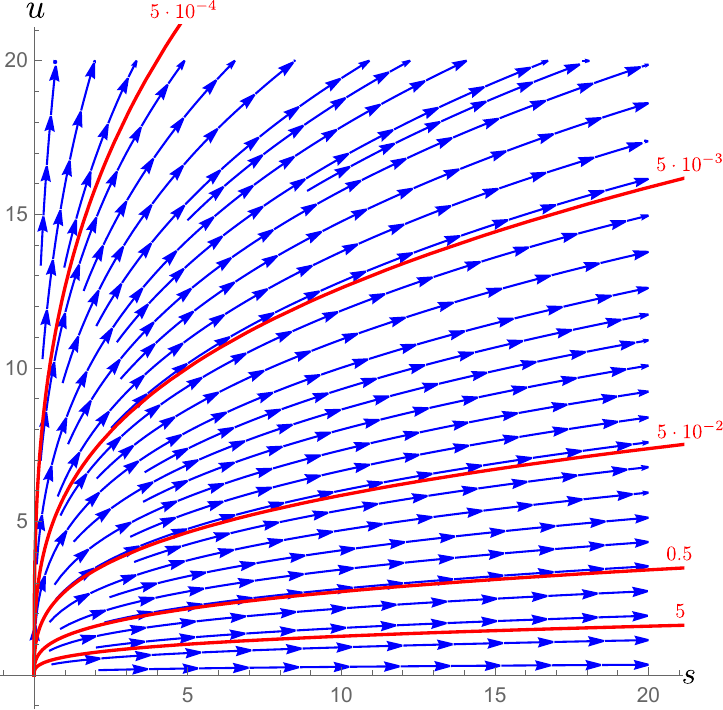}}
\subfigure[Case (II): $V_M=\frac{u}{s}+100\frac{s}{u^3}$]{\includegraphics[width=0.45\textwidth]{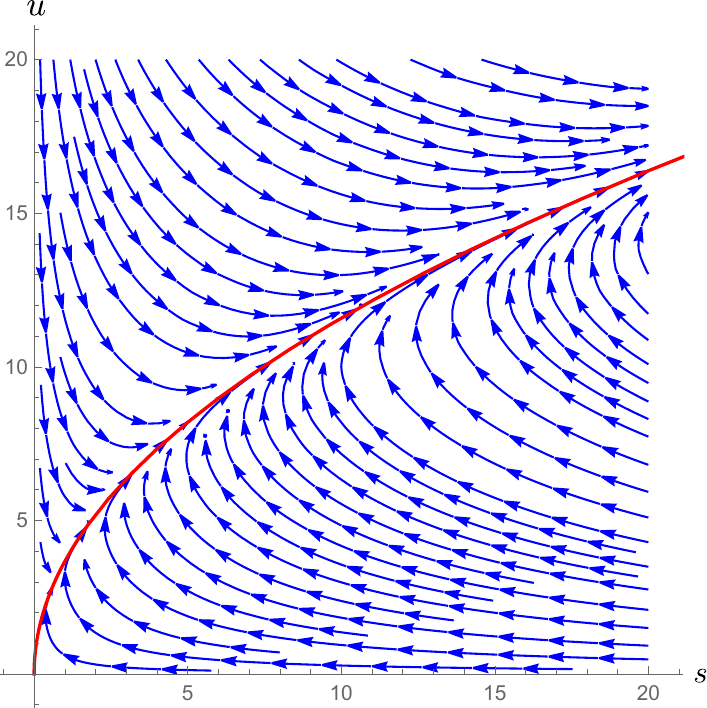}}
\caption{Examples of the two possible kinds of admissible potentials, corresponding to the examples shown in subsection \ref{sec:typeIIB}. In (a) there is a single dominant term, corresponding to $\vec{f}(\lambda)=(\alpha\lambda^3,\lambda)$ solutions, with different $\alpha>0$ solutions plotted. In (b) the two terms compete to give way to a single attractor, parameterized by $\vec{f}(\lambda)=\left(\frac{\lambda^2}{10\sqrt{2}},\lambda\right)$, sending the saxions to infinity. These two potentials correspond to a II$_{0,1}\rightarrow $V$_{2,2}$ singularity enhancement in the $\mathcal{R}_{12}$ growth sector (so that $d_s=1$ and $d_u=3$) of Asymptotic Hodge Theory of F-theory flux compactifications, corresponding to type IIB string theory at large volume.}
        \label{fig:two examples}
\end{figure}

In Figure \ref{fig:two examples} we show two examples corresponding to the two possible scenarios for gradient flow trajectories: (I) when there is a single dominant term, yielding a family of gradient flow trajectories; and (II) when several terms compete asymptotically, yielding a valley or attractor corresponding to a single asymptotic gradient flow trajectory. In other words, there are multiple solutions to \eqref{obt alpha} for the factor $\alpha_i$ defined in \eqref{eq:saxions-ansatz} in scenario (I) or a single solution for $\alpha_i$ in scenario (II). Throughout this paper, we will see that it is precisely the scenario (II) which can yield lower values of the dS coefficient potentially yielding accelerated expansion. This latter scenario had been missed so far in the literature.

The dS coefficient \eqref{def dS} along the gradient flow trajectory is computed as a result of the optimization problem in \eqref{eq:min-method}. For a gradient flow trajectory the tangent vector aligns with (minus) the vector $\vec{\partial} \log V$. Projecting in this direction gives the norm of that vector, this is, the dS coefficient. This way,
\begin{equation}\label{eq:gamma from lnV}
    \gamma_{\vec{f}}=\frac{\|{\nabla V}\|}{V}=\frac{\|\vec{\partial}V\|}{V}=-\frac{\partial_\lambda \ln V}{\|\dot{\vec{f}}(\lambda)\|}=-\frac{\partial_{\ln \lambda} \ln V}{|\vec{\beta}|},
\end{equation}
with $|\vec{\beta}|$ as in \eqref{norm beta}. This can be written as
\beq\label{dS gen}
    \gamma_{\vec{f}} = |\hat{\beta}^{i}  l^{\text{dom}}_i| = \frac{ |\beta^{i} l^{\text{dom}}_i|}{\sqrt{\frac12\sum_{j=1}^{{n}}(\beta^j)^2d_j}} \, ,
\eeq
where we have particularized to the case of the diagonal metric in the second step. As a consequence, the dS coefficient is nothing but the result to the optimization problem in \eqref{eq:min-method}, with ${\hat{\beta}}$ parameterizing the gradient flow trajectory \eqref{eq:saxions-ansatz} for the given potential. Moreover, the $l^{\text{dom}}_i$ entering here correspond to the saxionic exponents of the dominant term(s) of the potential that vanishes asymptotically along such given trajectory, i.e. with non zero $\beta^i l_i$.  

Notice that only the $\bm{l}\in\mathcal{E}$, and moduli space metric coming from the K\"{a}hler $K_{\rm cs}$ potential enter in $\gamma_{\vec{f}}$ expression (as in turn these fix $\vec{\beta}$). Furthermore, \eqref{dS gen} is well-defined in the sense that it does not depend in the parametrization of the trajectory sending the fields to infinity, as $\gamma_{\vec{f}}$ is invariant under $\vec{\beta}\rightarrow \mu\vec{\beta}$ for some $\mu>0$. As commented earlier, this allows for the natural normalized $\hat{\beta}=\frac{\vec{\beta}}{|\vec{\beta}|}$, which allows us to directly obtain the gradient flow solution through \eqref{eq:gamma from lnV} and can be integrated in terms of the traveled distance $D$ (for which $\|\dot{\vec{f}}(D)\|\equiv 1$) as
 \begin{equation}\label{eq:pot D}
      V|_{\vec{f}}(D)\sim e^{-\gamma_{\vec{f}}D}\;.
 \end{equation}

However, in the two moduli solution presented along this paper we will present the $\vec{\beta}$ solutions with integer components for clarity's sake (this is not necessarily always the case for more than two moduli, with irrational $\vec{\beta}$ easily found for $n\geq 3$). Recall also that one must replace $\Delta l_i\rightarrow l_i$ and $\Delta d_i\rightarrow d_i$ when applied to the potentials classified by Hodge Theory.

Although it will be explained in detail when considering specific examples in Section \ref{sec:examples}, let us give now some intuition to explain why the dS coefficient can be much lower in Scenario (II) above. When having several moduli growing at different rates, it is possible to have several terms of the potential exhibiting the same asymptotic growth along a given trajectory and therefore competing asymptotically. In that case, the gradient flow trajectory for the potential can be different (and have a smaller slope) that the gradient flow trajectory associated to each term of the potential independently. Furthermore, if the potential terms belong to $\mathcal{E}^{\rm rest}$, there can be cancellations in the numerator of \eqref{dS gen} since the exponents $l_i$ can take different signs.

\subsection{Comments on axion stabilization \label{brief stab ax}}

Finally, let us comment on the stabilization of the axions. The above discussion of the gradient flow trajectory relies on the axions being bounded so that the followed trajectories are geodesics, as well as remaining in a single growth sector  $\mathcal{R}_{i_1...i_{n}}$ (defined in \eqref{growth def}) for those potentials obtained through Asymptotic Hodge Theory on flux compactifications. Hence, the potential should prevent the axions from taking parametrically large field values. We will argue now that either the axions get stabilized at a finite value by the potential, or remain at leading order along a flat direction, playing no role on the gradient flow trajectory as $\lambda\rightarrow \infty$. To do so, we will show that the scalar potential has a minimum for the axions towards which they are attracted through gradient flow. 

Consider the general asymptotic form of the potential given in \eqref{eq:asymptotic-potential}. The axions appear only through the shift invariant functions $\|\rho_{\bm{l}}(G_4),\phi\|_\infty^2$. The gradient flow condition and \eqref{eqmot} require that $\partial^{\phi^i}V_M|_{\phi_0}=0$ for $\phi_0$ to be some axionic minimum of the potential. Furthermore, the mass matrix in the axionic coordinates must be semi-positive definite in a neighborhood around such point, containing the stabilization point, so that the later is stable. While a more complete derivation of this is present in Appendix \ref{stab ax}, we can summarize the argument here.

Recall that the axion dynamics are encapsulated through the $\|\rho_{\bm{l}}\|_{\infty}^2=\mathcal{K}^{\bm{l}}_{ij}\varrho^i_{\bm{l}}\varrho^j_{\bm{l}}$ functions, with $\mathcal{K}^{\bm{l}}$ constant, symmetric, semi-positive definite matrices and $\vec{\varrho}_{\bm{l}}$ depending only on the axions and the quantized fluxes. From \eqref{rho prop} one finds that $\partial_{\phi^j}\vec{\varrho}_{\bm{l}}=-\vec{\varrho}_{\bm{l}^{(j)}}$, with $  l^{(j)}_i=  l_i+2\delta_i^j$. This means that functions of lower degree in the axions will be more dominant for asymptotic values of the saxions since the exponent $l_i$ increases. It is then evident that those dominant terms in the $V_M$ potential, disregarding the asymptotic trajectory taken on the saxions or the (finite) value of the axions, will be accompanied by constant $\|\rho_{\bm{l}}\|_{\infty}^2$ functions. Otherwise, at least some of their derivative would not be constantly null and would be accompanying a more dominant term.
\begin{figure}[H]
\begin{center}		
		\begin{tikzcd}[column sep=scriptsize]
		\rho_{30}\ar[r,"-\partial_{b}" blue]\ar[d,"-\partial_{c}" Green] & \rho_{32}\ar[r,"-\partial_{b}" blue]\ar[d,"-\partial_{c}" Green] & \rho_{34}\ar[r,"-\partial_{b}" blue]\ar[d,"-\partial_{c}" Green]& \rho_{36}\ar[d,"-\partial_{c}" Green]\\
		\rho_{52}\ar[r,"-\partial_{b}" blue]& \rho_{54}\ar[r,"-\partial_{b}" blue]& \rho_{56}\ar[r,"-\partial_{b}" blue]& \rho_{58}
		\end{tikzcd}
\end{center}
\end{figure}
Consider the above diagram, representing the relations between the axion polynomials $\rho_{\bm{l}}$ of the different terms from a two moduli potential, with axions denoted by $(c,b)$. If two of them are connected by an arrow, the one at the end of it will always dominate asymptotically. As a consequence, the leading terms in the asymptotic regime are those at the end of the arrow chain, as further derivation in terms of the axions would go to 0, so that they do not depend on the axions.

Then the trajectory taken by the saxions will not depend on the axion stabilization points. Given that trajectory $s^i(\lambda)$, we can group the terms of $V_M$ along said trajectory by their order in $\lambda$. By the same arguments on the $\|\rho_{\bm{l}}\|_{\infty}^2$ functions, the first terms on $V_M$ not constant on the axions will be quadratic, so that the stabilization conditions will result on a set of linear equations on the axions. As it is argued in Appendix \ref{stab ax}, the output of this setup is always that some axions are stabilized at finite values while others might correspond to flat directions of the potential at this order in $\lambda$.

One then moves to the next subleading terms in the $\lambda\to\infty$ expansion. First of all, the axions that are stabilized in the previous step should be considered to be fixed in the corresponding finite values. This follows from the fact that any new contribution to their gradient flow will be irrelevant for $\lambda\rightarrow\infty$ with respect to the previous one. For the case of the axions that remained as flat directions, again they will appear at most linearly in the axion polynomials, and thus we have quadratic terms in the leftover axions. By the same reasoning, some of them will be stabilized at finite value or still remain as flat directions. Finally, one repeats this procedure until all terms in the potential are exhausted. We then conclude that the kind of potentials we are taking under consideration will give way to either stabilized (the structure of the $\|\rho_{\bm{l}}\|_{\infty}^2$ functions can be used to show that the mass matrix is asymptotically semi-positive definite) or along flat directions. This will also be checked in all the specific examples of Section \ref{sec:examples}. As it is proved in Appendix \ref{APP H damp ax}, in this last case axions along flat directions with an initial non-null velocity $\dot{\phi}^i(0)\neq 0$ are ultimately stopped at finite values by the Hubble damping.

Interestingly, this analysis shows that the asymptotic structure of the flux potential prevents highly turning gradient flow trajectories (as in axion monodromy models \cite{Silverstein:2008sg,McAllister:2008hb}) in which the axions travel for parametrically large distances, as if they appear in the potential they will get stabilized before leaving a growth sector. In the language of \cite{Calderon-Infante:2020dhm}, all trajectories approaching the infinite distance limit in these flux compactifications are asymptotically geodesic, so that the addition of the flux potential does not violate the Distance Conjecture (see also \cite{Baume:2016psm,Valenzuela:2016yny,Blumenhagen:2017cxt,Grimm:2020ouv} for previous works on the axion-saxion backreaction in this setup and its interplay with the Distance conjecture). This is also consistent with the results of \cite{Grimm:2022sbl}, which makes use of  tameness constraints to derive path-independent statements.

\section{Accelerated expansion at parametric control in flux compactifications\label{sec:examples}}

We proceed now to determine whether asymptotic accelerated expansion is possible in the context of 4d $\mathcal{N}=1$ supergravities arising from flux compactifications of string theory. We first review the no-go in \cite{Hellerman:2001yi,Rudelius:2021azq} that prevents asymptotic accelerated expansion in 4d $\mathcal{N}=1$ theories, to show that this no-go is not valid in general for more than one modulus when there are several terms of the potential that compete asymptotically. Secondly, we will compute explicitly the exponential rate of the potential along the asymptotic gradient flow trajectories for concrete examples, including Type IIA/B and F-theory flux compactifications. We will identify concrete potential examples with $\gamma\leq\sqrt{2}$ that could yield asymptotic accelerated expansion if assuming that Kähler moduli stabilization can be independently achieved.

\subsection{Evading no-go's in 4d $\mathcal{N}=1$ theories\label{Sec:SuperpotCond}}

It was argued in \cite{Rudelius:2021azq} that a scalar field rolling down a potential that asymptotes to a zero-energy supersymmetric minimum cannot yield an accelerating cosmology, based on the no-go of \cite{Hellerman:2001yi} for single field. However, we will show next that this argument does not apply in general when we have more than one scalar field, even if the trajectory is a one-dimensional gradient flow. This implies that, at the level of supergravity, there is no obstruction to get accelerated expansion in 4d $\mathcal{N}=1$ setups where the potential asymptotes to zero-energy at infinite distance.

Let us first review the argument of \cite{Hellerman:2001yi,Rudelius:2021azq} using the notation of Section \ref{SEC as pot}, where the potential \eqref{V from T} can be written as 
\beq
\label{VT}
V=\frac12\left(\|\nabla \cT\|^2-\frac32 \cT^2\right)
\eeq
with $\cT=2e^{K/2}|W|$. Recall that $K$ and $W$ are the K\"{a}hler potential and superpotential, respectively. It was shown in \cite{Lanza:2020qmt} that, whenever this tension satisfies that
\beq
\label{dTcT}
\|\nabla\cT\|=|\alpha|\cT
\eeq
(so the associated membrane of tension $\cT$ is said to be dilatonic \cite{Lanza:2020qmt}), then the dS coefficient $\gamma$ becomes proportional to the dilatonic factor as follows,
\beq
\label{dTT}
\gamma=\frac{\|\nabla V\|}{V}=\frac{2\|\nabla\cT\|}{\cT}=2|\alpha| \, .
\eeq

By replacing this into the scalar potential, one gets
\beq\label{V prop T2}
V=\frac14\left(\frac{\gamma^2}{2}-3\right)\mathcal{T}^2 \, .
\eeq
For a positive scalar potential, as explained in \cite{Rudelius:2021azq}, the dS coefficient is then lower bounded by $\gamma>\sqrt{6}$ in four dimensions, which is bigger than the strong bound \eqref{strong bound} and therefore  forbids accelerated expansion asymptotically. However, recall that this is only valid assuming that \eqref{dTT} holds.

The argument in \cite{Rudelius:2021azq} assumes that both $V$ and $\cT$ have the same gradient flow trajectory,  since it is assumed that in order to get $V\sim \exp(-\gamma\phi)$ one must impose $\mathcal{T}\sim \exp(-\frac{\gamma}{2}\phi)$, implying that \eqref{dTT} is indeed satisfied. However, as previously remarked in \cite{Lanza:2020qmt}, the condition \eqref{dTcT} (and consequently \eqref{dTT}) are not satisfied for general choices of fluxes. The superpotential (membrane tension) will be a sum of flux terms with different powers of the saxions, and will not satisfy in general that its derivative is proportional to $W$ ($\mathcal{T}$) itself. Since we are interested in the asymptotic regime, one could have thought that only one term will dominate in the end so that \eqref{dTT} is always satisfied, but this is not correct. As explained in Section \ref{SEC:gradflow}, there are cases in which several terms compete asymptotically as they have the same growth on $\lambda$, so that \eqref{dTT} does not hold asymptotically and $V$ and $T$ exhibit different gradient flow trajectories. These are precisely the cases in which the bound $\gamma>\sqrt{6}$ can be evaded.

Let's see this in more detail in some concrete examples. 
Consider, for instance, a two-modulus example with two contributions to the flux potential such that
\begin{subequations}
    \begin{align}
        W&=\rho_1(i s)^{\frac{d_s+l_s}{2}} (i u)^{\frac{d_u+l_u}{2}}+\rho_2 (i s)^{\frac{d_s+m_s}{2}} (i u)^{\frac{d_u+m_u}{2}} \, ,\\
        G&=\diag\left(\frac{d_s}{2s^2},\frac{d_u}{2u^2}\right) \, ,\\
        \mathcal{T}^2&=2^{2-d_s-d_u}\left(|\rho_1|^2s^{l_s}u^{l_u}+|\rho_2|^2s^{m_s}u^{m_u}+2\Re (\rho_1\bar{\rho_2})s^{\frac{l_s+m_s}{2}}u^{\frac{l_u+m_u}{2}}\cos\left(\frac{\pi}{4}(l_s+l_u-m_s-m_u)\right)\right) \, , \\
        V&=\frac{2^{-d_s-d_u}}{d_sd_u}\left((d_sl_u^2+d_ul_s^2-3d_sd_u)|\rho_1|^2s^{l_s}u^{l_u}+(d_sm_u^2+d_um_s^2-3d_sd_u)|\rho_2|^2s^{m_s}u^{m_u}\right.\notag\\
        &\quad\left.+2(d_sl_um_u+d_ul_sm_s-3d_sd_u)\Re (\rho_1\bar{\rho_2})s^{\frac{l_s+m_s}{2}}u^{\frac{l_u+m_u}{2}}\cos\left(\frac{\pi}{4}(l_s+l_u-m_s-m_u)\right)\right) \, .\label{V two mod}
    \end{align}
\end{subequations}

It is easy to see that in general the coefficients accompanying the saxion terms will be different for $\mathcal{T}^2$ and $V$, so that, even if they display the same power-like behavior, will not be proportional, as implied by \eqref{V prop T2}. This is only the case when just one of the terms dominate asymptotically, so that we can neglect the others (as in scenario (I) in Figure \ref{fig:two examples}). Assume, for example, that only the term associated to $\rho_1$ dominates asymptotically. In that particular case, one has
\begin{equation}
\label{alpha2}
  4\frac{\|\partial \mathcal{T}\|^2}{\mathcal{T}^2} =  \sum_{j=1}^{{n}}\frac{2l_j^2}{d_j}=\frac{2l_u^2}{d_u}+\frac{2l_s^2}{d_s} \, , 
\end{equation}
which must be bigger than 6 to get a positive $V$. One can then check that this factor becomes indeed equivalent to the square of dS coefficient in \eqref{dS gen} upon replacing
\beq
\frac{\beta^i}{\beta^j}=\frac{d_j}{d_i}\frac{l_i}{l_j} \, ,
\eeq
which are precisely the trajectories maximizing the dS coefficient and therefore yielding the gradient flow trajectory of steepest descent.

More generally, when two terms compete asymptotically as in scenario (II) in Figure \ref{fig:two examples}, the dS coefficient along the gradient flow is different than \eqref{alpha2}, thus violating \eqref{dTT}. Let us see this explicitly by
considering the following example, with $(l_s,l_u,m_s,m_u,d_s,d_u)=(1,-3,-2,2,1,3)$, so that
\begin{equation}
    V=\frac{1}{48}\left(3|\rho_1|^2\frac{s}{u^3}+7|\rho_2|^2\frac{u^2}{s^2}\right),\qquad \mathcal{T}^2=\frac{1}{4}\left(|\rho_1|^2\frac{s}{u^3}+|\rho_2|^2\frac{u^2}{s^2}\right),
\end{equation}
which are clearly positive and not proportional. Following the procedure described in Section \ref{SEC:gradflow}, one finds the following (unique) asymptotic gradient flow trajectories for both $V$ and $\mathcal{T}$ (which is the same as for $\mathcal{T}^2$):
\begin{equation}
    (s,u)_V(\lambda)=\left(\sqrt[3]{\frac{49|\rho_2|^2}{18|\rho_1|^2}}\lambda^5,\lambda^3\right),\qquad (s,u)_{\mathcal{T}}(\lambda)=\left(\sqrt[3]{\frac{7|\rho_2|^2}{6|\rho_1|^2}}\lambda^5,\lambda^3\right),
\end{equation}
which are indeed different. The dS coefficient is found to be $\gamma=2\sqrt{\frac{2}{13}}\approx0.78446<\sqrt{6}$. Both $c_4^{\rm TCC}=\sqrt{\frac{2}{3}}$ and $c_4^{\rm strong}=\sqrt{2}$ are violated. As the saxion evolution is dictated  by the gradient flow of $V$, they will follow $(s,u)_V\neq (s,u)_{\mathcal{T}}$, so that $\vec{\partial} \mathcal{T}$ will have tangent and normal components to the trajectory (as opposed to only tangent if the followed trajectory was $(s,u)_{\mathcal{T}}$). In this case $|\alpha|=\sqrt{\frac{146}{67}}\approx1.4762>\frac{\gamma}{2}$ (and bigger than $\sqrt{\frac{3}{2}}$ so $V>0$), thus showing that \eqref{dTT} is not valid in general. 

The above case was engineered to illustrate 
how 4d $\mathcal{N}=1$ potentials with more than two terms competing asymptotically do not satisfy the $\gamma>\sqrt{6}$ bound. Hence, if there is any obstruction for asymptotic accelerated expansion, this must come from quantum gravity considerations. In the rest of the section we will investigate whether there are potentials arising directly from string theory that also violate this bound. Some of the F-theory limits we will consider correspond to several type IIB String Theory scenarios, where the presence of K\"{a}hler moduli (which we will not take into account in this work) is responsible for the no-scale structure that cancels the terms proportional to $-|W|^2$ (equivalently $-\mathcal{T}^2$), thus making the scalar potential non-negative definite and trivializing the bound. However, through mirror symmetry one can go to type IIA String Theory limits, in which this is no longer true, and negative potentials could in principle be obtained from the perturbative flux superpotential. Even if it is difficult to generate accelerated expansion in IIA, we will at least find in Section \ref{sec:examples} some concrete examples violating the above bound of $\gamma>\sqrt{6}$, which exemplify that such bound can indeed be violated in string theory. As an example, we show here 
\begin{equation}
    V_M\sim\frac{1}{s^3}\left(A_{34}\frac{u}{s}+A_{52}\frac{s}{u^3}\right)\, ,
\end{equation}
which can be obtained from the following superpotential and K\"{a}hler potential
\begin{equation}
    W=\frac{\sqrt{521}}{4}i\left(\sqrt{3A_{34}}u+2\sqrt{A_{52}}s\right),\qquad K_{\rm cs}=-\log\left((\Phi^1+\overline{\Phi^1})^4(\Phi^2+\overline{\Phi^2})^3\right) \, ,
\end{equation}
with $\Phi^1=c+is$ and $\Phi^2=b+iu$. This is the eleventh row in Table \ref{table1}. Using that $\Delta d_s=1+3=4$ and $\Delta d_u=3$ we find a $\vec{\beta}=(2,1)$ trajectory and $\gamma_{\vec{f}}=7\sqrt{\frac{2}{19}}\approx 2.2711<\sqrt{6}$. Even if the above example does not give way to an accelerated expansion of the universe, it violates the supergravity no-go of \cite{Hellerman:2001yi,Rudelius:2021azq} as it has $\gamma<\sqrt{6}$.

\subsection{One modulus examples in string theory \label{sec:1 mod}}

From now on we will only consider explicit string theory examples. Let us begin with the easiest case of single-field limits in the complex structure moduli space of CY$_4$'s, in which only one modulus is sent to an infinite distance limit.

The types of infinite distance limits can be classified according to the action \eqref{nilpot} of the corresponding nilpotent monodromy operator on the limiting period vector
Following \cite{Grimm:2019ixq}, there are four general types of infinite distance limits, denoted by II, III, IV or V, depending on the value of the nilpotency order (or singularity degree) $d_i$ in \eqref{nilpot}. These four types correspond to $d_i=1,2,3,4$ respectively. To each of them, we can associate an integer $l$ (it will be an n-dimensional vector in the case of an n-moduli limit) that characterizes the asymptotic splitting of the flux lattice \eqref{lattice splitting}. The possible value of these integers $l$ for each limit can be found in \cite{Grimm:2019ixq} and plugged into \eqref{eq:asymptotic-potential} to get the asymptotic form of the flux potential. Let us consider two examples here and derive the dS coefficient in each case:

\begin{itemize}
\item Type II, for which $\Delta d=1$. The potential reads
\begin{equation}\label{pot1}
	V_{M} \sim \frac{(f-h \phi)^{2}}{s} - fh + h^2 s \, ,\qquad G_{ss} = G_{\phi\phi} = 2\partial_{S}\partial_{\bar{S}}K = \frac{1}{2s^2} \, ,
\end{equation}
where $\{f,h\}$ are the quantized fluxes.\footnote{This example can be obtained from the two moduli case (see Table \ref{table2}) after tuning to zero several fluxes (such as $h_3=h_2=h_1=h_0=f_0=A_{44}-A_{\rm loc}=0$) and stabilizing the second saxion and axion at $u_0=\sqrt{\frac{\sqrt{3}|f_4^2-2f_2f_6|}{3|f_2|}}$ and $b_0=-\frac{f_4}{f_2}$, with $f\equiv\frac{2\sqrt{2}}{3^{3/4}}|f_2|\sqrt{|f_2^2-2f_2f_6|}$.} For $h\neq 0$ the limit $s\to\infty$ is obstructed by the potential (in other words, $h^2s$ is a $\mathcal{E}^{\rm heavy}$ term), so we must set $h=0$. The potential gets a very simple form (with only one flux term in $\mathcal{E}^{\rm light}$) and the dS parameter is straightforwardly computed as $\gamma=\sqrt{2}$. Since $\partial_\phi V_M = 0$, the axion parameterizes a flat direction, but it will be stabilized at some finite value $\phi_0$ through Hubble damping or non-perturbative effects  as discussed in Appendix \ref{APP H damp ax}.

\item Type IV, for which $\Delta d=3$. The potential reads
\begin{equation}\label{IV 1mod}
    V_M=\frac{A_{1}}{s^3}+\frac{A_3}{s}+A_4+A_{5}s+A_7 s^3 \, ,
\end{equation}
In order to have $V_M \to 0$ as $s\rightarrow\infty$, we must set $A_4=A_5=A_7=0$ by turning off some fluxes. We then have the following two possibilities:
\begin{itemize}
    \item $A_3\neq 0$: In this case $V_M\sim \frac{A_3}{s}$, with the $A_1$ term (in which the axion enters) being subleading and stabilizing the axions at a finite values. The dS coefficient becomes $\gamma=\sqrt{\frac{2}{3}}$.
    \item $A_3=0$: In this case the potential contains a single term $V_M\sim \frac{A_1}{s^3}$, in which the axion does not appear. This results in $\gamma=\sqrt{6}$.
\end{itemize}

\end{itemize}

In order to embed these examples in 4d $\mathcal{N}=1$ compactifications of F-theory, we need to assume that there are more complex structure moduli that are not sent to the limit. This is because the CY$_4$ has to be elliptically fibered to take the F-theory limit, so it must have at least two moduli. This implies that the coefficients $A_{\bm{l}}$ will depend on the rest of the moduli, which will change the results if they do not get stabilized by the fluxes. Alternatively, we can embed these examples in 3d compactifications of M-theory on a CY$_4$ with a single complex structure modulus, so that the coefficientes $A_{\bm{l}}$ only depend on the fluxes and the axion. 

Notice that the deSitter coefficient only depends on the geometric data of the asymptotic limit, so it is the same in 3d or 4d. However, the de Sitter bounds \eqref{strong bound} and \eqref{TCC bound} depend on the space-time dimension, becoming stronger as the dimension decreases. This implies that, for example, the Type II asymptotic limit satisfies the Strong dS bound in 4d but would violate it in 3d. Analogously, the Type IV singularity with $A_3\neq 0$ violates both  $c_3^{\rm TCC}=\sqrt{2}$ and $c_3^{\rm strong}=2$ in 3d, while saturates the TCC in 4d; and the case with $A_3= 0$ satisfies all bounds in 3d and 4d. Hence, it seems easier to violate the bounds in 3d, but we will focus in 4d from now on for obvious phenomenological reasons.

In the next subsection we will consider two moduli limits, so that we can study full moduli stabilization of the complex structure sector in 4d $\mathcal{N}=1$ theories by simply considering Calabi-Yau's with two moduli. We will again find some potential counterexamples to the bounds which are at parametric control in the complex structure sector. However, they will not describe accelerated cosmologies until Kähler moduli stabilization is also achieved. This will be discussed in more detail in Section \ref{sec:no-scale}.

\subsection{Two moduli limits in string theory\label{sec:2 mod}}

We now consider the case of a two moduli limit in a $CY_4$, namely $\{\Phi^1=c+is,\Phi^2=b+iu\}$ with $s,u\rightarrow \infty$, such that $s$ grows faster than $u$ so that we are in the growth sector $\mathcal{R}_{12}$. We compactify F-theory in such $CY_4$ with fluxes to obtain a 4d $\mathcal{N}=1$ potential as described in Section \ref{sec: AHT}. The Kähler potential and the scalar potential take the general asymptotic form given in \eqref{asymp Kahler} and \eqref{eq:asymptotic-potential}, whose dependence on the two moduli $\Phi^1,\Phi^2$ are completely determined upon specifying the discrete data $(\Delta d_i, \Delta l_i)$ that characterize the type of asymptotic limit. The possible two-moduli asymptotic limits of $CY_4$'s were classified in \cite{Grimm:2019ixq}, where the asymptotic form of the scalar potential in each case was also derived. In total, there are 36 enhancement possibilities associated to 11 different potentials, but we will just focus on three cases in this paper, which are summarized in Table \ref{table summary}.
\begin{table}[h]
    \centering
    \begin{tabular}{|c|c|c|}
    \hline
    Type of limit&$(\Delta d_1,\Delta d_2)$&$(\Delta l_1,\Delta l_2)$\\
    \hline
    II$_{0,1}\rightarrow$V$_{2,2}$&(1,3)&(-1,-3), (-1,-1), (-1,1), (-1,3), (1,-3), (1,-1), (1,1), (1,3), (0,0)\\
    II$_{0,1}\rightarrow$III$_{0,0}$&(1,1)&(-1,-1), (-1,1), (1,-1), (1,1), (0,0)\\
    III$_{1,1}\rightarrow$V$_{2,2}$&(2,2)&(-2,-2), (-2,0), (0,-2), (-2,2), (2,-2), (0,0), (0,2), (2,0), (2,2)\\
    \hline
    \end{tabular}
    \caption{Summary of the three singularity enhancements studied, and the associated discrete data $(\Delta d_i, \Delta l_i)$ characterizing them. Notice that some of the $\bm{\Delta l}$ terms in the third column belong to $\mathcal{E}^{\rm heavy}$, and might need to be rendered null by setting some quantized fluxes to zero. Note that the first limit is equivalent to the IV$_{0,1}\rightarrow$V$_{2,2}$ enhancement just by simply exchanging $s^1\leftrightarrow s^2$ and the associated parameters.
    \label{table summary}}
\end{table}

For each potential $V_M$ we will consider all the possible asymptotic behaviors specified by a choice of certain (non-)vanishing fluxes.
As explained in Section \ref{brief stab ax}, in all cases the leading terms of the potentials will not depend on the axions. This will allow us to first find the gradient flow trajectory for the saxions and compute the dS coefficient without worrying about the axions. Afterwards, we show how the axions are either stabilized at some finite value along the gradient flow or remain as flat directions of the potential.

\subsubsection{Type IIB at large complex structure: II$_{0,1}\rightarrow$V$_{2,2}$} \label{sec:typeIIB}

The first F-theory limit to consider (first row in Table \ref{table summary}) corresponds to Type IIB orientifold compactification at weak string coupling and large complex structure. This can be derived from F-theory as follows.

Consider $Y_4$ to be elliptically fibered, denoting the complex structure modulus of the elliptic fiber as $S=C_0+ie^{-\phi_B}$ (such that $\phi_B$ corresponds with the 10-dimensional dilaton) and the two fiber one-forms as $\{\diff x,\,\diff y\}$. Moving to Sen's weak coupling limit \cite{Sen:1996vd} (i.e. $\phi_B \rightarrow-\infty$ so that ${\rm Im} S\gg 1$), we can split $G_4=H_3\wedge \diff y+F_3\wedge \diff x$, with $H_3$ and $F_3$ being the NS-NS and R-R Type IIB fluxes. One can then substitute this into \eqref{fluxpot} to find the Type IIB orientifold flux potential:
\begin{equation}\label{eq: IIB pot flux}
    V_{\rm IIB}=\frac{e^{3\phi_B}}{4(\mathcal{V}_s^B)^2}\left[e^{-\phi_B}\int_{Y_3}H_3\wedge\star H_3+e^{\phi_B}\int_{Y_3}F_3\wedge \star F_3-\int_{Y_3}F_3\wedge H_3\right] \, ,
\end{equation}
where the standard torus metric has been used and $\mathcal{V}_s^B=\mathcal{V}_be^{\frac{3}{2}\phi_B}$ corresponds to the volume of the CY$_3$ $Y_3$ arising in the orientifold limit in the 10-dimensional string frame. Note that now the Hodge norm in \eqref{eq: IIB pot flux} only depends on the complex structure moduli of $Y_3$. 

The Sen's weak coupling limit that we took corresponds geometrically to a type II singularity in the complex structure moduli space of the CY fourfold.
One can now enhance the type II singularity by sending another modulus to infinity, thus combining this weak coupling limit with another one in the complex structure moduli space of $Y_3$. Concretely, the enhancement II$_{0,1}\rightarrow$V$_{2,2}$ in the $\mathcal{R}_{12}$ growth sector corresponds to Type IIB string theory at the weak coupling and large complex structure limit \cite{Grimm:2019ixq}.

By plugging the discrete data in Table \ref{table summary} associated to this asymptotic limit into the flux potential \eqref{eq:asymptotic-potential}, one recovers the well known Type IIB flux potential:
\begin{equation} \label{eq:full-potential}
    V_M\sim \frac{A_{f_6}}{u^3s}+\frac{A_{f_4}}{us}+\frac{A_{f_2}u}{s}+\frac{A_{f_0}u^3}{s}+\frac{A_{h_0}s}{u^3}+\frac{A_{h_1}s}{u}+A_{h_2}us+A_{h_3}u^3s+A_{44}-A_ {\rm loc} \, ,
\end{equation}
where we have chosen the notation of the different terms to match with the RR $f_i$ and NS $h_i$ fluxes from a mirror Type IIA interpretation.
Following \cite{Grimm:2019ixq}, we are denoting $A_{ij}=A_{f}$, with $f$ being the flux appearing as a constant term in the axion polynomial $\rho_{ij}$, which are given by\footnote{Actually, the coefficients $A_f$ are not equal but only proportional to these axion polynomials since there can be an additional numerical factor or a function that depends on the other scalars not sent to infinity. However, this will not be important for the purposes of this paper so we will neglect it.}
\begin{subequations}
\begin{align} \label{eq:axion-polynomials}
    A_{f_6}&=\|\rho_{30}\|^2_\infty=\left(f_6\pm f_4b+\frac{1}{2}f_2b^2\pm\frac{1}{6}f_0b^3-h_0c+h_1cb-\frac{1}{2}h_2cb^2+\frac{1}{6}h_3cb^3\right)^2 , \\
    A_{f_4}&=\|\rho_{32}\|^2_\infty=\left(f_4\pm f_2b+\frac{1}{2}f_0b^2\pm h_1c\mp h_2cb\pm\frac{1}{2}h_3cb^2\right)^2 \, , \\
    A_{f_2}&=\|\rho_{34}\|^2_\infty=\left(f_2\pm f_0b-h_2 c+h_3cb\right)^2 \, , \\
    A_{f_0}&=\|\rho_{36}\|^2_\infty=\left(f_0\pm h_3 c\right)^2 \, , \\
    A_{h_0}&=\|\rho_{52}\|^2_\infty=\left(h_0-h_1b+\frac{1}{2}h_2b^2-\frac{1}{6}h_3b^3\right)^2 \, , \\
    A_{h_1}&=\|\rho_{54}\|^2_\infty=\left(h_1-h_2b+\frac{1}{2}h_3b^2\right)^2 \, , \\
    A_{h_2}&=\|\rho_{56}\|^2_\infty=\left(h_2-h_3b\right)^2 \, , \\
    A_{h_3}&=\|\rho_{58}\|^2_\infty=\left(h_3\right)^2 \, , \\
    A_{44}&=\|\rho_{44}\|^2_\infty\in\R \, , \\
    A_ {\rm loc}&=\langle G_4,G_4\rangle=\frac{\chi(Y_4)}{12}\in\R \, ,
\end{align}
\end{subequations}
where we have fixed the value of $A_ {\rm loc}$ to fulfill the tadpole condition \eqref{eq:tadpole}. 

Our goal is to determine the asymptotic gradient flow trajectory and the dS coefficient that this potential gives.
This, of course, will depend on what fluxes are turned on or off. Once a choice of fluxes is made, one can determine the gradient flow trajectory as explained in Section \ref{SEC:gradflow}. To compute the dS coefficient is necessary to identify the dominant terms of the potential along such gradient flow.  
This can be done in a pictorial way by following the argument in Section \ref{brief stab ax} (expanded in Appendix \ref{stab ax}): there is a notion of \emph{partial ordering} between the different terms, along which the axionic degree is lowered and the saxionic one grows, resulting in dominating terms that do not depend on the axions.
One can encode this information in a diagram that for the potential in \eqref{eq:full-potential} takes the form
\begin{figure}[H]
\begin{center}		
		\begin{tikzcd}[column sep=scriptsize]
		\rho_{30} \arrow[blue]{r} \arrow[Green]{d} & \rho_{32} \arrow[blue]{r} \arrow[Green]{d} & \rho_{34} \arrow[blue]{r} \arrow[Green]{d}& \rho_{36} \arrow[Green]{d}\\
		\rho_{52} \arrow[blue]{r}& \rho_{54} \arrow[blue]{r} & \rho_{56} \arrow[blue]{r}& \rho_{58}
		\end{tikzcd}
\end{center}
\end{figure}

\vspace{-15px}
Here blue rightwards arrows mean $-\partial_{b}$ in the axionic functions and a 2 degree increase in $u$ and green downwards ones mean $-\partial_{c}$ and a 2 degree increase on $s$. Further movement along those directions when there are no more arrows leads to zero. As discussed in Section \ref{brief stab ax}, the dominant terms are those at the end of the arrow sequence, in this case being just $A_{h_3}su^3=\|\rho_{58}\|^2_{\infty}su^3$. However, one needs to turn off those terms associated to $\mathcal{E}^{\rm heavy}$ which blow up along any direction to infinity. This can be done by setting $h_3=h_2=0$, so that it shortens the above sequences, giving rise to
\begin{figure}[H]
\begin{center}		
		\begin{tikzcd}[column sep=scriptsize]
		\rho_{30} \arrow[blue]{r} \arrow[Green]{d} & \rho_{32} \arrow[blue]{r} \arrow[Green]{d} & \rho_{34} \arrow[blue]{r} & \rho_{36} \\
		\rho_{52} \arrow[blue]{r}& \rho_{54} 
		\end{tikzcd}
\end{center}
\end{figure}
\vspace{-15px}
Hence, the dominant asymptotic behavior of the potential would now be given by 
\begin{equation}
    V_M\sim h_1^2\frac{s}{u}+f_0^2\frac{u^3}{s},
\end{equation}
responsible of determining the asymptotic trajectory. One could now turn off $h_1$ or $f_0$, making the partially ordered chain shorter and changing the dominant terms of the potential. Once the asymptotic trajectory is fixed, subleading terms will be responsible of (maybe partially) stabilizing the axions.

Notice that the diagram above is very useful also for counting the choices of fluxes giving rise to different asymptotic behaviors. In general, the polynomials $A_l$ cannot be turn off arbitrarily since they are interlinked by the fluxes. This must be done in a special order which can be easily read from the above diagrams; always starting from the end of the sequences such that we shorten them.
In Table \ref{table2} we give the results for all the cases featuring $\mathcal{E}^{\rm rest}$ and $\mathcal{E}^{\rm light}$ terms of the potential. Of these, only eight give rise to admissible gradient flows towards infinite distance. While almost all of these solutions result in non-accelerated universes (three of them saturate $c_4^{\rm strong}=\sqrt{2}$), we actually find a quadratic solution violating both $c_4^{\rm strong}$ and $c_4^{\rm TCC}$. Furthermore in some cases the $u$ saxion is stabilized at a point $u_0$ depending on the fluxes, with the dynamics effectively simplifying to a one modulus regime.

As it can be seen in the above diagrams, in the present potential only one or two terms can dominate. We will now present examples of both scenarios to highlight their differences, and show how to get the saxionic gradient flow and dS coefficient. For completeness, we also discuss how the axions get stabilized to finite values or remain as flat directions of the potential.

\begin{table}[h]
    \centering
    \resizebox{\textwidth}{!}{
    \begin{tabular}{|c||c|c|c||c|}
    \hline
    Potential &$\vec{\beta}$&$\gamma_{\vec{f}}$&$\chi$& Comments\\\hline
    
    $\frac{{A_{30}}}{su^{3}}+\frac{A_{32}}{us}+{{A_{34}}}\frac{u}{s}+A_{36}\frac{u^3}{s}+A_{52}\frac{s}{u^3}+A_{54}\frac{s}{u}+A_{44}-A_ {\rm loc}$& {(-2,-1)}&{-}&{-}&{\small {No admissible solution, $V_M|_{\vec{f}}\rightarrow\infty$}}\\
    
    $\frac{\color{lightgray}{A_{30}}}{su^{3}}+\frac{\color{Blue}A_{32}}{us}+{\color{lightgray}{A_{34}}}\frac{u}{s}+{ A_{36}}\frac{u^3}{s}+{ A_{52}}\frac{s}{u^3}+{ A_{44}-A_ {\rm loc}}$&(3,1) $\star$&$2\sqrt{\frac{2}{3}}$&$\frac{8}{3}w^{-1}$&{\small After axion stabilization $\gamma_{\vec{f}}:\sqrt{\frac{2}{3}}<c_4^{\rm strong}\rightarrow2\sqrt{\frac{2}{3}}$}\\
    
   $\frac{{A_{30}}}{su^{3}}+\frac{\color{lightgray}{A_{32}}}{us}+{\color{Blue}A_{52}}\frac{s}{u^3}+{ A_{34}}\frac{u}{s}+{ A_{54}}\frac{s}{u}+{ A_{44}-A_ {\rm loc}}$&(1,1) $\star$&$\sqrt{2}$&$4w^{-1}$&{\small Saturates $c_4^{\rm strong}$}\\
    $\frac{{\color{Blue}A_{30}}}{su^{3}}+\frac{{A_{32}}}{us}+{\color{lightgray}{A_{34}}}\frac{u}{s}+{\color{Blue}A_{36}}\frac{u^3}{s}$&(1,0)&$\sqrt{2}$&$2w^{-1}$&{\small Saturates $c_4^{\rm strong}$}, {\small {stabilized $u_0=u_0(A_{36},A_{32},A_{30})$, flat $c$}}\\
    $\frac{\color{lightgray}{A_{30}}}{su^{3}}+\frac{{\color{lightgray}{A_{32}}}}{us}+{\color{Blue}A_{34}}\frac{u}{s}+{\color{Blue}A_{52}}\frac{s}{u^3}$&(2,1)&$\sqrt{\frac{2}{7}}$&$w^{-1}$&{\small\textbf{Violates $\bm{c_4^{\rm TCC}}$}}\\
    $\frac{{A_{30}}}{su^{3}}+\frac{\color{lightgray}{A_{32}}}{us}+{\color{lightgray}{A_{52}}}\frac{s}{u^3}+A_{54}\frac{s}{u}$&{(-3,1)} &{-}&{-}&{{No admissible solution}}\\
    $\frac{\color{lightgray}A_{30}}{su^{3}}+\frac{\color{Blue}A_{32}}{us}+{\color{Blue}A_{52}}\frac{s}{u^3}$&(1,1)&$\sqrt{2}$&$4w^{-1}$&{{Flat direction $\rho_{30}(c,b)=0$}}\\
    $\frac{\color{Blue}A_{30}}{su^{3}}+\frac{{\color{lightgray}{A_{32}}}}{us}+{\color{Blue}A_{34}}\frac{u}{s}$&(1,0)&$\sqrt{2}$&$2w^{-1}$&{\small Saturates $c_4^{\rm strong}$}, {\small {stabilized $u_0=\left(3\frac{A_{30}}{A_{34}}\right)^{\frac{1}{4}}$, flat $c$} }\\
    $\frac{{\color{lightgray}{A_{30}}}}{su^{3}}+\frac{\color{Blue}A_{32}}{us}$&(3,1)&$2\sqrt{\frac{2}{3}}$&$\frac{8}{3}w^{-1}$&{\small {Flat $b$}}\\
    $\frac{{\color{lightgray}{A_{30}}}}{su^{3}}+A_{52}\frac{s}{u^3}$&{(-1,1)}&{-}&{-}&{\small {No admissible solution}, {flat $c$}}\\
    $\frac{\color{Blue}A_{30}}{su^{3}}$&(1,1)&$2\sqrt{2}$&$8w^{-1}$&{\small {Flat $c,b$}}\\
    \hline
    \end{tabular}}
    \caption{Solutions (in case they exist) of the different potentials obtained once axions are stabilized. We use $(s^1,s^2)=(s,u)$, and consider the enhancement II$_{0,1}\rightarrow$V$_{2,2}$ $(d_1=1,\,d_2=4)$, corresponding to weak coupling type IIB string theory at the large complex structure limit. There is also comment on the possible stabilization of saxions  and the axions being along flat directions. Those $A_{\bm{l}}$ functions colored in {\color{blue}blue} correspond to those dominating terms and those in {\color{lightgray}gray} are exponentially suppressed through axion stabilization. Note that in some cases (marked by $\star$) the (preliminary) leading terms determining $\vec{\beta}$ turn out to be constant when evaluated along the solution, and must be canceled by $A_{44}-A_{\rm loc}$ through an adequate fine tuning of the fluxes, with $\gamma_{\vec{f}}$ determined by subleading terms (which we now color in {\color{blue}blue}). The fourth column corresponds to the coefficient $\chi$ relating the decay rates of the potential and the leading tower becoming light, modulo $w=1,\,2,\,3$, which will be introduced in Section \ref{sec:distance conj}.}
    \label{table2}
\end{table}

\subsubsection*{One term example: $f_4\neq 0$}
Consider the case in which we set $h_3=h_2=h_1=h_0=f_0=f_2=0$ (as well as $A_{44}-A_{\rm loc}=0$ so that the potential vanishes asymptotically) and require $f_4\neq 0$, without further assumptions on the rest of fluxes. This corresponds to the ninth row in Table \ref{table2}. The partially ordered chain of terms is
\begin{figure}[H]
\begin{center}		
		\begin{tikzcd}[column sep=scriptsize]
		\rho_{30} \arrow[blue]{r} & \rho_{32}
		\end{tikzcd}
\end{center}
\end{figure}
\vspace{-20px}
\noindent resulting in a single term dominating,
\begin{equation} \label{eq:V-f4}
	V_M \sim \frac{f_{4}^{2}}{us} \, .
\end{equation}
Clearly, this potential can give rise to gradient flows such that $s,u\to\infty$ since all saxions appear with negative powers, so that the potential pushes them to infinity.

As expected, the leading term only depends on the saxion, and we can use it to obtain the asymptotic trajectory they will follow by solving the gradient flow equations \eqref{eq:saxionic-flow}, which take the form
\begin{equation}
    \frac{\partial_\lambda\log s}{\partial_{\lambda}\log  u}=\frac{\beta_s}{\beta_u}=\frac{3}{1}\frac{(-1)f_4^2(us)^{-1}}{(-1)f_4^2(us)^{-1}}=3.
\end{equation}
This gives rise to cubic trajectories, as $\vec{\beta}=(3,1)$ with $\vec{\beta}$ defined in \eqref{eq:saxions-ansatz}. As there is a single term dominating the potential, the gradient flow equations can be solve straightforwardly, without the need to consider the optimization problem in \eqref{eq:min-method}. One could try to obtain the $\vec{\alpha}=(\alpha_s,\alpha_u)$ through \eqref{obt alpha}, but it is immediate that they give way to a trivial condition, resulting in a continuous family of solutions given by
\begin{equation}
    (s,u)(\lambda)=(\alpha\lambda^3,\lambda)\qquad\text{with }\alpha>0
\end{equation}
up to $\lambda$ reparametrization. This is the example represented in Figure \ref{fig:two examples} (I) (for $f_4=1$). We notice that this gradient flow stays in the growth sector $\mathcal{R}_{12}$, i.e., $s\gg u$, so it is consistent.

One can now compute (either through the explicit expression \eqref{def dS} or directly using \eqref{dS gen}) the dS coefficient. The result is
\begin{align}
    \gamma&=\frac{\|\nabla V_M\|}{V_M}=\left.\frac{\sqrt{2s^2(-us^2)^{-2}+\frac{2u^2}{3}(-u^2s)^{-2}}}{(us)^{-1}}\right|_{(s,u)=(\alpha\lambda^3,\lambda)}=\frac{2\sqrt{6}}{3}\approx
    1.63299
\end{align}
This value is higher than $c_4^{\rm strong}$ and thus it neither violates any of the proposed bounds nor results in a universe in accelerated expansion.

Finally, let us discuss the behavior of the axions along the gradient flow. Plugging the solution to the saxions in the potential we get
\begin{equation}
    V_M(\lambda,a,b)=\frac{(f_6\pm f_4b)^2}{\alpha\lambda^6}+\frac{f_4^2}{\alpha\lambda^4}.
\end{equation}
From this it is immediate that
\begin{equation}
    \partial^cV_M\equiv 0,\qquad\partial^bV_M=\pm\frac{4f_4}{3 \alpha\lambda^2}(f_6\pm f_4 b),
\end{equation}
resulting in $c$ being a flat direction and $b$ stabilized to $b_0=\mp\frac{f_6}{f_4}$, which in turn makes the subleading term of the potential null. 

\subsubsection*{Two terms example: $f_2,h_0\neq 0$}
Consider now the potential we obtain after setting $h_1=h_2=h_3=f_0=A_{44}-A_{\rm loc}=0$ and require $f_2,h_0\neq 0$, with no extra conditions for other fluxes. This is the case corresponding with the fifth row in Table \ref{table2}. Now we have
\begin{figure}[H]
\begin{center}		
		\begin{tikzcd}[column sep=scriptsize]
		\rho_{30} \arrow[blue]{r} \arrow[Green]{d} & \rho_{32} \arrow[blue]{r} & \rho_{34}\\
		\rho_{52}
		\end{tikzcd}
\end{center}
\end{figure}
\vspace{-20px}
Notice that the partially ordered chain has two end points, resulting in two dominant terms for the potential:
\begin{equation} \label{eq:V-f2h0}
	V_{M} \sim f_{2}^{2} \frac{u}{s} + h_{0}^{2} \frac{s}{u^3} \, .
\end{equation}
Again, we want $V_{M}\to 0$ as $s,u\to\infty$. In this case this translates to $u\ll s \ll u^{3}$ along the would-be asymptotic gradient flow.  The gradient flow equations \eqref{eq:saxionic-flow} translate into
\begin{equation}
    \frac{\partial_\lambda\log s}{\partial_\lambda\log u}=\frac{\beta_s}{\beta_u}=3\frac{-f_2^2\frac{u}{s}+h_0^2\frac{s}{u^3}}{f_2^2\frac{u}{s}-3h_0^2\frac{s}{u^3}},
\end{equation}
which for $(\beta_s,\beta_u)=(\beta,1)$ and $(\alpha_s,\alpha_u)=(\alpha,1)$ (which we are allowed to do because of reparametrization invariance) simplify to
\begin{equation}
   \beta=3\frac{-f_2^2\alpha^{-1}\lambda^{1-\beta}+h_0^2\alpha\lambda^{\beta-3}}{f_2^2\alpha^{-1}\lambda^{1-\beta}-3h_0^2\alpha\lambda^{\beta-3}} \, .
\end{equation}

For $\lambda\to\infty$, we find three different regimes: $\beta\in[1,2)$, $\beta=2$ and $\beta\in(2,3]$. For the former and the latter case the equation completely fixes $\beta$, regardless the value for $\alpha$ or the fluxes, to $\beta=-3$ and $\beta=-1$ respectively.  
These trajectories move us away from the asymptotic regime as they send $s$ to small values and must therefore be discarded. The $\beta=2$ is the only admissible solution, which by \eqref{obt alpha} fixes $\alpha=\frac{\sqrt{5}}{3} \left|\frac{f_{2}}{h_{0}}\right|$, thus resulting in the following \emph{unique} asymptotic trajectory:
\begin{equation}
	(s,u)(\lambda) = \left(\frac{\sqrt{5}}{3} \left|\frac{f_{2}}{h_{0}}\right| \lambda^{2},\lambda\right)\,,
\end{equation}
again up to $\lambda$ reparametrization. The same solution is obtained through the  optimization problem \eqref{eq:min-method}.
This trajectory corresponds to the gradient flow represented in Figure \ref{fig:two examples} (II) (for $f_2=1$ and $h_0=10$). A difference with the previous case is that now there is an attractor for the saxions thanks to the competition between the two terms. Again, one can check that it stays in the $\mathcal{R}_{12}$ growth sector.

The dS coefficient along this asymptotic gradient flow is easily computed as in the previous case, yielding
\begin{equation}
	\gamma_{\vec{f}} =|\hat{\beta}^i\Delta l^{\rm dom}_i|=\frac{|\beta^i\Delta l^{\rm dom}_i|}{\sqrt{\frac{1}{2}\sum_{i=1}^2(\beta^i)^2\Delta d_i}}=\frac{|2\cdot(-1)+1\cdot 1|}{\sqrt{\frac{1}{2}(2^2\cdot 1+1^2\cdot 3)}}= \sqrt{\frac{2}{7}} \approx 0.534522 \,,
\end{equation}
which violates both $c_4^{\rm TCC}$ and $c_4^{\rm strong}$. As it will be later discussed, the potential under consideration only takes into account the complex structure moduli, ignoring the K\"{a}hler moduli, so unless these are somehow stabilized they will also contribute to the dS coefficient. Hence, this example will only yield an accelerating quintessence universe in String Theory if Kähler moduli are independently stabilized.

Once the saxion trajectory is obtained we discuss the axion stabilization. One finds that along the saxion trajectory
\begin{equation}\label{pot2 alonglambda}
    V_M(\lambda,c,b)=\frac{3}{\sqrt{5}}\left|\frac{h_0}{f_2}\right|(f_6\pm f_4 b+\frac{1}{2}f_2b^2-h_0c)^2\lambda^{-5}+\frac{3}{\sqrt{5}}\left|\frac{h_0}{f_2}\right|(f_4\pm f_2b)^2\lambda^{-3}+\frac{14\sqrt{5}}{15}|f_2h_0|\lambda^{-1} \, .
\end{equation}
One can then see that the $\mathcal{O}(\lambda^{-3})$ term stabilizes $b$ at leading order and then in turn the $\mathcal{O}(\lambda^{-5})$ does the same for $c$, resulting in
\begin{equation}\label{eq:ex bc}
    b_0=\pm\frac{f_4}{f_2},\qquad c_0=\frac{f_6}{h_0}-\frac{f_4^2}{2f_2h_0} \, .
\end{equation}
This results in the first two terms of \eqref{pot2 alonglambda}, corresponding to the subleading terms proportional to $\rho_{30}$ and $\rho_{32}$, to vanish asymptotically along the gradient flow.

\subsubsection{Type IIA on CY$_3$ at large volume\label{sec: IIA large V}}

Starting with the Type IIB potential \eqref{eq: IIB pot flux} on $Y_3$ at large complex structure, we can use mirror symmetry to map it to Type IIA theory on the mirror $\tilde{Y}_3$ at large volume \cite{Grimm:2004ua,Grimm:2005fa}. Using that the 4-dimensional dilatons $e^{D_i}=\frac{e^{\phi_i}}{\sqrt{\mathcal{V}_s^i}}$ (with $i=A,B$) are mapped from one to the other, and defining 
\begin{equation}
    s=e^{-\phi_A}\frac{\sqrt{\mathcal{V}_s^A}}{|\Omega^A|} \, ,\qquad u=(\mathcal{V}_s^A)^{\frac{1}{3}} \, ,
\end{equation}
then mirror symmetry identifies 
\begin{equation}
    e^{-\phi_B}\leftrightarrow s,\qquad \mathcal{V}_s^B\leftrightarrow |\Omega^A|^2,\qquad |\Omega^B|^2\leftrightarrow\mathcal{V}_s^A \, ,
\end{equation}
where we have defined the norm for the complex structure 3-forms as $|\Omega ^j|=i\int_{Y_3}\int\bar{\Omega}^j\wedge \Omega ^j$.

This way the $\frac{e^{3\phi_B}}{4(\mathcal{V}_s^B)^2}$ prefactor in $V_{\rm IIB}$ is mapped to $\frac{1}{4s^3|\Omega^A|^4}$ in $V_{\rm IIA}$. Notice that the dilaton dependence of the Type IIA potential includes an extra $\frac{1}{s^3}$ prefactor with respect to the  dilaton dependence of the Type IIB one. In order to obtain this extra prefactor from \eqref{cremmer}, we also have to modify the K\"ahler potential so that in Type IIA we have an effective $\Delta d_s=1+3=4$ and $\Delta d_u=3$. This will result in different values for the dS coefficient. Except for this additional prefactor, the potential looks the same than \eqref{eq:full-potential}, with $u$ now playing the role of the overall Kähler modulus in IIA while $s$ is still the 4d dilaton (see e.g. \cite{DeWolfe:2005uu}). Thus, the two-moduli limit II$_{0,1}\rightarrow$V$_{2,2}$ gets mapped to Type IIA at weak string coupling and large volume. Unlike in IIB, now all bulk moduli appear in the flux superpotential so that there are not additional contributions to $\gamma$ from those computed here. 

Notice that by performing the mirror symmetry to Type IIA, the no-scale structure property is lost, and thus through \eqref{VT} some of the terms of the potential might be negative. From \eqref{V two mod} one can see that only those $V_{\bm l}$ terms featuring $\frac{\|\nabla V_{\bm l}\|}{V_{\bm l}}\geq \sqrt{6}$ will be positive. For type IIA potentials derived from \eqref{eq:full-potential}, this means that coefficients $A_{54}$ and $A_{56}$ (which belong to $\mathcal{E}^{\rm rest}$ and thus could have been studied) will be negative. These corresponds to terms associated to geometric fluxes \cite{Marchesano:2020uqz}. Those potentials featuring $A_{54}$ or $A_{56}$ as leading terms will not feature gradient flows sending the saxions to asymptotic regions, and thus will not be considered. On the other hand, $A_{44}$ would vanish, but the $\frac{1}{s^3}$ term could still appear through $-A_{\rm loc}$, which will receive contributions from localised sources in the compactification manifold (and therefore can be either positive or negative depending on the type of sources).

The procedure to obtain the solutions is the same as in previous cases, so we will not elaborate more on that and simply present the results in Table \ref{table1}. Notice that all the obtained solutions fulfill the $c_4^{\rm strong}$ bound, and thus none of them result in accelerated expansion of the Universe. However, some of them violate the stronger supergravity no-go of \cite{Rudelius:2021azq} by exploiting the loophole discussed in Section \ref{Sec:SuperpotCond}. While for every term (apart from $-A_{\rm loc}$) $\frac{\|\nabla V_{\bm l}\|}{V_{\bm l}}\geq 6$, Scenario (II) with several terms competing asymptotically can still feature $\gamma_{\vec{f}}<\sqrt{6}$ (while keeping a positive potential). This is the case for the first, fourth, fifth, seventh and twelfth potentials from Table \ref{table1}. This last case is presented in further detail as an example in Section \ref{Sec:SuperpotCond}.

\begin{table}[h]
    \centering
    \resizebox{\textwidth}{!}{
    \begin{tabular}{|c||c|c|c||c|}
    \hline
    Potential &$\vec{\beta}$&$\gamma_{\vec{f}}$&$\chi$&Comments\\\hline
    
    $\frac{1}{s^3}\left(\frac{{\color{lightgray}{A_{30}}}}{su^{3}}+\frac{A_{32}}{us}+{\color{lightgray}{A_{34}}}\frac{u}{s}+{\color{Blue}A_{36}}\frac{u^3}{s}{\color{Blue}A_{52}}\frac{s}{u^3}{ -A_ {\rm loc}}\right)$&(3,1)&$3\sqrt{\frac{6}{13}}$&$6w^{-1}$&-\\

   $\frac{1}{s^3}\left(\frac{A_{30}}{su^{3}}+\frac{A_{32}}{us}+{{A_{34}}}\frac{u}{s}+A_{36}\frac{u^3}{s}{\color{Blue} -A_ {\rm loc}} \right)$&(1,0)&$\frac{3}{\sqrt{2}}$&$6w^{-1}$&{{Stabilized $u_0=\left(\frac{A_{32}}{3A_{36}}\right)^{\frac{1}{4}}$, flat $c$}}\\
    $\frac{1}{s^3}\left(\frac{A_{30}}{su^{3}}+\frac{\color{Blue}A_{32}}{us}+{{\color{lightgray}A_{34}}}\frac{u}{s}+{\color{Blue}A_{36}}\frac{u^3}{s}\right)$&(1,0)&$2\sqrt{2}$&$8w^{-1}$&{{Stabilized $u_0=u_0(A_{36}\neq 0,A_{32},A_{30})$, flat $c$}}\\
    $\frac{1}{s^3}\left(\frac{{\color{lightgray}{A_{30}}}}{su^{3}}+\frac{{\color{lightgray}{A_{32}}}}{us}+A_{34}\frac{u}{s}+{\color{Blue}{A_{52}}}\frac{s}{u^3}{\color{Blue} -A_ {\rm loc}}\right)$&(3,1)&$3\sqrt{\frac{6}{13}}$&$6w^{-1}$&-\\
    $\frac{1}{s^3}\left(\frac{{\color{lightgray}{A_{30}}}}{su^{3}}+\frac{{\color{lightgray}{A_{32}}}}{us}+{\color{Blue}A_{34}}\frac{u}{s}+{\color{Blue}{A_{52}}}\frac{s}{u^3}\right)$&(2,1)&$7\sqrt{\frac{2}{19}}$&$7w^{-1}$&-\\
    $\frac{1}{s^3}\left(\frac{A_{30}}{su^{3}}+\frac{{\color{lightgray}{A_{32}}}}{us}+A_{34}\frac{u}{s}{\color{Blue} -A_ {\rm loc}}\right)$&(1,0)&$\frac{3}{\sqrt{2}}$&$6w^{-1}$&{{Subleading $u$, flat $c$}}\\
    
    $\frac{1}{s^3}\left(\frac{\color{lightgray}A_{30}}{su^{3}}+\frac{{{A_{32}}}}{us}+{\color{Blue}A_{52}}\frac{s}{u^3}{\color{Blue} -A_ {\rm loc}}\right)$&(3,1)&$3\sqrt{\frac{6}{13}}$&$6w^{-1}$&{{Flat direction $\rho_{30}(c,b)=0$}}\\
    $\frac{1}{s^3}\left(\frac{\color{lightgray}A_{30}}{su^{3}}+\frac{{{\color{Blue}A_{32}}}}{us}+{\color{Blue}A_{52}}\frac{s}{u^3}\right)$&(1,1)&$5\sqrt{\frac{2}{7}}$&$10w^{-1}$&{{Flat direction $\rho_{30}(c,b)=0$}}\\
    
    $\frac{1}{s^3}\left(\frac{\color{Blue}A_{30}}{su^{3}}+\frac{{\color{lightgray}{A_{32}}}}{us}+{\color{Blue}A_{34}}\frac{u}{s}\right)$&(1,0)&$2\sqrt{2}$&$8w^{-1}$&{{Stabilized $u_0=\left(3\frac{A_{30}}{A_{34}}\right)^{\frac{1}{4}}$, flat $c$}}\\
    $\frac{1}{s^3}\left(\frac{{\color{lightgray}{A_{30}}}}{su^{3}}+\frac{A_{32}}{us}{\color{Blue} -A_ {\rm loc}}\right)$&(1,0)&$\frac{3}{\sqrt{2}}$&$6w^{-1}$&{{Subleading $u$, flat $c$}}\\
    $\frac{1}{s^3}\left(\frac{{\color{lightgray}{A_{30}}}}{su^{3}}+\frac{\color{Blue}A_{32}}{us}\right)$&(3,1)&$\sqrt{\frac{26}{3}}$&$\frac{26}{3}w^{-1}$&{Flat $b$}\\
    
    $\frac{1}{s^3}\left(\frac{{\color{lightgray}{A_{30}}}}{su^{3}}+{\color{Blue}A_{52}}\frac{s}{u^3}{\color{Blue} -A_ {\rm loc}}\right)$&(3,1)&$3\sqrt{\frac{6}{13}}$&$6w^{-1}$&{{Flat $b$}}\\
    $\frac{1}{s^3}\left(\frac{{\color{lightgray}{A_{30}}}}{su^{3}}+{\color{Blue}A_{52}}\frac{s}{u^3}\right)$&{(1,2)}&{$2\sqrt{2}$}&{-}&{No admissible solution}\\
    
    $\frac{1}{s^3}\left(\frac{A_{30}}{su^{3}}{\color{Blue} -A_ {\rm loc}}\right)$&(1,0)&$\frac{3}{\sqrt{2}}$&$6w^{-1}$&{{Subleading $u$, flat $c,b$}}\\
    $\frac{1}{s^3}\left(\frac{\color{Blue}A_{30}}{su^{3}}\right)$&(1,1)&$\sqrt{14}$&$14w^{-1}$&-\\
    $\frac{1}{s^3}\left({\color{Blue} -A_ {\rm loc}}\right)$&(1,0)&$\frac{3}{\sqrt{2}}$&$6w^{-1}$&{{Flat $u,c,b$}}\\
    
    \hline
    \end{tabular}}
    \caption{Solutions (in case they exist) of the different potentials obtained in Type IIA at large volume and weak coupling. It can be obtained from mirror symmetry of the enhancement II$_{0,1}\rightarrow$V$_{2,2}$ in IIB. Even if $d_1=1,\,d_2=4$, we need to include an extra 3 in $\Delta d_1$ to account for the $s^{-3}$ contribution upon mirror symmetry. In those relevant cases, we distinguish between $ -A_ {\rm loc}$ null and different from 0. Those $A_{\bm{l}}$ functions colored in {\color{blue}blue} correspond to those dominating terms and those in {\color{lightgray}gray} are exponentially suppressed through axion stabilization. The fourth column corresponds to the coefficient $\chi$ relating the decay rates of the potential and the leading tower becoming light, modulo $w=1,\,2,\,3$, which will be introduced in Section \ref{sec:distance conj}.}
    \label{table1}
\end{table}

As commented in several examples of Tables \ref{table2} and \ref{table1}, for some potentials $u$ stabilizes to some finite value $u_0$ which depends on the fluxes. This happens when the terms that would be considered leading push $u$ to lower values, for which subleading terms become relevant and can compensate and cancel the first term. As an example, in the ninth potential from Table \ref{table1}, $V_M=\frac{A_{30}}{s^4u^3}+A_{34}\frac{u}{s^4}$, one has $\partial_u V_M\propto -3A_{30}u^{-4}+A_{34}$, which results in the stabilization point $u_0=\left(3\frac{A_{30}}{{A_{34}}}\right)^{\frac{1}{4}}$ and $\beta_2=0$. As $A_{34}$ is independent of the axions and for $A_{30}$ $b$ is already stabilized ($c$ not appearing in the potential), then the $u_0$ value only depends on the quantized fluxes. Similar arguments can be used to determine where $u$ is stabilized in the rest of potentials.

\subsubsection{Another limit: F-theory II$_{0,1}\rightarrow$III$_{0,0}$ enhancement\label{sec:dif1}} 
Let us consider now a different asymptotic limit corresponding to the second example in Table \ref{table summary}, namely the II$_{0,1}\rightarrow$III$_{0,0}$ enhancement (thus having $\Delta d_1=\Delta d_2=1$). This corresponds to another type IIB weakly coupled string theory limit  which is not the large complex structure limit (since it is still a type II singularity enhanced to a different singularity, in this case of type III). Again, in order to have $V\rightarrow 0$, all fluxes in $\mathcal{E}^{\rm heavy}$ must vanish. The remaining possible flux choices and their associated solutions are depicted in Table \ref{table3}. Since we are considering an enhancement to a lower degree singularity, this potential has less possibilities than the one in Table \ref{table2}.

Only one flux configuration of the $V_M$ potential results in an admissible solution, though it does not violate any of the proposed bounds and would give rise to a non-accelerated Universe. This is somehow expected from looking at the expression for the dS coefficient \eqref{dS gen}, since higher singularity degrees $d_i$ seem to produce a smaller value for $\gamma$. 

\begin{table}[h]
    \centering
    \resizebox{\textwidth}{!}{
    \begin{tabular}{|c||c|c|c||c|}
    \hline
    Potential &$\vec{\beta}$&$\gamma_{\vec{f}}$&$\chi$& Comments\\\hline
    
    $ \frac{{\color{lightgray}{A_{32}}}}{us}+{ A_{34}}\frac{u}{s}+{ A_{54}}\frac{s}{u}+{ A_{44}-A_ {\rm loc}}$&(1,1) $\star$&0&-&{{Flat $s,u$ after axion stab, $\gamma_{\vec{f}}:2\rightarrow 0$}}\\
    $ \frac{{\color{lightgray}{A_{32}}}}{us}+{\color{Blue}A_{34}}\frac{u}{s} $&{(1,-1)}&{$2$}&-&{\small{No admissible solution}, {flat $c$}}\\
    $ \frac{{\color{lightgray}{A_{32}}}}{us}+{\color{Blue}A_{54}}\frac{s}{u} $&{(-1,1)}&{$2$}&-&{\small{No admissible solution}, {flat $c$}}\\
    $ \frac{\color{Blue}A_{32}}{us}$&(1,1) &$2$&$4w^{-1}$&{Flat $c,b$}\\
    
    \hline
    \end{tabular}}
    \caption{Solutions (in case they exist) of the different potentials obtained after axion stabilization in the limit II$_{0,1}\rightarrow$ III$_{0,0}$ $(d_1=1,\,d_2=2)$, corresponding to a type IIB string theory weak coupling limit away from large complex structure. In those relevant cases, we distinguish between $A_{44}-A_ {\rm loc}$ null and different from 0. There is also comment on the possible stabilization of saxions (paying special attention to the possibility they stabilize at 0 value) and the axions being along flat directions. Note that as this potential is symmetric under $s\leftrightarrow u$ those ``non admissible solutions'' are actually the solution to $V$ in the $\mathcal{R}_{21}$ growth sector, in which the potential is the same. Those $A_{\bm{l}}$ functions colored in {\color{blue}blue} correspond to those dominating terms and those in {\color{lightgray}gray} are exponentially suppressed through axion stabilization. Note that in some cases (marked by $\star$) the (preliminary) leading terms determining $\vec{\beta}$ turn out to be constant when evaluated along the solution, and must be canceled by $A_{44}-A_{\rm loc}$ through an adequate fine tuning of the fluxes, with $\gamma_{\vec{f}}$ determined by subleading terms (which we now color in {\color{blue}blue}). The fourth column corresponds to the coefficient $\chi$ relating the decay rates of the potential and the leading tower becoming light, modulo $w=1,\,2,\,3$, which will be introduced in Section \ref{sec:distance conj}.}
    \label{table3}
\end{table}
\subsubsection{Another limit: F-theory III$_{1,1}\rightarrow$V$_{2,2}$ enhancement\label{sec:dif2}}
As we saw above, singularities of less order give potentials with less possibilities and that, generically, yield higher dS coefficients. Thus, it is interesting to consider enhancements with higher degree singularities. This forces us to go beyond the Type IIB weak coupling limit, since it corresponds to the infinite distance singularity of smallest degree. For example, let us take the III$_{1,1}\rightarrow$V$_{2,2}$ singularity enhancement (so that $\Delta d_s=\Delta d_u=2$), which is the third example in Table \ref{table summary}. This does not have a Type IIB interpretation since we are not at weak string coupling anymore, and corresponds to a very different limit in the complex structure moduli space of F-theory on a CY$_4$. 

From Table 5.3 in \cite{Grimm:2019ixq}, we have the following partially ordered chain of terms:
\begin{figure}[H]
\begin{center}		
		\begin{tikzcd}[column sep=scriptsize]
		\rho_{20} \arrow[blue]{r} \arrow[Green]{d} & \rho_{22} \arrow[blue]{r} \arrow[Green]{d} & \rho_{24} \arrow[Green]{d}\\
		\rho_{42} \arrow[blue]{r} \arrow[Green]{d} & \rho_{44} \arrow[blue]{r} \arrow[Green]{d} & \rho_{46} \arrow[Green]{d}\\
		\rho_{64} \arrow[blue]{r} &\rho_{66}\arrow[blue]{r}& \rho_{68}
		\end{tikzcd}
\end{center}
\end{figure}
\vspace{-20px}
Now, as the $\rho_{46}$, $\rho_{66}$ and $\rho_{68}$ terms are part of $\mathcal{E}^{\rm heavy}$ (respectively or order $s^2$, $u^2$ and $s^2u^2$) we must set them to zero by turning off the corresponding fluxes. We are thus in a situation in which the dominant terms are those associated to $\rho_{24}$, $\rho_{44}$\footnote{Notice that in the previous cases this term was in a different ``connected component'' of the chain, and thus its value was independent from the rest. This is no longer the case.} and $\rho_{64}$. This is the starting point for Table \ref{table4}, in which we present the results for all the possible flux choices giving rise to different potentials vanishing asymptotically. The result for $\gamma$ is shown in the Table for each case.  Interestingly, there is one case that violates the Strong dS bound \cite{Rudelius:2021azq}, so let us explain this example in more detail.

\begin{table}[h]
    \centering
    \resizebox{\textwidth}{!}{
    \begin{tabular}{|c||c|c|c||c|}
    \hline
    Potential &$\vec{\beta}$&$\gamma_{\vec{f}}$&$\chi$& Comments\\\hline
    
    $ \frac{\color{Blue}A_{20}}{s^2u^2}+\frac{{\color{lightgray}{A_{22}}}}{s^2}+\frac{{\color{lightgray}{A_{42}}}}{u^2}+{ A_{24}}\frac{u^2}{s^2}+{ A_{64}}\frac{s^2}{u^2}+{ A_{44}-A_ {\rm loc}} $&(1,1) $\star$&$2\sqrt{2}$&$8w^{-1}$&-\\
     $ \frac{\color{Blue}A_{20}}{s^2u^2}+\frac{{\color{lightgray}{A_{22}}}}{s^2}+\frac{{\color{lightgray}{A_{42}}}}{u^2}+{ A_{24}}\frac{u^2}{s^2}+{ A_{64}}\frac{s^2}{u^2}-{ A_ {\rm loc}} $&(1,1) $\star$&$2\sqrt{2}$&$8w^{-1}$&-\\
    
      $ \frac{\color{lightgray}{A_{20}}}{s^2u^2}+\frac{{\color{lightgray}{A_{22}}}}{s^2}+\frac{\color{Blue}A_{42}}{u^2}+{\color{Blue}A_{24}}\frac{u^2}{s^2}$&(2,1)&$\frac{2}{\sqrt{5}}$&$2w^{-1}$&\textbf{Violates $\bm{c_4^{\rm strong}}$}\\
      $ \frac{\color{lightgray}{A_{20}}}{s^2u^2}+\frac{\color{Blue}A_{22}}{s^2}+\frac{{\color{lightgray}{A_{42}}}}{u^2}+{\color{Blue}A_{64}}\frac{s^2}{u^2}$&{{(1,2)}}&{{$\frac{2}{\sqrt{5}}$}}&{{$2w^{-1}$}}&{Change of growth sector, }\textbf{Violates $\bm{c_4^{\rm strong}}$}\\
         
   $ \frac{\color{Blue}A_{20}}{s^2u^2}+\frac{\color{lightgray}{A_{22}}}{s^2}+\frac{\color{lightgray}{A_{42}}}{u^2}+{ A_{44}-A_ {\rm loc} }$&(1,1)&$2\sqrt{2}$&$8w^{-1}$&-\\
    $ \frac{{\color{lightgray}{A_{20}}}}{s^2u^2}+\frac{\color{Blue}A_{22}}{s^2}+\frac{\color{Blue}A_{42}}{u^2}$&(1,1)&$\sqrt{2}$&$4w^{-1}$&Saturates $c_4^{\rm strong}$\\
     $ \frac{{\color{lightgray}{A_{20}}}}{s^2u^2}+\frac{\color{Blue}A_{22}}{s^2} $&(1,0)&2&$4w^{-1}$&{\small{{Flat $u,\,c$}}}\\
     $ \frac{{\color{lightgray}{A_{20}}}}{s^2u^2}+\frac{\color{Blue}A_{42}}{u^2} $&{{(0,1)}}&{2}&{{$4w^{-1}$}}&{Change of growth sector, {Flat $s,\,{b}$}}\\
     $ \frac{\color{Blue}A_{20}}{s^2u^2}$&(1,1)&$2\sqrt{2}$&$8w^{-1}$&{Flat $c,b$}\\
    
    \hline
    \end{tabular}}
    \caption{Solutions (in case they exist) of the different potentials after axion stabilization, in the F-theory limit corresponding to the III$_{1,1}\rightarrow$V$_{2,2}$ $(d_1=2,\,d_2=4)$   enhancement. In this case $A_{44}$ is not a constant, but rather appears also in the axionic chains and thus cannot take non-zero values after turning off too many fluxes. There is also comment on the possibility of the axions being along flat directions. Note that as this potential is symmetric under $s\leftrightarrow u$, the potential is the same under both $\mathcal{R}_{12}$ and $\mathcal{R}_{21}$, so that we should not worry about the solution being in one or another regime. Those $A_{\bm{l}}$ functions colored in {\color{blue}blue} correspond to those dominating terms and those in {\color{lightgray}gray} are exponentially suppressed through axion stabilization. Note that in some cases (marked by $\star$) the (preliminary) leading terms determining $\vec{\beta}$ turn out to be constant when evaluated along the solution, and must be canceled by $A_{44}-A_{\rm loc}$ through an adequate fine tuning of the fluxes, with $\gamma_{\vec{f}}$ determined by subleading terms (which we now color in {\color{blue}blue}). The fourth column corresponds to the coefficient $\chi$ relating the decay rates of the potential and the leading tower becoming light, modulo $w=1,\,2,\,3$, which will be introduced in Section \ref{sec:distance conj}.}
    \label{table4}
\end{table}

Consider the example in the third row of Table \ref{table4}, which has $A_{64}=A_{44}-A_{\rm loc}=0$ and therefore the following chain:
\begin{figure}[H]
\begin{center}		
		\begin{tikzcd}[column sep=scriptsize]
		\rho_{20} \arrow[blue]{r} \arrow[Green]{d} & \rho_{22} \arrow[blue]{r}&\rho_{24} \\
		\rho_{42}
		\end{tikzcd}
\end{center}
\end{figure}
\vspace{-20px}
This corresponds to the following potential
\begin{equation}
    V_M=\frac{A_{20}}{s^2u^2}+\frac{A_{42}}{u^2}+\frac{A_{22}}{s^2}+A_{24}\frac{u^2}{s^2} \, .
\end{equation}

Notice, from the diagram, that the dominant terms are those associated to $\rho_{24}$ and $\rho_{42}$. These, as always, do not depend on the axions so we treat them as constants.

Solving \eqref{eq:min-method} and \eqref{obt alpha} it is easy to see that the gradient flow of the saxions go to infinity along the following attractor:
\begin{equation}
    (s,u)(\lambda)=\left(\sqrt{\frac{2A_{24}}{A_{42}}}\lambda^2,\lambda\right) \, .
\end{equation}
This trajectory stays in the growth sector $\mathcal{R}_{12}$. The dS coefficient is computed to be $\gamma=\frac{2}{\sqrt{5}}\approx0.89443$, which violates $c_4^{\rm strong}$ but not $c_4^{\rm TCC}$, thus resulting in an accelerated expansion of the Universe that nonetheless satisfies the TCC bound \eqref{TCC bound}. 
As in previous examples, this requires that the K\"ahler moduli are stabilized, otherwise they give an additional contribution to the dS coefficient.

As for the axions, we first notice that the $\rho_{22}$ term always dominates over the $\rho_{20}$ one. One can easily see from the diagram that $\rho_{22}$ will be linear on the axion $b$ and independent of $c$. This term will be responsible of stabilizing $b$ to the value $b_0$ such that $\rho_{22}=0$. Similarly, we read from the diagram that $\rho_{20}$ is linear in $c$ and quadratic in $b$. However, the previous term has already fixed $b=b_0$, so we end up with a term that is only linear in $c$ and trivially stabilizes it to the value $c_0$ such that $\rho_{20}=0$.\footnote{If we set $\rho_{42}=\tilde h_1$ and $\rho_{24}=\tilde f_1$ then $\rho_{22}=-\tilde f_1b+\tilde f_0$ and $\rho_{20}=\frac{1}{2}\tilde f_1b^2-\tilde f_0b-\tilde h_1c+\tilde h_0$, so that $(b_0,c_0)=\left(\frac{\tilde f_0}{\tilde f_1},\frac{1}{\tilde h_1}\left(\frac{\tilde f_0^2}{2\tilde f_1}-\tilde h_0\right)\right)$}

\subsection{Summary of results and comparison with Swampland bounds \label{sec: gen bounds}}

We now turn to compare our results with previous bounds found in the literature for CY's \cite{Cicoli:2021fsd,Bastian:2020egp} as well as the Swampland bounds on the potential \cite{Bedroya:2019snp,Rudelius:2021azq}. 
As explained in Section \ref{Sec:SwamplandBounds}, it was proposed in \cite{Rudelius:2021azq} that string theory does not allow for accelerated expansion asymptotically, implying the bound \eqref{strong bound} on the dS coefficient which we repeat here for convenience,
\beq
\gamma_{}\geq c_d^{\rm strong}=\frac{2}{\sqrt{d-2}} \, .
\label{boundac}
\eeq 
This results in $\gamma\geq \sqrt{2}$ in four dimensions.
The same conclusion was also claimed in \cite{Cicoli:2021fsd} after checking some examples in perturbative string theory. However, in the previous Section, we have found potential counterexamples to these bounds which might lead to asymptotic accelerated expansion in string theory. These have some caveats, which we discuss next, but at the very least exemplify that the question of whether string theory allows for asymptotic accelerated expansion is far from settled, since the bound \eqref{boundac} can be violated at the level of the flux potential of a CY compactification.\\

There are two key differences between our work and previous literature that explain why we find, for first time, potential counterexamples to \eqref{boundac} in the realm of flux CY compactifications. First, we allow for various terms of the potential to compete asymptotically along the gradient flow trajectory, a scenario that had been missed and that presents a loophole for the supergravity no-go in \cite{Hellerman:2001yi,Rudelius:2021oaz}. Secondly, we study the gradient flow trajectories along more general infinite distance limits that go beyond the canonical large volume and weak string coupling point, extending this way previous results \cite{Hertzberg:2007wc,Garg:2018zdg, ValeixoBento:2020ujr,Andriot:2020lea,Andriot:2022xjh,Cicoli:2021fsd}. In the following, we will elaborate on these two key differences and summarize the potential counterexamples that we have found.

Recall the two scenarios outlined in Figure \ref{fig:two examples}. Scenario (I) consists of a single term of the potential dominating asymptotically. In this case, the no-go of \cite{Rudelius:2021azq} applies and no accelerated expansion is possible.
Some bounds (implicitly assuming this Scenario (I)) were also given in \cite{Bastian:2020egp} for the case of Calabi-Yau threefolds, in agreement with \cite{Rudelius:2021azq}. In fact, we will extend these bounds in Section \ref{sec:bounds convex hull} to show that Scenario (I) can never yield accelerated expansion. 
On the other hand, in Scenario (II), two or several terms compete asymptotically along the gradient flow trajectory, so that they generate a single valley approaching the infinite distance point. The dS coefficient can get lower in this case, evading previous bounds and potentially yielding accelerated expansion.
While in previous works \cite{Rudelius:2021azq,Bastian:2020egp} the studied potentials featured a single dominant term belonging to $\mathcal{E}^{\rm light}$ (i.e. Scenario (I))), our potential counterexamples come precisely from considering the possibility of several dominant terms as in Scenario (II). When the dominant terms belong to $\mathcal{E}^{\rm rest}$, there seems to be no lower bound for the dS coefficient at the level of the flux potential, as it will become more clear in Section \ref{sec:bounds convex hull}. 

By checking concrete examples in Section \ref{sec:examples}, we have found two cases that violate \eqref{boundac}.  The first one (fifth row in Table \ref{table2}) corresponds to IIB at weak coupling and large complex structure. It can even evade the TCC bound, having $\gamma =\sqrt{\frac{2}{7}}\approx0.5345$. Even if it gets smaller than the TCC bound, it is only marginally compatible with experimental cosmological bounds on our universe (see e.g. \cite{Raveri:2018ddi, Heisenberg:2018yae, Akrami:2018ylq}), since the strongest bounds imply $\gamma\simleq 0.51-0.54$. The other example (third row in Table \ref{table4}) does not have a type IIB interpretation, but rather a different limit in the F-theory complex structure moduli space of a CY$_4$ , for which the $\gamma=\frac{2}{\sqrt{5}}\approx0.89443$ value evades the Strong dSC but not the TCC bound.

As mentioned, the other key difference with respect to previous works \cite{Hertzberg:2007wc,Garg:2018zdg, ValeixoBento:2020ujr,Andriot:2020lea,Andriot:2022xjh,Cicoli:2021fsd} is that we go beyond the large volume and weak coupling limit of perturbative Type IIA string theory (mirror dual to perturbative Type IIB at large complex structure). 
From the discussion in section \ref{Sec:SuperpotCond}, it is clear that there is a priori no obstruction to violate \eqref{boundac} at the level of the flux potential, and it seems a matter of finding some infinite distance limit with the appropriate discrete data $(\Delta d_i,\Delta l_i)$ to generate a valley along which $\gamma$ gets small.

We want to remark again that the weak string coupling limit is just one type of infinite distance limit of many others existing in string theory. In this paper, we have considered two further asymptotic limits beyond weak coupling/large complex structure, but there are eight more possibilities arising as 2-moduli limits classified in \cite{Grimm:2019ixq}, and many more if considering more moduli. And this is without even going beyond Calabi-Yau compactifications.
Notice that each possible infinite distance limit corresponds to a different asymptotic regime that will provide solutions at parametric control, since we can follow the potential infinitely far away. Hence, proving the absence of asymptotic accelerated expansion is a daunting task if attempted to be shown on a case by case basis, even if restricted to CY compactifications. In this paper, we are far from achieving such goal, but we consider at least other infinite distance limits beyond weak coupling in Type II, which allow us to already provide some potential examples of accelerated expansion. Notice also that the weak coupling limit has the lowest possible singularity degree $\Delta d_1=1$, so it seems a good idea to go beyond weak coupling if one is looking for decreasing $\gamma$ as $\Delta d_i$ appears in the denominator (see \eqref{eq:gamma from lnV}). Going beyond weak coupling was indeed essential to find the counterexample with $\gamma=\frac{2}{\sqrt{5}}$ in Table \ref{table4}. To our surprise, we were also able to realize scenario (II) and find an asymptotic valley along which $\gamma =\sqrt{\frac{2}{7}}$ in the complex structure moduli space of perturbative Type IIB (but not of Type IIA\footnote{ It is worth noticing that, even if the Type IIA flux potential studied in \cite{Cicoli:2021fsd} is indeed the same as our example in Section \ref{sec: IIA large V}, we still get different results. This is because we have computed the dS coefficient along gradient flow trajectories, and they differ from some of the trajectories considered in \cite{Cicoli:2021fsd}.
We believe that gradient flows are the most natural trajectories to consider, since they solve the equations of motion asymptotically, and we expect that any other trajectory will eventually become a gradient flow asymptotically.}).
However, as we will see in Section \ref{sec:distance conj}, this second example is in tension with the mass scale for the tower of states predicted by the Distance Conjecture, while the first one seems consistent.

Now, the caveats. In order to consider other types of infinite distance limits, we have studied examples in the complex structure moduli space of F-theory Calabi-Yau fourfolds with $G_4$ flux. This has the advantage that it allows us to explore more exotic limits, but the disadvantage that K\"ahler moduli do not get stabilized by fluxes. Therefore, to promote our examples to fully controlled accelerated cosmologies one first needs to stabilize somehow  the K\"ahler moduli (which is a problem on its own) while keeping some complex structure moduli rolling down the potential. As detailed in Section \ref{sec:no-scale}, the K\"ahler moduli needs to get stabilized because otherwise the runaway on the overall volume is too large to yield accelerated expansion  \cite{Burgess:2022nbx,Cicoli:2021fsd}. In any case, this is also a requirement of any phenomenologically viable quintessence model, since the experimental bounds coming from 5th forces are very strong if the quintessence field is the overall volume (as it affects the value of the Planck scale), while they are much weaker otherwise.\\

To sum up, the flux potential in Calabi-Yau compactifications does not automatically satisfy the Strong dS conjecture \eqref{strong bound}
(not even the TCC \eqref{TCC bound}). This had been missed before because it is crucial to consider gradient flow trajectories in which two or more terms in $V$ compete asymptotically to generate a valley. Hence, there is a priori no obstruction to yield accelerated expansion asymptotically. However, in practice, the explicit potential counterexamples to the Strong dS conjecture that we have found, have caveats. More concretely, we are focusing on a subsector of the field space, staying agnostic about K\"ahler moduli stabilization, since we were just interested in whether the slope of the flux potential can be made smaller than what previous bounds suggested \cite{Rudelius:2021azq,Cicoli:2021fsd,Bastian:2020egp}. Hence, these examples are not enough to conclude one way or another. It could very well be that, in the end, string theory forbids asymptotic accelerated expansion, but this is going to occur in a more convoluted way which requires to study ingredients beyond the flux potential.

\section{A Convex Hull Formulation for the dS Conjecture\label{sec: convex hull and SDC} }

In this section, we will reformulate the Strong dS conjecture as a convex hull scalar WGC for membranes, similar to the convex hull condition for the Distance conjecture \cite{Calderon-Infante:2020dhm}. This will allows us to provide bounds for the dS coefficient in certain scenarios. We will also check the consistency of our results with the presence of the tower of states predicted by  the Distance conjecture \cite{Ooguri:2006in}.

\subsection{Convex Hull  dS conjecture\label{sec: Convex hull}}

One can further rewrite our results in a language much closer to the one used in the formulation of the Convex Hull WGC \cite{Cheung:2014vva, Palti:2017elp} and Convex Hull SDC \cite{Calderon-Infante:2020dhm}. In the same way that the Distance conjecture can be written as a Convex Hull condition for the Scalar WGC for particles (see explanation in \cite{Calderon-Infante:2020dhm}), we will see here that the Strong de Sitter conjecture can be written as a Convex Hull condition for the Scalar WGC for membranes.

Let us first take the unit tangent vector of the trajectory in field space $\hat T$ to an orthonormal basis. This is, given an $n$-vein $e^{a}_{i}$ for the metric in field space we write
\begin{equation}
    t^a = e^{a}_{i} \, \hat{T}^i \, .
\end{equation}
Particularizing for the field space metric in \eqref{eq:field-metric} and parametrizing the trajectory as in \eqref{eq:saxions-ansatz} this yields
\begin{equation} \label{beta to t transform}
    t^i = \sqrt{\frac{d_i}{2}} \, \beta^i \quad \text{(no sum over } i \text{)}\, .
\end{equation}
Notice that $t^2=t^it^j\delta_{ij}=|\vec{\beta}|^2$ as defined in \eqref{norm beta}. In a similar manner, for each term $V_{\bm{l}}$ of the potential $V=\sum_{\bm{l}\in\mathcal{E}}V_{\bm{l}}$ we define its \emph{dS ratio} as
\begin{equation}\label{l to mu transform gen}
    \mu^a_{\bm{l}} = - \delta^{ab} e_{b}^{i} \, \frac{\partial_i V_{\bm{l}}}{V_{\bm{l}}} \, ,
\end{equation}
where we are using the inverse $n$-vein $e_{b}^{i}$ to also take it to the orthonormal basis. This particularizes for the case of the metric \eqref{eq:field-metric} and potential \eqref{eq:pot} to
\begin{equation}\label{l to mu transform}
    \mu^i_{\bm{l}} = - \sqrt{\frac{2}{d_i}} \, l_i \quad \text{(no sum over } i \text{)}\, \, .
\end{equation}
Notice that ${\beta}^{i} l_{i} = -\vec{t} \cdot \vec{\mu}_{\bm{l}}$, with the dot representing the usual Cartesian scalar product. From \eqref{dS gen}, it is then immediate that the de Sitter coefficient reads
\begin{equation} \label{eq:dS-flat-coordinates}
    \gamma_{\vec{f}}=-\hat{\beta}^{i} l_{i}^{\rm dom} = \, \hat{t} \cdot \vec{\mu}^{\rm dom} \, ,
\end{equation}
with $\hat{t}^2=1$. This is just a particular realization of the general fact that $\gamma_{\vec{f}}$ is the projection of the dominant dS ratio(s) $\vec{\mu}^{\rm dom}$ along the direction $\hat{t}$. In this language, the optimization problem in \eqref{eq:min-method} can be written as:
\begin{equation} \label{eq:min-method-2}
	\gamma_{\vec{f}}=\max_{\hat{t} \in \mathbb{S}^{{n}}} \left\{ \min_{\left\{ \vec{\mu} \right\}} \left\{\hat{t} \cdot \vec{\mu} \right\} \right\} \quad \text{with} \quad \mathbb{S}^{n} = \left\{\hat{t}\in\mathbb{R}^{n}: |\hat{t}|^{2} = \sum_{i=1}^{n} (\hat{t}^i)^2=1 \right\} \, .
\end{equation} 
One can show\footnote{
By definition,
\begin{equation}
    {\rm Hull}(\{\vec{\mu}\})=\left\{\sum_{\bm{l}}\alpha_{\bm l}\vec{\mu}_{\bm l}:\sum_{\bm l}\alpha_{\bm l}=1,\; \alpha_{\bm l}\geq 0\right\},\qquad {d_{\rm{Eucl}}}(\vec{0},{\rm{Hull}}(\{\vec{\mu}\}))=\inf_{\vec{x}\in {\rm{Hull}}(\{\vec{\mu}\})}|\vec{x}| \, .
\end{equation}
We will assume that $\vec{0}\not\in {\rm Hull}(\{\vec{\mu}\})$ (the other case, in which ${d_{\rm{Eucl}}}(\vec{0},{\rm{Hull}}(\{\vec{\mu}\}))=0$, will be covered later). As ${\rm{Hull}}(\{\vec{\mu}\})$ is a convex set not containing the origin, the above infimum exists and corresponds to some unique point belonging to ${\rm{Hull}}(\{\vec{\mu}\})$, which we will denote by $\vec{x}_0=\sum_{\bm l}\alpha_{\bm l}^{(0)}\vec{\mu}_{\bm l}\neq \vec{0}$. Defining $\hat{t}_0=\frac{\vec{x}_0}{|\vec{x}_0|}\in\mathbb{S}^n$, then ${d_{\rm{Eucl}}}(\vec{0},{\rm{Hull}}(\{\vec{\mu}\}))=\hat{t}_0\cdot\vec{x}_0\geq \hat{t}\cdot\vec{x}_0$ for all $\hat{t}\in\mathbb{S}^n$. Then it is immediate that
\begin{equation}
    {d_{\rm{Eucl}}}(\vec{0},{\rm{Hull}}(\{\vec{\mu}\}))=\max_{\hat{t}\in\mathbb{S}^n}\{\hat{t}\cdot \vec{x}_0\} \, .
\end{equation}
On the other hand, $\hat{t}_0\cdot \vec{x}_0\leq \frac{\vec{x}}{|\hat{x}|}\cdot \vec{x}$ for any $\vec{x}\in {\rm{Hull}}(\{\vec{\mu}\}))$. This translates in 
\begin{equation}
    {d_{\rm{Eucl}}}(\vec{0},{\rm{Hull}}(\{\vec{\mu}\}))=\max_{\hat{t}\in\mathbb{S}^n}\left\{
        \min_{\alpha_{\bm{l}}}\left\{\sum_{\bm{l}}\alpha_{\bm{l}}\hat{t}\cdot\vec{\mu}_{\bm{l}}
    \right\}\right\}\, ,
\end{equation}
with $\sum_{\bm l}\alpha_{\bm l}=1$ and $\alpha_{\bm l}\geq 0$. Now, given some fixed $\hat{t}\in \mathbb{S}^n$ it follows that $\min_{\alpha_{\bm{l}}}\left\{\sum_{\bm{l}}\alpha_{\bm{l}}\hat{t}\cdot\vec{\mu}_{\bm{l}}
    \right\}=\min_{\vec{\mu}_{\bm{l}}}\{\hat{t}\cdot\vec{\mu}_{\bm l}\}$, as this corresponds to setting $\alpha_{\bm{l}}=0$ for those $\bm{l}$ such that $\hat{t}\cdot\vec{\mu}_{\bm{l}}>\min_{\vec{\mu}_{\bm{l}}}\{\hat{t}\cdot\vec{\mu}_{\bm l}\}$. This way we end up with ${d_{\rm{Eucl}}}(\vec{0},{\rm{Hull}}(\{\vec{\mu}\}))=\max_{\hat{t} \in \mathbb{S}^{{n}}} \left\{ \min_{\left\{ \vec{\mu} \right\}} \left\{\hat{t} \cdot \vec{\mu} \right\} \right\}=\gamma_{\vec{f}}$, as we wanted to show.
}
that in fact $\gamma_{\vec{f}}={d_{\rm{Eucl}}}(\vec{0},{\rm{Hull}}(\{\vec{\mu}\}))$, the Euclidean distance from the origin to the convex hull of the $\{\vec{\mu}_{\bm l}\}_{{\bm{l}}\in\mathcal{E}}$. The dominant terms $\{\vec{\mu}^{\rm dom}\}$ will be those spanning the minimal convex hull at the closest distance, i.e. $\gamma_{\vec{f}}={d_{\rm{Eucl}}}(\vec{0},{\rm{Hull}}(\{\vec{\mu}\}))={d_{\rm{Eucl}}}(\vec{0},{\rm{Hull}}(\{\vec{\mu}^{\rm dom}\}))$. The tangent vector $\hat{t}$ corresponding to the asymptotic gradient flow trajectory becomes the direction along which the convex hull is reached first. Notice that if $\hat{t} \cdot \vec{\mu}_{i}^{\rm dom}=0$ then the potential is flat in this subspace. 
We will explain later how to proceed in this special case, and  consider for the moment the general case with ${d_{\rm{Eucl}}}(\vec{0},{\rm{Hull}}(\{\vec{\mu}\}))>0$.

In this language, the scenario (I) in which a single term dominates the potential corresponds to ${\rm Hull}(\{\vec{\mu}^{\rm dom}\})=\vec{\mu}^{\rm dom}$ (the convex hull of a single point is the point itself) and $\hat{t}_0$ along the direction joining it with the origin. The dS coefficient is given by the distance from the origin to $\vec{\mu}^{\rm dom}$, $\gamma_{\vec{f}}=|\vec{\mu}^{\rm dom}|$. On the other hand, in the scenario (II) we have several terms $\vec{\mu}_{\bm{l}}^{\rm dom}$ that compete along the gradient flow. This means that $\hat{t} \cdot \vec{\mu}_{\bm{l}}^{\rm dom}=\hat{t} \cdot \vec{\mu}_{\bm{m}}^{\rm dom}$ for all dominant terms so that the dS coefficient is then given by $\gamma_{\vec{f}}=\hat{t} \cdot \vec{\mu}_{\bm{l}}^{\rm dom}$. In fact, one gets the same result if evaluated along any point of the convex hull of the dominant terms.

We must recall that, as there is a unique closest point between $\vec{0}$ and a convex set such as ${\rm Hull}(\{\vec{\mu}\})$, then the asymptotic gradient flow trajectory $\hat{t}$ (and thus $\vec{\beta}$) is uniquely determined, as it is also shown in Appendix \ref{APP1} through different means.

\begin{figure}[htb]
\centering
\includegraphics[width=0.75\textwidth]{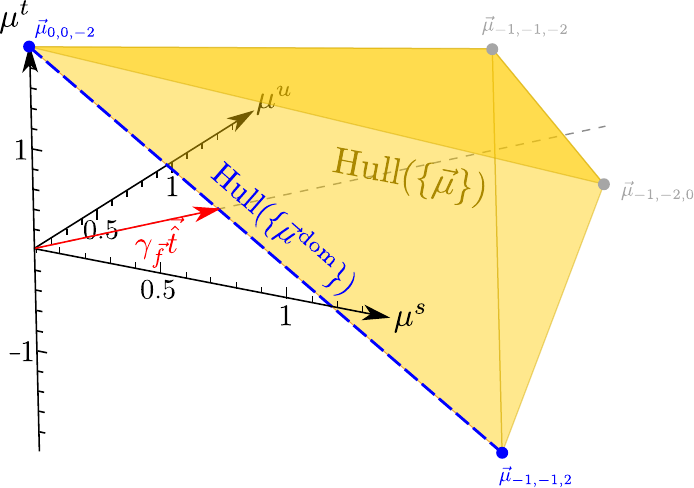}
\caption{Example of how the convex hull prescription solves the problem for $n=3$ saxions $s,\, u$ and $t$. The potential $V=\frac{1}{sut^2}+\frac{1}{su^2}+\frac{1}{t^2}+\frac{t^2}{su}$ and we take $(d_s,d_u,d_t)=(1,1,2)$, so that $\{\vec{\mu}\}=\{(\sqrt{2}, \sqrt{2}, 2), (\sqrt{2}, 2\sqrt{2}, 0), (0, 0, 2), (\sqrt{2}, \sqrt{2}, -2)\}$. It is easily found that the closest point to the origin is $\left(\frac{2 \sqrt{2}}{5},\frac{2 \sqrt{2}}{5},\frac{2}{5}\right)$, belonging to ${\rm Hull}(\{\vec{\mu}_{0,0,-2},\vec{\mu}_{-1,-1,2}\})$. We find that its length is $\gamma_{\vec{f}}=\frac{2}{\sqrt{5}}$, and doing the inverse \eqref{beta to t transform} transformation, we recover a $\vec{\beta}=(2,2,1)$ trajectory. Note that $\gamma_{\vec{f}}\,\hat{t}$ is indeed perpendicular to ${\rm Hull}(\{\vec{\mu}^{\rm dom}\})$, but the different axis ratios might make this difficult to perceive.}
\label{fig:convex hull3}
\end{figure}

This way, we are led to propose the following reformulation for the optimization problem in \eqref{eq:min-method-2}, in which now the dS coefficient is given by
\begin{equation}\label{eq:convexhull-dist}
    \gamma_{\vec{f}}=\min\left\{d_{\rm{Eucl}}\left(\vec{0},{\rm Hull}(\{\vec{\mu}\})\right)\right\}.
\end{equation}
This geometric interpretation also identifies the gradient flow trajectory, since its unit tangent vector $\hat{t}$ is the one pointing towards the minimum distance point to the convex hull.
An example of this can be found in Figure \ref{fig:convex hull3} for a three moduli case. 
Notice that, in order for the saxions not to be sent to zero (away from the asymptotic regime), the solution must fulfill $\hat{t}^i\geq 0$. Furthermore, in the cases we are working in some growth sector, \eqref{check growth sector}  imposes an analogous conical restriction on the admissible values of $\hat{t}$.

Using this, one can also reformulate the asymptotic dS conjecture in the following way:

\paragraph{Convex Hull dS conjecture:} Given an asymptotic scalar potential $V=\sum_{\bm{l}\in \mathcal{E}} V_{\bm l}$, the asymptotic de Sitter conjecture with $\frac{\| \nabla V\|}{V}\geq c_d$ will be satisfied if the convex hull of all the de Sitter ratios $\vec{\mu}_{\bm l}$ in \eqref{l to mu transform gen} lie outside the ball of radius $c_d$.\\

\begin{figure}[htbp]
\centering
\subfigure[Satisfies dSC, $V_l^{\rm dom}\in \mathcal{E}^{\rm light}$\label{a}]{\includegraphics[width=0.32\textwidth]{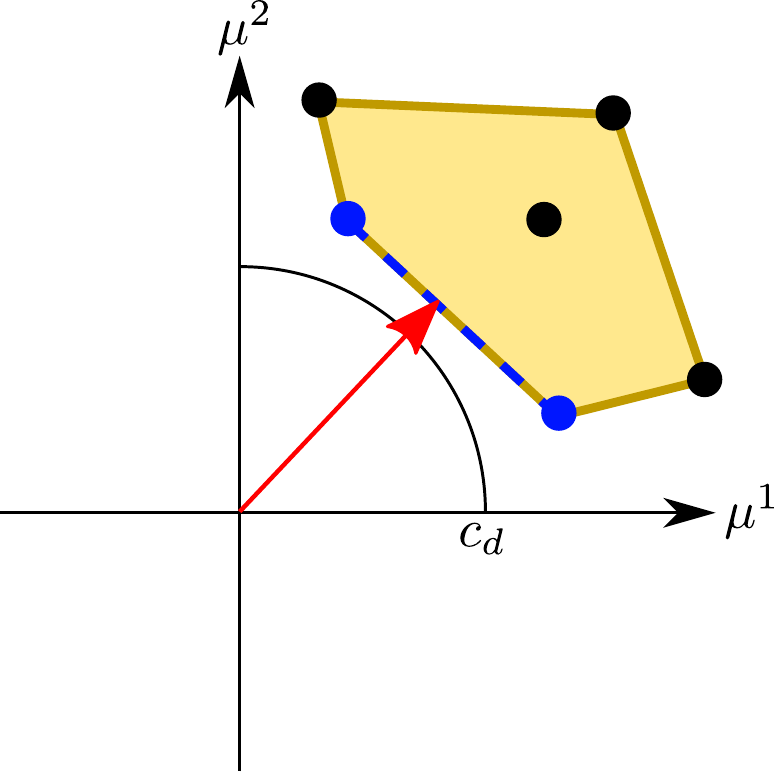}}
\subfigure[\label{b}Satisfies dSC, $V_l^{\rm dom}\in\mathcal{E}^{\rm light}$]{\includegraphics[width=0.32\textwidth]{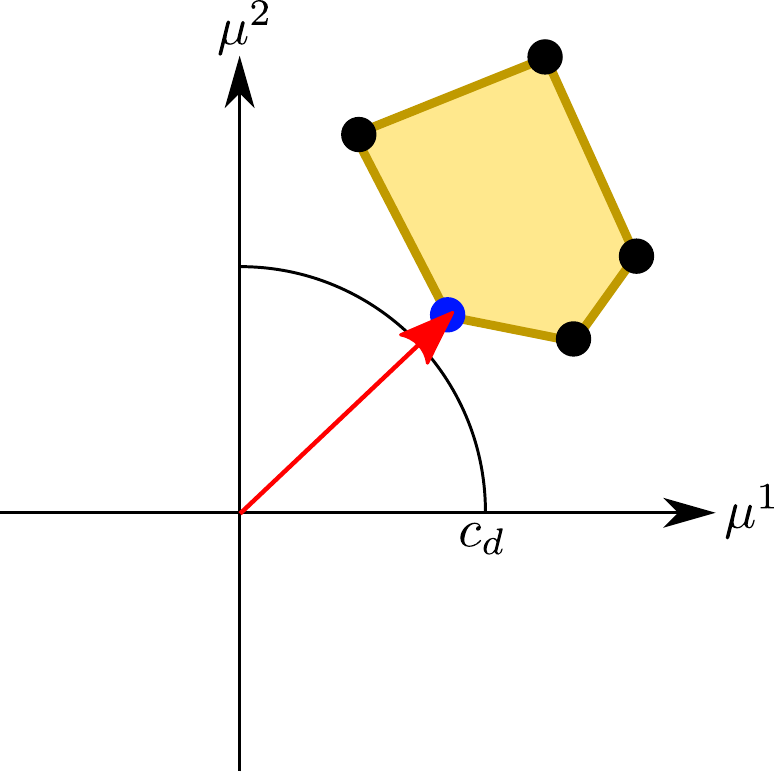}}
\subfigure[\label{c}Satisfies dSC, $V_l^{\rm dom}\in\mathcal{E}^{\rm rest}$]{\includegraphics[width=0.32\textwidth]{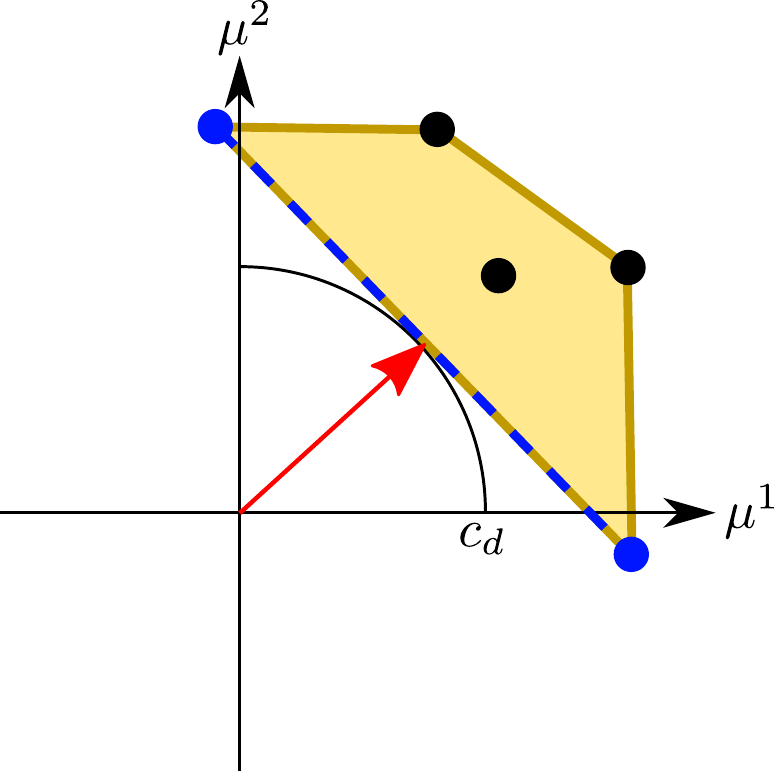}}
\subfigure[\label{d}Violates dSC, $V_l^{\rm dom}\in\mathcal{E}^{\rm rest}$]{\includegraphics[width=0.32\textwidth]{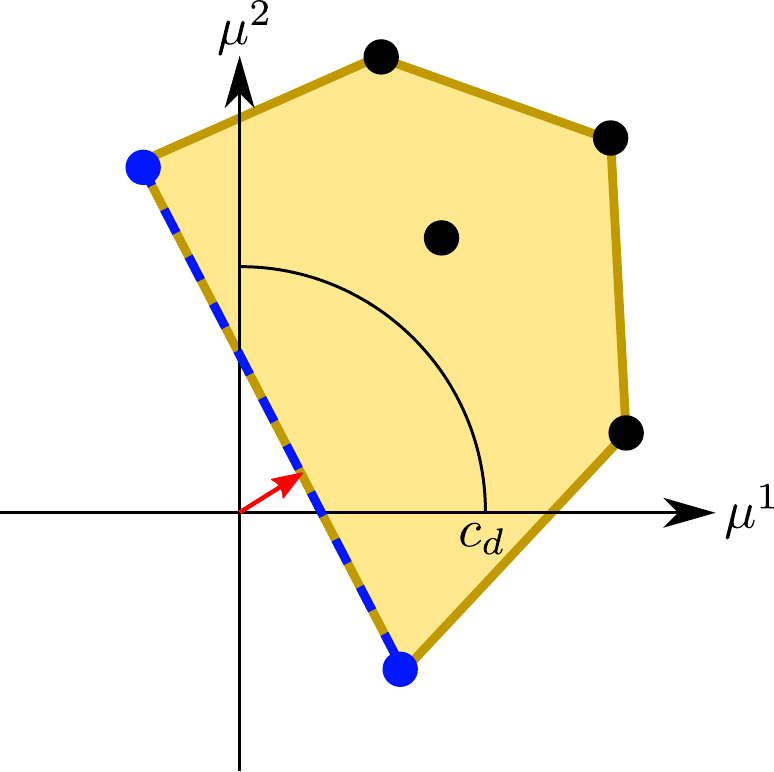}}
\subfigure[\label{e}Violates dSC, $V_l^{\rm dom}\in \{\mathcal{E}^{\rm light},\, \mathcal{E}^{\rm rest}\}$]{\includegraphics[width=0.32\textwidth]{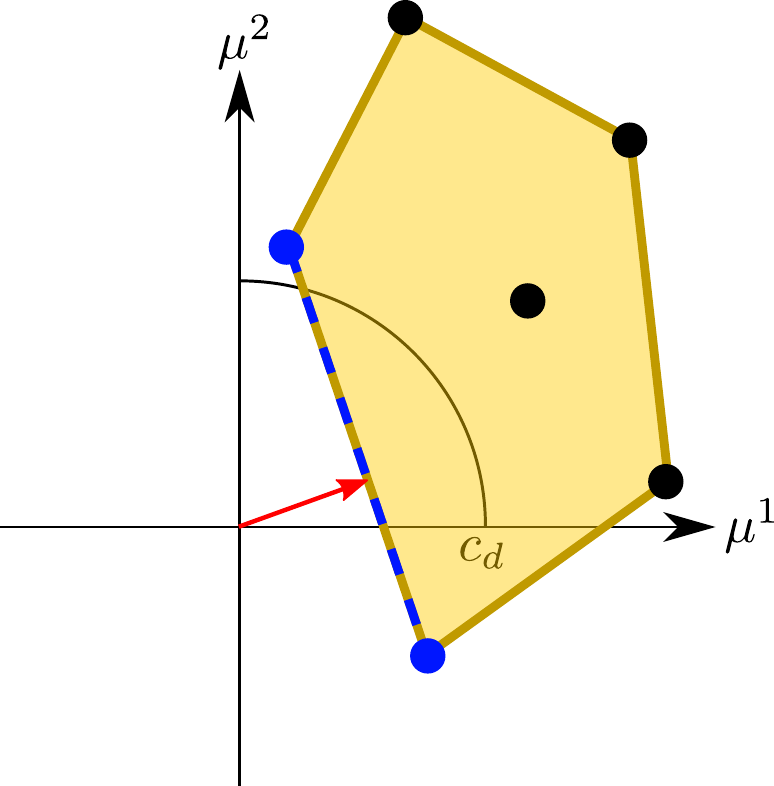}}
\subfigure[\label{f}Non-admissible asymptotic solution, $V_l^{\rm dom}\in\{\mathcal{E}^{\rm light},\ \mathcal{E}^{\rm rest}\}$]{\includegraphics[width=0.32\textwidth]{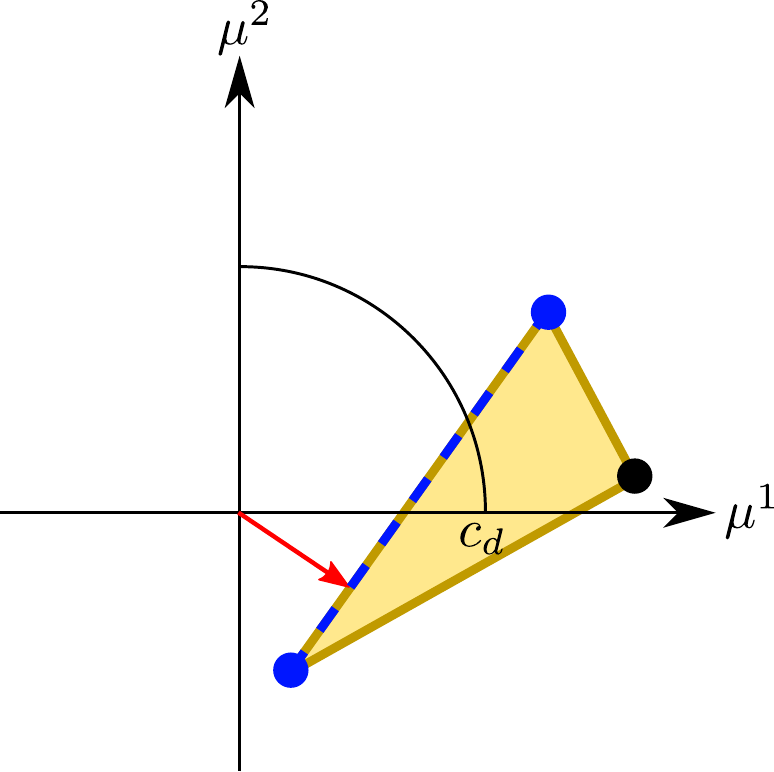}}
\caption{Toy model examples satisfiying or violating the Convex Hull deSitter conjecture. Each point is a term of the potential, such that the dominant ones are in blue. The associated convex hull is filled in yellow. The $\vec{t}$ direction determining the gradient flow is denoted by the red arrow. The result for $\gamma_{\vec{f}}=\frac{\| \nabla V\|}{V}$ is the shortest distance to the convex hull from the origin (ie. the length of the red arrow).
}
\label{fig:convex hull ex}
\end{figure}

\begin{figure}[htbp]
\centering
\subfigure[ $\gamma_{\vec{f}}<c_4^{\rm TCC}$]{\includegraphics[width=0.45\textwidth]{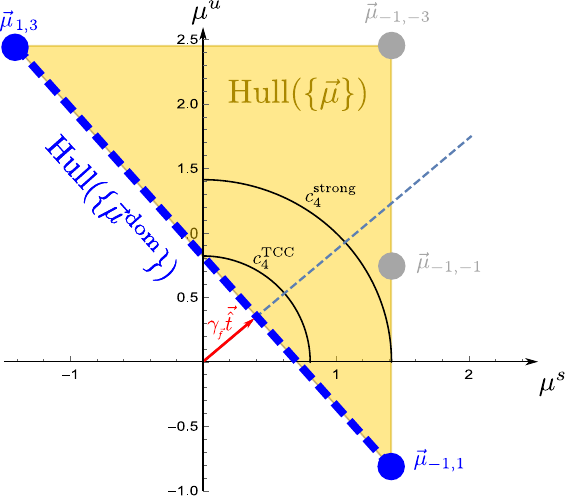}}
\subfigure[ $\gamma_{\vec{f}}>c_4^{\rm strong}$]{\includegraphics[width=0.35\textwidth]{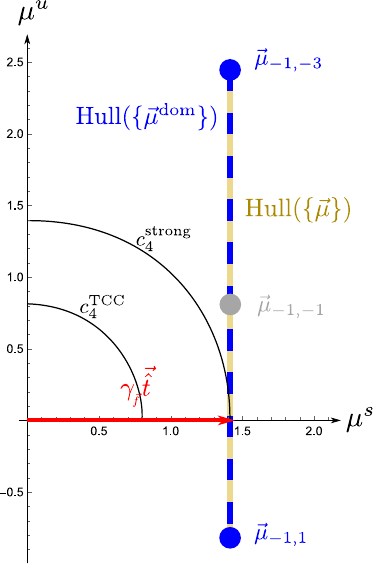}}
\caption{Two examples of the Convex Hull procedure to obtain the gradient flow solutions and the dS coefficient, for two moduli potentials, both taken from Table \ref{table2}. Figure (a) corresponds to the fifth potential, with the dominant terms being $A_{34}\frac{u}{s}$ and $A_{52}\frac{s}{u^3}$. The closest minimal convex hull is ${\rm Hull}(\{\vec{\mu}_{-1,1},\vec{\mu}_{1,-3}\})$, to which the distance from the origin is $\gamma_{\vec{f}}=\sqrt{\frac{2}{7}}$ along a $\vec{t}=(2,\sqrt{3})$ direction, then translating to $\vec{\beta}=(2,1)$. Notice that the other two terms, shown in gray, corresponding to $A_{30}\frac{1}{su^3}$ and $A_{32}\frac{1}{us}$, are rendered 0 by axion stabilization. Figure (b) corresponds to the eighth potential of said table, where $u$ is stabilized and after axion stabilization the potential is dominated by $\frac{A_{30}}{su_0^3}$ and $A_{34}\frac{u_0}{s}$ (actually the only terms left after axion stabilization suppress the $\vec{\mu}_{-1,-1}$ term, in gray). The convex hull of these is a segment parallel to the $u$ axis, and thus it is reached first through the $\vec{t}=(1,0)$ (then $\vec{\beta}=(1,0)$) direction, at a distance $\gamma_{\vec{f}}=\sqrt{2}$.
}
\label{fig:convex hull1}
\end{figure}

Hence, the Strong dS conjecture forbidding asymptotic acceleration is equivalent to the above convex hull condition where the ball is fixed at a radius $c_4^{\rm strong}=\sqrt{2}$ in four dimensions. In Figure \ref{fig:convex hull ex} we draw some toy model examples satisfying (\ref{a},\ref{b},\ref{c}) or violating (\ref{d},\ref{e}) the CH dS conjecture for some arbitrary $c_d$. If $c_d=c_d^{\rm strong}$ the examples in \ref{d} and \ref{e} lead to asymptotic accelerated expansion, while \ref{f} does not have an admissible asymptotic solution since the trajectory move us away from the asymptotic regime (so the conjecture does not apply). In addition, Figure \ref{fig:convex hull1} shows a couple of concrete examples coming from asymptotic flux compactifications studied in Section \ref{sec:examples}.  

It is interesting that the convex hull version of dS conjecture requires the convex hull to be \emph{outside} of the ball, while both the Convex Hull WGC \cite{Cheung:2014vva} and Convex Hull SDC \cite{Calderon-Infante:2020dhm} impose that the convex hull \emph{contains} the ball (or extremal region in more generality). Notice that while the latter requires the existence of some superextremal state for any charge direction, the former can be thought as imposing superextremality to all states. This is because the leading terms of the potential are those that have a smaller de Sitter ratio, so it is important that they lie outside of the ball if one wants to forbid accelerated expansion.  This analogy has indeed a microscopic interpretation, since the deSitter ratios correspond to the scalar charge to mass ratio of the membranes that are charged under the fluxes of the potential. To see this, let us take a look back at the formulation of the scalar potential as the physicial charge of a BPS membrane \eqref{V from T} whose quantized charges are the fluxes, so that $Q^2=2V$, and its tension is proportional to the superpotential $\cT=2e^{K/2}|W|$ (as explained in Section \ref{SEC as pot} and in more detail in \cite{Lanza:2020qmt}). Each term of the flux potential is a monomial on the saxions, so that it satisfies \eqref{dTcT} and \eqref{dTT} implying that
\beq
\vec{\mu}_{\bm l}=- 2\vec{e}^{\,i} \, \frac{\partial_i \cT_{\bm{l}}}{\cT_{\bm{l}}} \, .
\eeq
This is precisely the scalar charge to mass ratio of the membrane, which measures the balance between the scalar force interaction and the gravitational force (see \cite{Lanza:2020qmt}). A scalar version of the WGC for membranes would imply that the scalar force must be stronger than the gravitational force, so that\footnote{From \cite{Lanza:2020qmt} we find this is equivalent to 
\begin{equation}
    |\vec{\mu}_{\bm l}|=2\frac{\|\nabla \cT_{\bm l}\|}{\cT_{\bm{l}}}\geq 2\sqrt{\frac{p(p-d+2)}{d-2}},
\end{equation}
for parallel $(p-1)$-branes in a $d$-dimensional spacetime. For 2-branes $p=3$ in a 4-dimensional spacetime, $|\vec{\mu}_{\bm l}|>\sqrt{6}$.
} $|\vec{\mu}_{\bm l}|>\sqrt{6}$, which is larger than $c_4^{\rm strong}=\sqrt{2}$. Note that this $\sqrt{6}$ is the same factor required in \eqref{V prop T2} in order to have $V>0$ in single-term potentials.
Hence, the Strong de Sitter conjecture coincides with a convex hull condition for a scalar WGC applied to membranes, in which \emph{all} membranes generating the flux potential must experience a sufficiently strong scalar interaction (so that the convex hull lies outside the ball of radius $\sqrt{2}$).
This formulation also makes manifest that one could violate the strong dS bound and yield accelerated expansion without violating the usual WGC (or its scalar version) in which only \emph{some} membranes need to satisfy the bound.  

\begin{figure}[htbp]
\centering
\subfigure[ $\vec{0}\in({\rm Hull}(\{\vec{\mu}\}))^\circ$]{\includegraphics[width=0.35\textwidth]{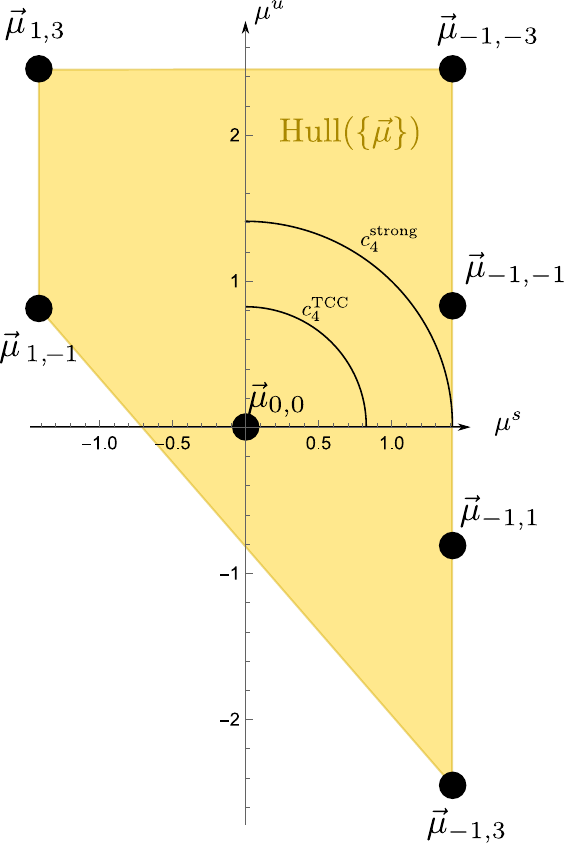}}
\subfigure[ $\vec{0}\in\partial{\rm Hull}(\{\vec{\mu}\})$]{\includegraphics[width=0.4\textwidth]{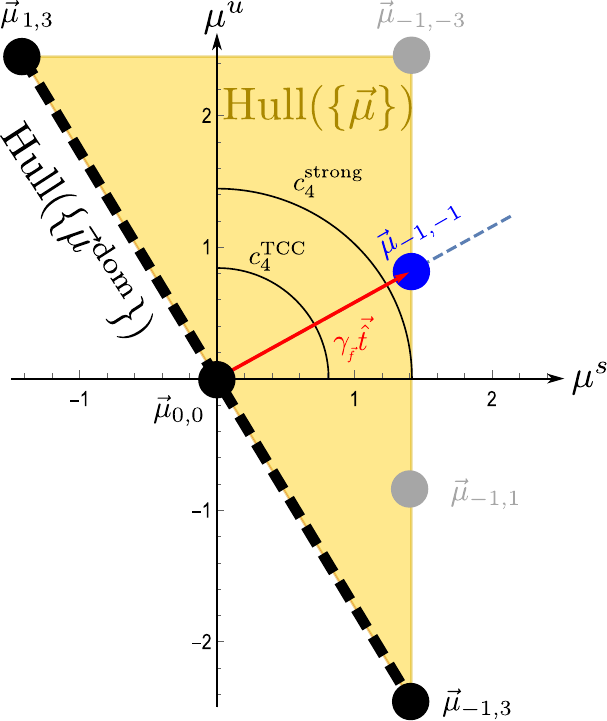}}
\caption{Two more examples of the Convex Hull procedure to obtain the gradient flow solutions, in this case corresponding to the described ``problematic'' cases, taken from Table \ref{table2}. Figure (a) corresponds to the first potential and as it is immediate to see $\vec{0}$ is in the interior of ${\rm Hull}(\{\vec{\mu}\})$. This means that along every direction the scalar potential blows up and thus the scalars will be located at some minimum, whose expression we will not obtain. The closest ``boundary'' of the convex hull, that joining $\vec{\mu}_{1,-1}$ and $\vec{\mu}_{-1,3}$ is at distance $\sqrt{\frac{2}{7}}$ and direction $\vec{t}=(-2,-\sqrt{3})$ ($\vec{\beta}=(-2,-1)$), along which the potential blows up fastest. Figure (b) corresponds to the second potential of said table, dominated by $A_{36}\frac{u^3}{s}$, $A_{52}\frac{s}{u^3}$ and $A_{44}$. The convex hull of these, which is a segment, contains the origin, and thus will correspond to a constant along the $\vec{t}=(3,\sqrt{3})$ (then $\vec{\beta}=(3,1)$) direction. 
After the terms $A_{30}\frac{1}{su^3}$ and $A_{34}\frac{u}{s}$ (shown in grey) are also canceled (through axion stabilization), the only term left is $A_{32}\frac{1}{us}$ (shown in blue), with $\{\vec{\mu}_{-1,-1}\}$ being its own convex hull. Then $\gamma_{\vec{f}}$ corresponds to the distance along the $\vec{t}$ direction from the origin to the $\gamma_{\vec{f}}$ projection. Notice that in this case said point is contained in that direction, though in general this is not the case.}
\label{fig:convex hull2}
\end{figure}

As promised, let us finally discuss a case in which one has to proceed a bit more carefully with this criterion. This happens when $\vec{0}\in {\rm Hull}(\{\vec{\mu}\})$ (the convex hull crosses or contains the origin) and as such the distance is zero. This can correspond to either of the following two cases which are depicted in Figure \ref{fig:convex hull2}:
\begin{itemize}
    \item The origin is in the interior of the convex hull. This means that the potential has terms that blow up in the asymptotic limit, so the trajectories get obstructed and there is no admissible solution.
    
    \item The origin is at the border of the convex hull. This corresponds to the scenario (II) illustrated in Figure \ref{fig:convex hull2}. The leading terms compete to drive the gradient flow towards a flat direction (or subspace of directions). As we saw before, these corresponds to the direction(s) that are orthogonal to the side of the convex hull containing the origin. In this case we must consider subleading terms if they are present. This means projecting the points $\left(\{\vec{\mu}\}-\{\vec{\mu}^{\rm dom}\}\right)$ onto the hyperplane of fixed directions and solving the convex hull criterion in \eqref{eq:convexhull-dist} on it. When doing this, one has to be careful with terms that disappear from the potential after axion stabilization. In addition, one should impose that the leading terms cancel so that the potential goes to zero asymptotically. This requires a level of fine tuning that may  not be achievable.
\end{itemize}

\subsection{Deriving dS bounds using the Convex Hull formulation\label{sec:bounds convex hull}}

Interestingly, we can use this convex hull formulation of the dS conjecture to derive  some lower bounds for $\gamma_{\vec{f}}$ in the different scenarios, as well as to understand the features of the potential that allow for  violation of the Strong dS bound. \\

\noindent\textbf{Scenario (I):}\\
We first consider scenario (I) from Figure \ref{fig:two examples}, with a single term of the potential dominating asymptotically. This term must belong to $\mathcal{E}^{\rm light}$, so that the associated $\vec{\mu}_{\bm{l}^{\rm dom}}$ is in the positive quadrant. Now, as the convex hull of $\vec{\mu}_{\bm{l}^{\rm dom}}$ is the vector itself, one gets $\gamma_{\vec{f}}=\sqrt{\sum_{i=1}^n({\mu}^i_{\bm{l}^{\rm dom}})^2}$, which for diagonal metrics reduces to
    \begin{equation}
    \gamma_{\vec{f}}=\sqrt{2\sum_{i=1}^{{n}}\frac{( l_i)^2}{ d_i}} \, .
\end{equation}
Now, the closest $\vec{\mu}_{\bm{l}^{\rm dom}}$ (and thus lowest $\gamma_{\vec{f}}$) will be obtained for $l^{\rm dom}_i=-1$ for all $i$. Furthermore, for flux potentials in CY$_4$ studied through asymptotic Hodge theory, we have $\Delta d_1+\cdots+\Delta d_{n}=d_{n}\leq 4$, so that $(\Delta d_1)^{-1}+...+(\Delta d_{n})^{-1}\geq \frac{n^2}{d_{n}}\geq\frac{n^2}{4}$, yielding
\begin{equation}\label{eq: bound single light new}
    \gamma_{\vec{f}} \geq \sqrt{2\sum_{i=1}^{n}\frac{1}{\Delta d_i}}\geq\sqrt{\frac{2}{d_{n}}}n \geq \frac{n}{\sqrt{2}} \, .
\end{equation}
Notice that for $n\geq 2$ then $\gamma_{\vec{f}}\geq c_4^{\rm strong}$, preventing any accelerated expansion of the universe. Nevertheless, for $n=1$  our bound allows for values smaller than the strong dS for $d_1=2\,,3\,,4$ and TCC bounds for $d_1=4$ in 4d  . We note however that $d_1\neq 4$ for potentials coming from F-theory since an elliptically CY fourfold does not have single field limits with $d_1=4$. Thus, for elliptically fibered CY's one gets $\gamma_{\vec{f}} \geq \sqrt{\frac{2}{3}}$, preserving the TCC bound \cite{Bedroya:2019snp}. Some examples with $n=1$ and $d_1=1,3$  were discussed in Section \ref{sec:1 mod}. For Type IIA, though, $\gamma_{\vec{f}}\geq \sqrt{6}$ because otherwise the dominant term of the potential becomes negative as explained in Section \ref{sec: IIA large V}. Recall that this bound is only valid for Scenario (I) in which a single term dominates. \\

\noindent\textbf{Scenario (II):}\\

In this case, two or several terms compete asymptotically along the gradient flow trajectory, so that they generate a single valley approaching the infinite distance point. Let us consider two possibilities: either all the competing terms belong to $\mathcal{E}^{\rm light}$ or we also consider terms in $\mathcal{E}^{\rm rest}$

as discussed in Section \ref{SEC:gradflow}.
\begin{itemize}
\item  $V_l\in\mathcal{E}^{\rm light}$:
Here, one can still derive a lower bound for $\gamma_{\vec{f}}$. In the convex hull formulation, note that ${\rm Hull}(\{\vec{\mu}_{\bm l}^{\rm dom}\})$ will span at most a codimension one hyperplane of that will cut the $\mu^i$ axis at $\mu^i_0$. Then said hyperplane (and thus the closest convex hull) will be at a distance\footnote{We recall that a hyperplane $a_px_p+...+a_1x_1+a_0=0$ is at a distance $\frac{|a_py_p+...+a_1y_1+a_0|}{\sqrt{a_1^2+...+a_p^2}}$ from the point $(y_1,...,y_p)$.}
\begin{equation}
    \gamma_{\vec{f}}=\left(\sum_{i=1}^n(\mu^i_0)^{-2}\right)^{-\frac{1}{2}}
\end{equation}
from the origin. Now, the lower bound for $\gamma_{\vec{f}}$ will be given by taking the lowest possible values of $\{\mu^i_0\}_{i=1}^n$, which will correspond to putting the individual terms $\{\vec{\mu}_{\bm l}^{\rm dom}\}$ themselves on the axes. This is equivalent to say that $\beta^i$ is $0$ or $1$ for all saxions, so that the gradient flow solution is given by $\vec{\beta}=(1,...,1)$ when taking into account only the saxions contributing.
Using that, for diagonal metrics, $\mu^i_{\bm{l}} = - \sqrt{\frac{2}{d_i}} \,l_i$ and that $l_i\geq 1$, we get 
\begin{equation}
    \gamma_{\vec{f}}\geq \left(\frac{2}{\sum_{i=1}^{{n}} d_i}\right)^{\frac{1}{2}} \, .
\end{equation}
Again, using the $\{\Delta d_i\}_{i=1}^n$ properties for diagonal metrics obtained from asymptotic Hodge theory, this implies $\gamma_{\vec{f}}\geq \frac{1}{\sqrt{2}}$ for a CY$_4$ or  $\gamma_{\vec{f}}\geq \sqrt{\frac{2}{3}}$ if it is elliptically fibered. 
We thus arrive to a lower bound for $\mathcal{E}^{\rm light}$ potentials that allows for violation of the Strong bound \eqref{boundac}. 
Note that for Type IIA, as it is required that $|\vec{\mu}_{\bm l}|>\sqrt{6}$ in order to have positive potentials, through a similar argument as thus above we arrive to $\gamma_{\vec{f}}\geq\sqrt{\frac{6}{n}}$ for those potentials only featuring $\mathcal{E}^{\rm light}$ terms. For $n=1,\, 2$ or 3 this translates in no accelerated expansion being possible.

\item $V_l\in \mathcal{E}^{\rm rest}$:
For the final possibility, we have competing terms in scenario (II) with at least one of them belonging to $\mathcal{E}^{\rm rest}$, so that some saxions are in the numerator and others in the denominator of the potential. This implies that $\left|\sum_{i=1}^{n}\beta^i\Delta l_i\right|$ in \eqref{dS gen} can be smaller than the value of $|\beta^i\Delta l_i|$ for each individual term due to possible cancellations thanks to having different signs for $\Delta l_i$. How small it can get, it is not clear. In the convex hull language, this is equivalent to having the $\{\vec{\mu}_{\bm l}^{\rm dom}\}$ points at arbitrary values, not necessarily close to the origin, but such that their convex hull is (see Figure \ref{d}).

\end{itemize}

This convex hull formulation also allows us to understand why, even in IIA in which there is non no-scale structure and all moduli are stabilized, we can violate the supergravity no-go \cite{Hellerman:2001yi,Rudelius:2021azq}, as discussed in section \ref{Sec:SuperpotCond}. Even if each potential term satisfies $|\mu_l|> \sqrt{6}$ (since otherwise such term becomes negative in the scalar potential), the convex hull of all these points could still be at a distance $\gamma_{\vec f}\leq \sqrt{6}$ closer to the origin. In other words, even if all the individual terms of the potential are outside the ball, the convex hull could still cut the ball and violate the bounds. And this is consistent with keeping $V>0$.

Finally, recall that we are assuming, for simplicity, a diagonal field metric (to leading order) to derive the above bounds, but we expect them to hold for non-diagonal metric ones as well, as was the case for CY$_3$'s \cite{Bastian:2020egp}. However, it would be interesting to check this. Notice that for $n>4$, the non-diagonal component of the field metrics cannot be neglected anymore since there will always be some modulus with $\Delta d_i=0$. We will not delve more into this because, even by restricting to the set of diagonal metrics, we were already able to show that the Strong dS bound \eqref{strong bound} can be violated at the level of the flux potential.

\subsection{Relation to the Distance Conjecture\label{sec:distance conj}}
Since we are in an asymptotic regime of the field space, there will be an infinite tower of states becoming massless at the infinite field distance limit. In the following, we will check that the presence of this tower is consistent with our previous results.

The mass scale of the tower behaves as
\begin{equation}
m(D)\simeq m(0)\exp\left\{-\alpha D\right\}
\end{equation}
as the geodesic distance $D\rightarrow \infty$.
The existence of this tower is expected by the Distance Conjecture, and has been identified in a plethora of string compactifications (see e.g. \cite{Grimm:2018ohb,Grimm:2018cpv,Corvilain:2018lgw, Lee:2018urn, Font:2019cxq, Marchesano:2019ifh,Lee:2019wij} for CY manifolds) and in the context of AdS/CFT \cite{Baume:2020dqd,Perlmutter:2020buo}. This evidence also include 4d $\mathcal{N}=1$ theories \cite{Lanza:2020qmt,Lanza:2021udy,Klaewer:2020lfg,Cota:2022yjw} as the ones we are considering. In fact, it has been proposed that the asymptotic runaway of the potential is a consequence of the presence of the tower \cite{Ooguri:2018wrx}, implying a correlation between the dS coefficient and the exponential rate of the tower. 

Recall from the discussion in Section \ref{SEC:gradflow} that the asymptotic gradient flow trajectories forced by the flux potential are geodesics of the moduli space without potential. Along them, the potential behaves as \eqref{eq:pot D}, i.e,
\begin{equation}
    V(D)\sim V(0)\exp\left\{-\gamma D\right\}
\end{equation}
along the trajectory  $s^i(D)=s^i(0) \exp\{\hat{\beta}^iD\}$ as $D\rightarrow \infty$ (equivalently \eqref{eq:saxions-ansatz} for $\lambda\rightarrow\infty$). Here, $\gamma$ is given in \eqref{dS gen}, and the geodesic distance $D$ is related to the parameter $\lambda$ by
\begin{align}
    D(s(\lambda_0),s(\lambda))=\int_{\lambda_0}^\lambda\sqrt{G_{t^i\bar{t}^j}(\tau)\dot{t}^i(\tau)\dot{\bar{t}}^j(\tau)}\diff \tau
     =|\vec\beta|\log\left(\frac{\lambda}{\lambda_0}\right),
\end{align}

where the norm of $\beta$ is taken with the field space metric, see \eqref{norm beta}.

Since both the potential and the tower mass decay exponentially on the field distance, we have
\beq\label{eq:V m}
V\sim m^\chi
\eeq
with $\chi>0$, which is confirmed in flux string compactifications and also motivated by the generalized Distance Conjecture \cite{Lust:2019zwm}. This indeed implies that the dS coefficient is related to the exponential rate $\alpha$ of the leading tower by
\beq
\gamma = \chi \alpha \, .
\eeq
Therefore, a bound on $\alpha$ translates into a bound on $\gamma$ and vice versa. One could wonder then whether a small value of $\gamma$ that allows for accelerated expansion is compatible with bounds on the exponential rate of the tower. In this subsection, we explain how this could be possible.\\

First, let us check the exponential rate for the tower of states in all previous F-theory examples.
Supported by all known string theory examples, it was proposed in \cite{Lanza:2020qmt,Lanza:2021udy} that there is always a BPS axionic string  becoming tensionless at every infinite distance limit of 4d $\mathcal{N}=1$ theories. This is known as the Distant axionic string conjecture and the strings were denoted as EFT strings. The string tension behaves as
\beq
T=\Mpl^2e^i\ell_i(\lambda) \, ,
\eeq
where $\ell_i=-\frac12\frac{\partial K}{\partial s^i}$ are the dual saxions and $e^i$ are the quantized string charges. Along the gradient flow trajectory $s^i(\lambda)=\alpha^i \lambda^{\beta^i}$, the tension
decays exponentially on the proper field distance with an exponential rate which is purely determined by geometric data,
\beq
T(\lambda)\simeq T(0)\exp(-\alpha_s D)\quad \text{with } \alpha_s=\hat{\beta}_{\rm max}=\frac{\beta_{\rm max}}{|\vec\beta |}
\eeq
where $\beta_{\rm max}$ corresponds to the saxion that grows the fastest along such trajectory and therefore identifies the string decaying fastest.
Even if this string already provides a candidate for the tower of the Distance Conjecture, there could be other (non-BPS) towers of particles becoming light faster. However, it was found in \cite{Lanza:2020qmt} that, in all known string theory examples, the leading tower always satisfies
\beq
m^2\simeq \Mpl^2A\left(\frac{T}{\Mpl^2}\right)^w \quad \text{with } w=1,2,3.
\eeq
This was promoted to the Integral Scaling conjecture in \cite{Lanza:2022zyg}. Combining all this, we get that the potential and the leading tower are related as \eqref{eq:V m} with
\beq\label{eq:chi def}
\chi=\frac{\gamma}{\alpha}=\frac{2}{w}\frac{\gamma}{\alpha_s}=\frac{2}{w}\frac{|\beta^i\Delta l_i^{\text{dom}}|}{\beta^1} \, ,
\eeq

where $\beta_{\rm max}=\beta^1$ in the notation of working in the  growth sector $\mathcal{R}_{1...}$. We have computed the value of $\chi$ (mod the value of $w$) for all previous examples and included the result in the Tables \ref{table1} to \ref{table4}. To determine $w$ we would need to study the non-BPS towers present in each example, which goes beyond the scope of this work. But we can use the exponential rate of the string modes (i.e. assuming $w=1$) as a lower bound, since at the very least we will have the BPS string becoming tensionless. From Tables \ref{table2} to \ref{table4} we find $\chi$ values ranging from $\chi=w^{-1}$ (fifth row in Table \ref{table2}, with $\gamma_{\vec{f}}=\sqrt{\frac{2}{7}}<c_4^{\rm TCC}$) and $\chi=2w^{-1}$ (third and fourth rows in \ref{table4}, both with $c_4^{\rm TCC}<\gamma=\frac{2}{\sqrt{5}}<c_4^{\rm strong}$) to $\chi=14w^{-1}$ (second to last row in Table \ref{table1}). Notice that apart from the first three mentioned cases (which violate some of the proposed dS bounds), $\chi>2w^{-1}$, with $\chi$ larger for higher $\gamma$ values, as inferred from \eqref{eq:chi def}.

Intuitively, if $\chi$ becomes too small, the EFT will break down since the tower of states becomes lighter than the potential energy. So what is the minimum value of $\chi$ such that the EFT remains trustworthy? Associated to the tower of states, there is a Quantum Gravity cut-off at which the EFT drastically breaks down and no local QFT description is valid anymore. This is the species scale $\Lambda_{\rm sp}$ \cite{Dvali:2007wp,Dvali:2010vm} and corresponds to the mass scale of the tower in the case of a string perturbative limit (ie. when $w=1$) or to the Planck scale of the higher dimensional theory in a decompactification limit. In the latter case, $\Lambda_{\rm sp}\sim m^{\frac{N}{N+2}}> m$ where $N$ is the number of dimensions that decompactify. For the EFT to be valid, we need to require at least that the Hubble scale remains below the cut-off, $H\sim V^{1/2}\lesssim \Lambda_{\rm sp}$, which is always satisfied as long as 
\beq
\chi\geq \frac{2N}{N+2} \, .
\eeq
Hence, there is a priori some room to get accelerated expansion without the tower invalidating the EFT, since $\chi$ can be as small as $\frac{2}{3}$ in the best case scenario (i.e. when $N=1$). 

Since the tower of states generically contains higher spin fields (either string modes or KK copies of the graviton), we could impose an even stronger bound coming from the Higuchi bound \cite{Higuchi:1986py}. This would imply that $m^2> V$, assuming that the Higuchi bound for dS remains unchanged in a quintessence scenario. This in turn implies that $\chi\geq 2$ as proposed in \cite{Montero:2022prj} (see also \cite{Rudelius:2022gbz}). Depending on the value of $w$, this could still be satisfied by all examples of the previous section (including the one violating the strong dS bound in Table \ref{table4}) except for the example violating the TCC in Table \ref{table2}. This could be a hint that this example is somehow sick, although there is a priori nothing obviously wrong with it at the level of the flux potential. Recall that it corresponds to the large complex structure limit of perturbative Type IIB  with $f_2$ RR and $h_0$ NS flux. 
This is interesting since, either we have the first example violating the TCC or it can be used to identify flux potentials at parametric control  that still have no UV completion in quantum gravity.

We should finally note that asymptotic accelerated expansion is  not compatible with the lower bound for the exponential rate of the Distance conjecture proposed in \cite{Etheredge:2022opl} if assumed $\chi\geq  2$, as explained in that same paper. However, it is still compatible if one only imposes the weaker (but more solid) condition that $H\leq \Lambda_{\rm sp}$.

\section{Open avenues and caveats to get accelerated expansion\label{sec:open avenues}}

\subsection{K\"{a}hler moduli stabilization\label{sec:no-scale}}

Through out this paper we have seen some examples in which the TCC \cite{Bedroya:2019snp} and strong dS \cite{Rudelius:2021azq} bounds are violated. As we discussed, the caveat is that in all these cases we are assuming the K\"{a}hler moduli to be somehow stabilized. In this section we show that this is indeed necessary since, otherwise, they give an extra contribution to the dS coefficient that forbids accelerated expansion.

Consider both the contribution to the K\"{a}hler potential from the K\"{a}hler ($T^a$) and complex structure moduli ($N^\mu$), i.e. $K=K_{\rm T}+K_{\rm cs}$ at leading order. If the superpotential does not depend on the K\"{a}hler moduli (as occurs in F-theory/IIB when only considering fluxes), there is a partial no-scale structure for the K\"{a}hler moduli:
\begin{align}
   \partial_a W &= 0 \, , \\
    K_{T}^{a\bar{b}} (\partial_a K_T) (\partial_{\bar{b}} K_T) &= {\bm{n}} \, ,
\end{align}
where ${\bm{n}}$ depends on the type of asymptotic limit. For instance, if we are in perturbative Type IIB at large volume, then ${\bm{n}}=d=3$.
 Plugging this into \eqref{cremmer} one gets
\begin{equation} \label{eq:no-scale-VfromW}
    V = e^{K_T} \, V_{\rm cs}\qquad\text{with  } V_{\rm cs}=e^{K_{\rm cs}}\left(K_{\rm cs}^{\mu\bar{\nu}}D_\mu W\bar{D}_{\bar{\nu}}\bar{W}-(3-{\bm{n}})|W|^2\right) \, .
\end{equation}
The potential then has a contribution from the K\"{a}hler moduli, only depending on its K\"{a}hler potential, and another from the complex structure, that takes the form of \eqref{cremmer} but with $(3-{\bm{n}})$ instead of $3$ in the negative term. Using this and the block diagonal form of the field space metric one can easily show that
\begin{equation}
    \frac{\gamma^2}{2} = \frac{\gamma_{\rm T}^{2}}{2} + \frac{\gamma_{\rm cs}^{2}}{2} \, .
\end{equation}
This essentially tells us that the dS coefficient gets two separate contributions, one from the K\"{a}hler and another from the complex structure sector. 

The K\"{a}hler moduli contribution can be easily computed to be
\begin{equation}
    \frac{\gamma_{\rm T}^{2}}{2} = K_{\rm T}^{a\bar{b}} (\partial_a K_{\rm T}) (\partial_{\bar{b}} K_{\rm T}) = {\bm{n}} \, ,
\end{equation}
thus showing that their (partial) no-scale structure is automatically fixing their contribution to the dS coefficient. Hence, the contribution from the runaway of the Kähler moduli (if they are not stabilized) places a bound on the dS coefficient of the form
\begin{equation}
    \gamma \geq \sqrt{2{\bm{n}}}\, .
\end{equation}
This forbids accelerated expansion even  if the Kähler moduli run towards an infinite distance limit with ${\bm{n}}=1$. For the case of Type IIB at large (overall) volume, this bound precisely coincides with the no-go in \cite{Rudelius:2021azq} since ${\bm{n}}=3$.

It is then important to avoid or reduce the contribution to the dS coefficient coming from the K\"{a}hler sector. As already pointed out in Section \ref{sec: gen bounds}, this is also a requirement of any phenomenologically viable quintessence model, since the quintessence field cannot be the overall volume due to experimental constraints on 5th forces. There are therefore two possibilities. The first option is to stabilize somehow the K\"{a}hler moduli, while leaving the runaway on the complex structure moduli, so that $\gamma_T=0$. This will probably require to get an inverted hierarchy in which the running complex structure modulus remains lighter than the Kähler moduli, maybe along the lines of \cite{Demirtas:2021nlu,Bastian:2021hpc}. The second possibility is to add some potential for the Kähler moduli in such a way that, even if it does not cure the runaway, it reduces at least the slope so that $\gamma_T<\sqrt{2}$; either by having different competing terms exploiting the loophole in Section \ref{Sec:SuperpotCond} and/or by going to different asymptotic limits. In the context of F-theory/Type IIB, both options require to add further ingredients beyond the perturbative flux potential, so we will not explore them here. One could also go to Type IIA by using mirror symmetry, where even the Kähler moduli can be stabilized by fluxes. However, when doing this at the large complex structure limit, there is an extra overall factor of the dilaton arising in the Einstein frame (see Section \ref{sec: IIA large V}) that changes the value of the dS coefficient in such a way that there are not examples yielding accelerated expansion anymore (although there are examples violating the supergravity no-go of \cite{Rudelius:2021azq}). If one could perform mirror symmetry at other asymptotic limits of the moduli space beyond weak string coupling (e.g. at the enhancement of Section \ref{sec:dif2}), it might happen that new examples violating the strong dS bound (which take into account the Kähler moduli) appear. However, it is not known how to do this and what would be the additional overall factor of the dilaton arising in those cases\footnote{If one naively assumes that it is the same factor as in the large complex structure limit, ie. the potential is the same except for an overall $\frac{1}{s^3}$ factor, one can indeed obtain examples with $\gamma\leq \sqrt{2}$ in Type IIA.}.

\subsection{More moduli\label{SEC: more mod}}
In previous sections we have seen that the strong bounds on the dS coefficient can be violated when having more than one modulus such that several terms compete asymptotically. We have also found concrete 2-moduli examples realizing such scenario in the complex structure moduli space of a Calabi-Yau fourfold.
One could then wonder if we could make the value of the dS coefficient $\gamma_{\vec{f}}$ arbitrarily low by having a very large number of moduli. 
This option is specially powerful when considering terms in $\mathcal{E}^{\rm rest}$. In principle, one could take advantage of having different signs of $l^{\text{dom}}_i$ to make $|\beta^i l^{\text{dom}}_i|$ small with respect to $|\vec{\beta}|$, thus resulting in a small dS coefficient in \eqref{dS gen}. Na\"ively, working with $\beta^i\sim \mathcal{O}(1)$, this could be easy to achieve by taking a large number $n$ of moduli, as $|\vec{\beta}|\sim\mathcal{O}(n)$ while having $|\beta^i l^{\text{dom}}_i|\sim\mathcal{O}(1)$. In the language of Section \ref{sec: Convex hull}, this would mean finding a set of points $\{\vec{\mu}\}$, not necessarily close to $\vec{0}$, but such that their convex Hull is as close as possible to the origin (without intersecting it). 

However, when trying to actually embed this into a Quantum Gravity framework, such as using Asymptotic Hodge Theory for flux compactifications, one quickly finds that this is much harder than expected. As previously discussed, for a CY fourfold compactification, the maximum number of moduli one can send to infinity while keeping a diagonal moduli space metric (as in \ref{eq:met}) is $n=4$, corresponding to the II$\rightarrow$ III$\rightarrow$ IV$\rightarrow$ V enhancement \cite{Grimm:2019ixq}. For $n>4$, some of the moduli will have $\Delta d_i=0$ and subleading corrections will need to be considered so that the moduli space metric is non-singular (see Appendix \ref{toy model} for an example in which this is discussed). For the resulting non-diagonal moduli space metrics $|\vec{\beta}|$ will no longer correspond to a ellipsoid, but as inferred from \eqref{app met}, in those cases $s^{i_0}$ is not featured in the leading term of $K$ (or equivalently a subleading term that goes as the leading term along the trajectory), then $\beta^{i_0}$ will not enter in $|\vec{\beta}|$, which will grow as the number of moduli featured in the leading terms of the K\"{a}hler potential. As $n$ grows, a smaller proportion of the moduli number will enter in $|\vec{\beta}|$.

Hence, we do not expect that  $\gamma_{\vec{f}}$ will decrease parametrically with the number of fields $\hat n$, although it would be interesting to check how small it can get.

For instance, for $n=3,4$ there are some limits in which $G$ is still diagonal, so that \eqref{dS gen} still applies and we expect therefore a suppression given by the number of moduli. The associated potentials should be relatively easy to study, with some solutions likely violating the strong dS bound.

\subsection{Beyond parametric control}
In the analysis of the dS coefficient along the previous sections we have always worked in the strictly asymptotic limit, $\lambda\rightarrow\infty$, where all the subleading terms to the constant $\gamma_{\vec{f}}$ value can be neglected. One could wonder if there is any chance that the dS coefficient value is lowered when subleading corrections are included for finite $\lambda$, even if accelerated expansion could then only occur for a finite time and not parametrically. As we will show, including the subleading corrections when going beyond the strict asymptotic limit does not seem to help to lower $\gamma_{\vec{f}}$.

\subsubsection{Subleading terms of the potential}
We first consider the effect that the subleading terms of the potential can have on $\gamma_{\vec{f}}$. To see this, we can rewrite \eqref{def dS} out of the $\lambda\rightarrow \infty$ limit. Following the notation from Appendix \ref{stab ax}, that divides the potential depending on its order in $\lambda$, we get
\begin{align}
    \displaystyle
    \gamma_{\vec{f}}(\lambda)&=-\frac{
    \sum_{\mathcal{E}^{(j)}}\left\{\hat{\beta}^i  l_i\lambda^{\beta^i  l_i}\sum_{\bm{l}\in \mathcal{E}^{(j)}}A_{\bm{l}}\right\}}
    {\sum_{\mathcal{E}^{(j)}}\left\{\lambda^{\beta^i  l_i}\sum_{\bm{l}\in \mathcal{E}^{(j)}}A_{\bm{l}}\right\}},
\end{align}
where we sum over the different terms that appear for each $\mathcal{E}^{(i)}$, with the dominant one being $\mathcal{E}^{(0)}$, of order $\lambda^{\beta^il_i^{\rm dom}}$, and so on. It is then immediate that $\lambda^{\beta^i  l_i^{\text{dom}}}<\lambda^{\beta^i  l_i^{\text{sub}}}<...$, and so on. We recall that for potentials obtained through Asymptotic Hodge Theory all the $A_{\bm{l}}$ are positive apart from $A_{\bm{4}}\equiv A_{\bm{4}}-A_ {\rm loc}$, which can have any sign. We then have
\begin{equation}\label{corr sub}
    \gamma=\underbrace{|\hat{\beta}^il_i^{\rm dom}|}_{\gamma_{\vec{f}}}+\frac{\sum_{\mathcal{E}^{(1)}}A_{\bm{l}}}{\sum_{\mathcal{E}^{(0)}}A_{\bm{l}}}\underbrace{\hat{\beta^i}(l_i^{\rm dom}-l_i^{\rm sub})\lambda^{-{\beta^i}(l_i^{\rm dom}-l_i^{\rm sub})}}_{>0}+\mathcal{O}(\text{sub-subdom.})
\end{equation}

Then the sign of the subleading corrections will depend on that of $\frac{\sum_{\mathcal{E}^{(1)}}A_{\bm{l}}}{\sum_{\mathcal{E}^{(0)}}A_{\bm{l}}}$. If we want to have a positive potential, then its dominating terms must be positive, so that $\sum_{\mathcal{E}^{(0)}}A_{\bm{l}}>0$. One could think that having $\sum_{\mathcal{E}^{(0)}}A_{\bm{l}}<0$ for low values would allow us to lower $\gamma$ from the asymptotic $\gamma_{\vec{f}}$. However, as discussed in Appendix \ref{APP1} when this term becomes sufficiently relevant this will end up being a tachyonic trajectory. This tension between lowering the dS coefficient and the presence of tachyons is reminiscent of the refined dS conjecture \cite{Garg:2018reu,Ooguri:2018wrx}. It would be interesting to quantify this relation as a way of testing this refined version and to check how much one can lower the dS coefficient while avoiding the presence of the tachyon.

For potentials coming from Asymptotic Hodge Theory applied on flux compactifications we recall that the only possible negative $A_{\bm{l}}$ is $A_{\bm{4}}\equiv A_{\bm{4}}-A_{\rm loc}$, which however appears as a constant factor that does not contribute to the dS coefficient. Nevertheless, as discussed in Section \ref{sec: IIA large V}, when going from Type IIB to Type IIA string theory limits through Mirror Symmetry we find negative terms that do contribute to the dS coefficient. In this sense, this setup might be more promising for lowering the value of the dS coefficient beyond parametric control.

\subsubsection{Some comments on corrections for trajectories out of the attractor}

Apart from considering corrections to the dS coefficient coming from the subleading terms of the asymptotic potential, one could consider another type of corrections that appear for trajectories that have not reached an attractor yet. 

Let us first discuss the case of the axions. 
As discussed in Section \ref{brief stab ax} (see also Appendix \ref{stab ax}), all the axions entering the scalar potential $V_M$ are asymptotically stabilized on a minimum $\phi_0$ dictated by the $\|\rho_{\bm{l}}\|_{\infty}^2$ functions accompanying subleading terms of the potential (the other option being that they parameterize flat directions of the potential). 
One could then consider a gradient flow trajectory that has not reached this attractor for the axions yet, and see whether the dS is lowered in this way. However, we can easily argue that the corrections from being away of this attractor (while still in the asymptotic regime for the saxions) are positive.

First of all, we can use that the metric in moduli space does not have mixed components between saxions and axions to split their contribution to the dS coefficient as follows
\begin{equation}
    \gamma^2=\frac{\|\nabla^{(s)}V_M\|^2+\|\nabla^{(\phi)}V_M\|^2}{V_M^2} \, .
\end{equation}
Now we remember that, when the saxions are taken to infinity, the axions do not enter in the saxionic contribution to the dS coefficient. This is related to the leading terms of the potential being always independent of the axions as shown in Section \ref{brief stab ax}. Therefore, in the asymptotic regime we have
\begin{equation}
    \gamma^2=\gamma_{\phi_0}^2+\frac{\|\nabla^{(\phi)}V_M\|^2}{V_M^2}\geq \gamma_{\phi_0}^2 \, ,
\end{equation}
where we have noted that the saxionic contribution can be identified as the value of the dS coefficient when the axions are in their attractor $\gamma_{\phi_0}$, i.e., where $\|\nabla^{(\phi)}V_M\|=0$. It is then easily seen that any deviation from the axion attractor makes the dS coefficient larger.

Let us stress that this only works asymptotically when the saxions are taken to infinity, thanks to the saxions contribution to the dS coefficient being independent of the axions values in that limit. In the interior of the moduli space, even though the axions contribution to the dS coefficient gets larger when out of the their attractor, the saxion contribution (that now depends on the axions values) could become in principle smaller.

Now we turn to the case of saxions. In the cases in which we find attractors for them, when we use $s^i(\lambda)=\alpha^i\lambda^{\beta^i}$ we are disregarding subleading terms that describe how the trajectory approaches the attractor. We expect this corrections also to be positive. This has been checked in several examples, and can be understood intuitively in the following way: The saxions attractor is generated in scenario (II) by the competition of two terms $V=V_1+V_2$ (more terms can appear for $n>2$). More precisely, there is a cancellation between both contributions to the gradient of the potential. This is, each term has a component which is tangent to the attractor flow and another which is orthogonal, being the latter the ones that cancel out at the locus of the attractor. If one slightly moves away from the attractor loci this cancellation becomes imprecise, thus giving an extra contribution to the dS coefficient and making its value larger.

\section{Summary and conclusions\label{sec:summ and conc}}

In this paper we have explored whether string theory flux compactifications allow for an asymptotic accelerated expansion of the universe at parametric control. This is important, not only for obvious phenomenological reasons, but also to determine the fate of asymptotic observables in string theory.

The asymptotic form of the flux scalar potential is highly constrained due the axionic shift symmetries and exhibit an exponentially decreasing behavior in terms of the proper field distance towards the infinite distance limit. For concreteness, we have focused on 4d $\mathcal{N}=1$ compactifications, and shown that the asymptotic trajectories are gradient flows that become asymptotically geodesics of the moduli space without potential. This forbids highly turning axionic trajectories for a parametrically large distance. To determine whether accelerated expansion is possible, one must check whether the exponential rate (denoted as de Sitter coefficient) satisfies $\gamma\leq \sqrt{2}$ in four dimensions along its gradient flow. We provide a general systematic recipe to determine this gradient flow and the de Sitter coefficient for general flux potentials. Interestingly, this recipe can be pictorially represented as a convex hull condition like other Swampland conjectures as follows:\\

\emph{Given an asymptotic scalar potential $V=\sum_{\bm{l}\in\mathcal{E}} V_{\bm{l}}$, it will not allow for accelerated expansion along its gradient flow (meaning $\gamma\geq \sqrt{2}$) if the convex hull of all de Sitter ratios $\{\vec{\mu}_{\bm{l}}\}_{\bm{l}\in\mathcal{E}}$ in \eqref{l to mu transform gen} for each potential term $\{V_{\bm{l}}\}_{\bm{l}\in\mathcal{E}}$ is outside of the ball of radius $\sqrt{2}$.}\\

The intuition behind this condition is clear, as the minimum distance of the convex hull to the origin corresponds to the de Sitter coefficient $\gamma$. Interestingly, this geometric formulation of the de Sitter conjecture is equivalent to a sort of Convex Hull Scalar WGC applied to the membranes generating the flux potential in which \emph{all} membranes must satisfy the bound.

We can then distinguish two scenarios, depending on whether a single or several terms of the potential exhibit the same asymptotic growth and dominate asymptotically. We show that the first scenario of a single term dominating (which happens when the closest point of the convex hull to the origin is a vertex) can never yield accelerated expansion. However, accelerated expansion becomes in principle possible in the second scenario when there are several terms competing asymptotically (meaning that the closest point of the convex hull to the origin is in an edge), since the de Sitter coefficient can be made much smaller than the one corresponding to each potential term separately. This can occur whenever we have several moduli sent to a limit at different rates. This second scenario had not been considered yet in the literature and presents a loophole to the argument against accelerated expansion given in \cite{Hellerman:2001yi,Rudelius:2021azq}. There, it was argued that asymptotic accelerated expansion is forbidden in 4d $\mathcal{N}=1$ supergravities, but this assumes that the scalar potential and the superpotential share the same gradient flow, which does not occur in the second scenario.

To make our analysis more concrete and test the general procedure in particular examples, we have focused on F-theory flux compactifications on Calabi-Yau fourfolds. The goal was to compute explicitly the de Sitter coefficient in string theory examples going beyond what was done in previous works, which were always focused on the particular asymptotic limit associated to weak string coupling and large volume \cite{Hertzberg:2007wc,Garg:2018zdg, ValeixoBento:2020ujr,Andriot:2020lea,Andriot:2022xjh,Cicoli:2021fsd}. To do so, we considered different limits in the complex structure moduli space of F-theory, which can be classified using Asymptotic Hodge Theory. The possible asymptotic flux potentials arising in two-moduli limits were classified and computed in \cite{Grimm:2019ixq}. We choose three of them (the first one corresponding to the canonical perturbative Type IIA/B at large volume/large complex structure, the second to a different complex structure point and third  a pure F-theoretical one without perturbative Type II description) and compute the de Sitter coefficient for all possible combinations of fluxes. Already within this set of examples, we find potentials that exhibit $\gamma\leq \sqrt{2}$, going beyond previous expectations \cite{Bastian:2020egp}. Since all these limits contain BPS strings that become tensionless asymptotically, we check that the tower of string modes (satisfying the Distance Conjecture) is of order or above the Hubble scale. We also discuss how the story changes with more moduli or taking into account corrections beyond the strict asymptotic limit. The former could help to decrease even more $\gamma$ (but not parametrically), while the corrections do not seem to help.

Our examples with $\gamma\leq \sqrt{2}$ are interesting potential candidates to yield accelerated expansion asymptotically at parametric control. The caveat is that we are only focusing on the complex structure moduli space, being agnostic about Kähler moduli stabilization. However, only if the overall volume modulus gets somehow stabilized, while the runaway on the complex structure moduli persists, the de Sitter coefficient computed in this paper will be the final one; otherwise, it will receive an additional contribution from the runaway of the Kähler moduli that could prevent accelerated expansion\footnote{In order to have a small contribution from the Kähler moduli (without stabilizing them) that does not spoil the accelerated expansion one should go beyond large volume, so that $\gamma$ gets smaller, but this would bring us away from the supergravity regime.}. Notice that the overall volume needs to get stabilized, not only to keep a small $\gamma$, but also for phenomenological reasons in our universe, since otherwise it would yield a 5\textsuperscript{th} force which is ruled out experimentally. Hence, to construct a realistic quintessence model, one should stabilize the overall volume anyway while moving in a different direction, like a complex structure modulus. Intuitively, one would think that this is easier to achieve if the complex structure field is lighter than the overall volume. Such inverted hierarchies have been found for certain flux choices \cite{Demirtas:2021nlu,Bastian:2021hpc}, so it would be interesting to explore the viability of such scenario further. Since such analysis requires to go beyond the flux potential (as Kähler moduli are not stabilized by fluxes in F-theory), we leave it for future work. Of course, it could also be that after studying full moduli stabilization, it turns out that accelerated expansion is forbidden asymptotically. This could have implications for de Sitter holography and the existence of asymptotic observables in expanding cosmologies. But it is too early to know.

\section*{Acknowledgments}

We would like to thank  Thomas Grimm, Damian van de Heisteeg, Luis Ib\'a\~nez, Fernando Marchesano, Miguel Montero, David Prieto, Tom Rudelius and  \'Angel Uranga for useful discussions. The authors acknowledge the support of the Spanish Agencia Estatal de Investigacion through the grant “IFT Centro de Excelencia Severo Ochoa 
CEX2020-001007-S and the grants PGC2018-095976-B-C21 and PID2021-123017NB-I00, funded by MCIN/AEI/10.13039/ 501100011033 and by ERDF A way of making Europe. The work of I.V. is partly supported by the grant RYC2019-028512-I from the MCI (Spain).
The work of I.R. is supported by the Spanish FPI grant No. PRE2020-094163. The work by J.C. is supported by the FPU grant no. FPU17/04181 from the Spanish Ministry of Education.

\appendix

\section{Tachyon-free potentials\label{APP1}}
One important aspect of the obtained gradient flow solutions is that we find no tachyons.
To check this, we evaluate the mass matrix $M_{ab}=\nabla_a\partial_b V$ for some generic power-like potential as that from \eqref{eq:pot} along some trajectory $s^j=\alpha^j\lambda^{\beta^j}$ (not necessarily the gradient flow solution) to find:
\begin{align}
    M_{s^js^k}|_{\vec {f}}&=\frac{1}{\alpha^j\alpha^k\lambda^{\beta^j+\beta^k}}\sum_{\bm{l}\in\mathcal{E}}A_{\bm{l}}l_jl_i\lambda^{\beta^k l_k}=\sum_{\bm{l}\in\mathcal{E}}\lambda^{\beta^k l_k}\underbrace{\left[\frac{ l_j}{\alpha^j\lambda^{\beta^j}}A_{\bm{l}}\frac{ l_k}{\alpha^j\lambda^{\beta^k}}\right]}_{N^{(\bm{l})}_{jk}(\lambda)},
\end{align}
Notice that $\lambda^{\beta^i  l_i}$ are positive for all $\bm{l}\in\mathcal{E}$. We thus have a positive linear combination of matrices with entries
\begin{equation}
    N^{(\bm{l})}_{jk}(\lambda)=\frac{  l_j}{\alpha^j\lambda^{\beta^j}}A_{\bm{l}}\frac{  l_k}{\alpha^j\lambda^{\beta^k}} \, .
\end{equation}
Notice that if $A_{\bm{l}}>0$ and we consider $\mathbb{R}$ as a vector space and $\{\frac{  l_i}{\alpha^i\lambda^{\beta^i}}\}_{i=1}^{n}$ a set of vectors within it, then $N^{(\bm{l})}$ are Gram matrices associated to the inner product $a\cdot_{A_{\bm{l}}}b=abA_{\bm{l}}$, and as such they are semi-positive definite. Now, if this is the case for all $\bm{l}\in\mathcal{E}$ then $M|_{\vec{f}}$ is the sum of semi-positive definite matrices, and such it is semi-positive definite, and any considered trajectory (be it the gradient flow solution or not) is non-tachyonic. Finally, consider the case that $A_{\bm{l}}<0$ (we recall that for potentials coming from Asymptotic Hodge Theory this can only happen for $A_{\bm{4}}-A_{\rm loc}$, while for those obtained for Type IIA through mirror symmetry $A_{44}$, $A_{54}$ and $A_{56}$ were also negative). This can only happen if said term is subleading, as otherwise the potential would not be able to send the scalar fields to infinity. Considering $\lambda$ high enough, the Gram matrices associated to the dominant terms (which are indeed semi-definite positive) will make $M|_{\vec{f}}$ semi-definite positive, and thus stable. We thus conclude that every potential with admissible solutions does not contain tachyonic trajectories along which the saxions can go to infinity.

This result is, in particular, useful for the following. One could worry that the optimization problem \eqref{eq:min-method} to determine the gradient flow trajectory gives more than one solution. Indeed, one could expect to find different asymptotic gradient flow solutions for the sectors defined by the asymptotic behavior of the rhs of \eqref{eq:saxionic-flow}. Notice that having more than one asymptotic solution sending saxions to infinity requires having a $n-1$ dimensional repeller separating the two sectors corresponding to these different behaviors. The directions normal to this repeller would be tachyonic (this is, $\hat{N}^\intercal M\hat{N}<0$ for $\hat{N}$ some normal vector to this interface, with $M_{ab}=\nabla_a\partial_b V$ the mass matrix of the potential). However as proven above, all trajectories taking some saxions to infinity are non tachyonic, as the dominant terms in the potential must have $A_{\bm{l}}>0$, and thus correspond to semi-positive definite contributions to $M$. 
Through this argument we conclude that the potentials taken under consideration have at most a unique kind of trajectories given by $\vec{\beta}$ (different $\vec{\alpha}$ values are indeed possible, as discussed), and thus the solution obtained from the optimization problem \eqref{eq:min-method} is the global solution.

The above discussion only applies for the cases in which we have a diagonal metric (more complicated metrics will involve a less simple structure for the Christoffel symbols which can spoil the stability and uniqueness of asymptotic trajectories). As a toy model, one can consider the scalar and Kähler potentials $V=\frac{a}{s^2}+\frac{b}{u^2}$ and $K_{\rm cs}=-\log(s^2+Bu^2s)$ in the $\mathcal{R}_{12}$, for which it is not difficult to find $\left(\left(\frac{a}{2b}\right)^{\frac{1}{6}}\lambda,\lambda\right)$ and $(B\lambda^2,\lambda)$ as asymptotic trajectories, with the first being stable and the second tachyonic.

\section{A toy model example with $\Delta d_i=0$\label{toy model}}
In the previous sections we have assumed in our results that $\Delta d_i\neq 0$ for all $i=1,...,n$. This allows us to consider only the dominant term inside $K_{\rm cs}$ and ignore possible subleading corrections in the saxions $\{s^i\}_{i=1}^{n}$, \cite{Grimm:2019ixq}. When some $\Delta d_i=0$ (as is always the case for $n>4$) subleading polynomial terms must be considered. This complicates the form of the moduli space metric $G$, as the appearance of new terms inside the logarithm of the $K_{\rm cs}$ expression results in a non diagonal metric, significantly increasing the difficulty of computing the dS coefficient $\gamma$. While expression \eqref{eq:gamma from lnV} is general and does not depend on the form of the moduli space metric, only for a diagonal hyperbolic metric one obtains \eqref{dS gen}. Even though it is not the ultimate goal of this paper to generally study the dS coefficient when subleading terms in the metric are taken under consideration (which does not necessarily affect the result when all the saxions appear in the dominating term but will always be needed for $n>4$), the dS coefficient can still be computed, as we will show through the following toy model example.

From the potentials obtained in \cite{Grimm:2019ixq}, we consider a III$\rightarrow$ III enhancement in a CY 4-fold with $n=2$, denoting the saxions by $(s^1,s^2)=(s,u)$, corresponding with $\Delta d_s=2$ and $\Delta d_u=0$ in the growth sector $\mathcal{R}_{12}$. The associated potential is given by
\begin{equation}
    V_M=\frac{A_{22}}{s^2}+\frac{A_{42}}{u^2}+A_{46} u^2+A_{64}s^2+A_{44}-A_{\rm loc},
\end{equation}
where we must set $A_{64}=A_{46}=A_{44}-A_{\rm loc}=0$ in order for admissible solutions to exist. Up to constant terms, we will have that $K_{\rm cs}\sim-\log(s^2+f(s,u))$, with $f(s,u)\simleq \mathcal{O}(s^2)$ on $\mathcal{R}_{12}$. As a toy model, we will consider $f(s,u)=Bus$, with $B>0$ (other choices, such as $f(s,u)=Bu$, do not feature a solution sending at least one saxion to infinity). Using the metric
\begin{equation}\label{metric non d}
    G=\left(
\begin{array}{cc}
 \frac{1}{(B u+s)^2}+\frac{1}{s^2} & \frac{B}{(B u+s)^2} \\
 \frac{B}{(B u+s)^2} & \frac{B^2}{(B u+s)^2} \\
\end{array}
\right)
\end{equation}
one obtains after solving the gradient flow equations that asymptotically the trajectory of the saxions is given by $\vec{\beta}=(1,1)$. Even though this implies the subleading terms in the K\"{a}hler potential are actually of the same order of the dominant term, the solution is admissible, as it is in $\mathcal{R}_{12}$. Using \eqref{eq:gamma from lnV} the dS coefficient is found to be $\gamma_{\vec{f}}=\sqrt{2}$, which saturates the $c^{\rm strong}_4$.

The non-null Christoffel symbols associated to the moduli space metric \eqref{metric non d} are
\begin{equation}
    \tilde{\Gamma}^s_{ss}=-\frac{1}{s},\quad \tilde{\Gamma}^u_{ss}=\frac{u}{s^2+Bsu},\quad\tilde{\Gamma}^u_{su}=\tilde{\Gamma}^u_{us}=-\frac{1}{s+Bu},\quad \tilde{\Gamma}^u_{uu}=-\frac{B}{s+Bu} \, .
\end{equation}
It is straightforward to check that the solutions of the kind $s=s_0 e^D$ and $u=u_0 e^D$ fulfill the geodesic equations, so that even if the metric is not of the diagonal hyperbolic kind, the gradient flow solutions are still geodesic curves.

Even though we have not made direct use of it, in this example we have the following metric on $\vec{\beta}$:
\begin{equation}\label{app met}
    |\vec{\beta}|=\left\{\begin{array}{ll}
         \sqrt{2}|\beta^1|& \text{if }\beta^1\geq\beta^2 \\
         \sqrt{(\beta^1)^2+(\beta^2)^2}&\text{if }\beta^2\geq\beta^1
    \end{array}\right.
\end{equation}
Notice that the region defined by $\{|\vec{\beta}|=1\}$ is no longer an ellipsoid, as it is not bounded in the $\beta^2\rightarrow-\infty$ region, corresponding with the trajectories for which $u$ ceases to be relevant.\\
As a curiosity, the non-diagonal term of the metric results in multiple terms with different signs in $\vec{\partial} V_M$:
\begin{equation}
    -\vec{\partial}V_M=\left(\frac{2}{s}-\frac{2 s^2}{B u^3},\frac{2 \left(\frac{2 s^2}{B^2}+\frac{2 s u}{B}+u^2\right)}{u^3}-\frac{2}{B s}\right),
\end{equation}
which results in an attractor for $B>0$. This was not the case when the moduli space metric was diagonal (note that all the terms are from $\mathcal{E}^{\rm light}$). Note that the solution also exists for $B<0$.

In the above example the leading and subleading terms of the K\"{a}hler potential ended up being of the same $\lambda$ order under a $\vec{\beta}=(1,1)$ solution. However, in general it could be the case that the subleading terms (in some cases more than one would have to be considered) along the solution are indeed much smaller than the dominant one, as the potential and the moduli space metric are on principle independent, so that the terms containing some $s^{i_0}$ are all subleading. In that case when evaluating the $\|\dot{\vec{f}}\|$ norm, the component associated with $\dot{s}^{i_0}(\lambda)$ will become irrelevant in $\|\dot{\vec{f}}\|$ as $\lambda\rightarrow \infty$. This way $\|\dot{\vec{f}}\|$ will be smaller in those cases in which some of the saxions are irrelevant in the K\"{a}hler potential evaluated on the solution, thus resulting in a higher $\gamma_{\vec{f}}$ value. We thus conclude that when trying to find solutions with low $\gamma$ values (as we are interested in obtaining accelerating universes), then we should look for those in which all the saxions appear in equally dominating terms in $K$.

It was proposed in Section \ref{SEC: more mod} that one could lower the value of $\gamma$ if a sufficiently high number $n$ of moduli where considered, as this could naïvely allow for a large value of $\|\dot{\vec{f}}\|$ while keeping $\partial_{\lambda}V_M(\vec{f}(\lambda))$ controlled. However, it seems that in general this is not very likely to occur, as it would be necessary to feature all the saxions in terms of the same order as the dominant one, which will be more difficult to occur as $n$ grows.

\section{Hubble damping for the axions \label{APP H damp ax}}
In this appendix, we consider in more detail the case in which the axions are located along a flat direction of the potential (for example if some of them do not appear in the $V_M$ expression, $\partial_{\phi^i} V_M\equiv 0$). In principle this could mean that if the initial velocity of the axion is not null, ${{\phi}^i}'(t_0)\neq 0$, it could drift towards infinity, thus leaving the growth sector regime. We will now show that in general we expect this not to be the case. In this situation the axion equation of motion is no longer \eqref{eqmot2}, but from the full  e.o.m. \eqref{eqmot}, one finds
\begin{equation}
    {\phi^a}''+\left[(d-1)H-2\frac{{s^a}'}{s^a}\right]{\phi^a}'=0\;,
\end{equation}
where we recall that $t$ is the proper cosmological time and make use of the fact that the only non-vanishing Christoffel symbols\footnote{We thank Filippo Revello for pointing out a non-vanishing Christoffel symbol that was missing in a previous version of this paper, thus changing the equation of motion for the axion.} of the moduli space metric are\footnote{While this expression for the Christoffel symbols is only valid for diagonal matrices, we expect that similar conclusions can be found for more general metrics.}
\begin{equation}
    \Tilde{\Gamma}^{s^a}_{\phi^a\phi^a}=-\Tilde{\Gamma}^{s^a}_{s^as^a}=-\Tilde{\Gamma}^{\phi^a}_{s^a\phi^a} = \frac{1}{s^a}\;.
\end{equation}
For initial velocity ${\phi^a}'(t_0)\neq 0$ one can show that
\begin{equation}
    {\phi^a}'(t)={\phi^a}'(t_0)\frac{s^a(t)^2}{s^a(t_0)^2}e^{-(d-1)\int_{t^0}^tH(\tau)\diff \tau}\;.
\end{equation}
The condition then for the axion to remain finite (and thus within our growth sector regime) is then given by
\begin{equation}
    \phi^a(t)-\phi^a(t_0)={\phi^a}'(t_0)\int_{t_0}^{\infty}\frac{s^a(t)^2}{s^a(t_0)^2}e^{-(d-1)\int_{t_0}^tH(\tau)\diff \tau}\diff t <\infty\;,
\end{equation}
which is equivalent to $s^a(t)^2\exp\left\{-(d-1)\int_{t_0}^tH(\tau)\diff \tau\right\}<\mathcal{O}(t^{-1})$. Hence, as long as the exponential suppression given by the Hubble damping dominates over the $s^a\to\infty$ enhancement, the axion will remain finite along the trajectory. Performing a more careful analysis is beyond the scope of this paper, but it would be interesting for future research.\footnote{See indeed \cite{Revello:2023hro} for a posterior publication identifying a solution in which the axion is sent to infinity with a steep enough potential.}

One must keep in mind that the comments above are only needed in order for the Hubble damping term to damp the axions to finite values when they are along flat directions. If the axions are stabilized at a potential minimum no requirements on the Hubble parameter are needed. Furthermore, we have only considered that our fields are subject to classical dynamics, without considering quantum effects such as instantonic corrections to the potential, which could further stabilize some axions even if they are located along a flat direction.

\section{Axion stabilization \label{stab ax}}

In this appendix we demonstrate that axions in the context of F-theory flux compactifications will always either get stabilized at a finite point $\phi_0$ or remain as flat directions in the asymptotic regime, so they do not enter into the analysis of the asymptotic trajectories. We will first show that in the case axions enter the scalar potential $V_M$, a unique minimum exists, which will in turn determine the stabilization point (not necessarily coinciding). Later in Section \ref{sec:how fast axions} we will determine the value of said stabilization point and how fast it is approached in terms of the saxionic geodesic distance.

We will start by recalling that $\|\rho_{\bm{l}}\|_{\infty}^2=\mathcal{K}^{\bm{l}}_{ij}\varrho^i_{\bm{l}}\varrho^j_{\bm{l}}$ (no sum over $\bm{l}$), where $\mathcal{K}^{\bm{l}}$ are constant, symmetric, semi-positive definite matrices and $\vec{\varrho}_{\bm{l}}$ are vector functions of the axions and the quantized fluxes. Furthermore, from \eqref{rho prop}, $\partial_{\phi^j}\vec{\varrho}_{\bm{l}}=-\vec{\varrho}_{\bm{l}^{(j)}}$, with $\Delta l^{(j)}_i=\Delta l_i+2\delta_i^j$. It is then immediate that for every path sending all the saxions to infinity the $\bm{l}^{(j)}$ term (unless it is null) will dominate\footnote{For trajectories in which some of the saxions remain finite the two terms might be of the same order, and thus when moving in the chain complex along the direction of said saxion with $\beta_i=0$ we will go to a term of the same order (if the saxion $s^i$ is stabilized) or only marginally more dominant (if $s^i$ grows in a subleading way).} over $\bm{l}$ for all $\bm{l}\in\mathcal{E}$ and $j=1,...,n$. However, the fact that there is only a finite number of $\bm{l}$ terms in $\mathcal{E}$ and that $\deg_{\phi^k}\|\rho_{\bm{l}^{(k)}}\|^2_{\infty}=\deg_{\phi^k}\|\rho_{\bm{l}}\|^2_{\infty}-2$ means that at some point the following chain must reach a zero term:
\begin{equation}\label{chain}
    \|\rho_{\bm{l}}\|^2_{\infty}\rightarrow\|\rho_{\bm{l}^{(k_1)}}\|^2_{\infty}\rightarrow\|\rho_{\bm{l}^{(k1,k2)}}\|^2_{\infty}\rightarrow...\rightarrow\|\rho_{\bm{l}^{(k_1,...,k_j)}}\|^2_{\infty}\rightarrow 0
\end{equation}
The terms in which the above chains end will depend on the non-zero quantized fluxes, so that making more of them null might shorten the length of the chain, but never make it longer. As a result, this translates in $\partial_{\phi^i}\|\rho_{\bm{l}^{(k_1,...,k_j)}}\|^2_{\infty}=0$ for all $i=1,...,n$, so that $\|\rho_{\bm{l}^{(k_1,...,k_j)}}\|^2_{\infty}$ does not depend on the axions but only on the quantized fluxes. Similarly, $\vec{\varrho}_{\bm{l}^{(k_1,...,k_j)}}$ and $\vec{\varrho}_{\bm{l}^{(k_1,...,k_{j-1})}}$ contain only linear terms in the axions.

We can use the \eqref{chain} chain relations to partition $\mathcal{E}=\cup_{i=0}^{N} \mathcal{E}^{(i)}$, so that $\bm{l}\in\mathcal{E}^{(0)}$ are the terms associated to the last non-null $\|\rho_{\bm{l}}\|_\infty^2$, $\mathcal{E}^{(1)}$ the following ones, and so on. We see now that given a potential $V_M$, those terms dominating for every path taken towards infinity (and thus those which will determine the trajectory provided we are in the asymptotic region and the axions are finite) will be those from $\mathcal{E}^{(0)}$, independent of the axions $\{\phi^i\}_{i=1}^{n}$. Then, once we know the trajectory followed by the saxions, we can further partition $\mathcal{E}^{(i)}=\cup_{j= 0}^{N_i}\mathcal{E}^{(i,j)}$, so that terms of $\mathcal{E}^{(i,j)}$ dominate over those of $\mathcal{E}^{(i,j+1)}$ and so on. This way we have the following finer chain over the followed saxionic trajectory:
\begin{equation}
    \mathcal{E}^{(N,N_N)}\rightarrow\mathcal{E}^{(N,N_{N-1})}\rightarrow \cdots \rightarrow\mathcal{E}^{(N,0)}\rightarrow\mathcal{E}^{(N-1,N_{N-1})}\rightarrow \cdots \rightarrow \mathcal{E}^{(1,0)}\rightarrow \mathcal{E}^{(0,N_0)}\rightarrow \cdots \rightarrow\mathcal{E}^{(0,0)}
\end{equation}
with the terms belonging to $\mathcal{E}^{(i,j)}$ being of order $\mathcal{O}(\lambda^{(\beta^k\Delta l_k)_{(i,j)}})$,  with $(\beta^k\Delta l_k)_{(0,0)}>(\beta^k\Delta l_k)_{(0,1)}>...>(\beta^k\Delta l_k)_{(N,N_N)}$.\\

\begin{figure}[htbp]
\centering
\includegraphics[width=0.60\textwidth]{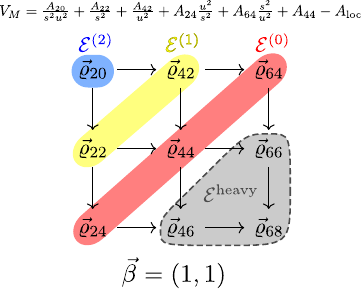}
\caption{First potential from Table \ref{table4} and the associated $\mathcal{E}=\cup_ i\mathcal{E}^{(i)}$ partition.}
        \label{Fig:chain}
\end{figure}

An example of this partition for a specific potential appears in Figure \ref{Fig:chain}. We then have that the most dominating terms in which the axions appear (before the need of considering any specific trajectory) will be those from $\mathcal{E}^{(1)}$. Now notice that for these terms
\begin{equation}
    \partial_{\phi^k}\|\rho_{\bm{l}}\|^2_{\infty}=-2\mathcal{K}^{\bm{l}}_{ij}\varrho^i_{\bm{l}}\varrho^j_{\bm{l}^{(k)}}\qquad\text{(no sum over $\bm{l}$)},
\end{equation}
with $\vec{\varrho}_{\bm{l}}$ and $\vec{\varrho}_{\bm{l}^{(k)}}$ being linear and independent on the axions, respectively. As a result $\partial_{\phi^k}\|\rho_{\bm{l}}\|^2_{\infty}$ is a linear function on the axions for all $\bm{l}\in\mathcal{E}^{(1)}$. Now, for this level of the chain, the extreme condition reads
\begin{equation}\label{opt cond}
    \sum_{\bm{l}\in \mathcal{E}^{(1)}}\partial_{\phi^k}\|\rho_{\bm{l}}\|^2_{\infty}\prod_{i=1}^{n}(s^i)^{{\Delta l_i}}=\sum_{i=0}^{N_1}\lambda^{(\beta^j\Delta l_j)_{(1,i)}}\sum_{\bm{l}\in\mathcal{E}^{(1,i)}}\partial_{\phi^k}\|\rho_{\bm{l}}\|^2_{\infty}=0\qquad\forall k=1,...,n \, .
\end{equation}

Now, because in the $\lambda\rightarrow\infty$ limit the above equation is dominated by the $\mathcal{E}^{(1,0)}$ terms, we can go step by step and first consider the following interesection of hyperplanes
\begin{equation}\label{Pi10}
    \Pi^{(1,0)}=\bigcap_{k=1}^{n}\left\{\sum_{\bm{l}\in\mathcal{E}^{(1,0)}}\partial_{\phi^k}\|\rho_{\bm{l}}\|^2_{\infty}=0\right\} \, .
\end{equation}
The stabilization point(s) $\phi_0$ will then correspond precisely to the intersection of the $n$ hyperplanes defined by each minimization condition. Now, given a system of linear equations such as the one above, one has to study the existence of solutions. We can rewrite it as $\{M^{(1,0)}\phi+\Phi^{(1,0)}=0\}$, with
\begin{align}
    M_{ij}^{(1,0)}&=\sum_{\bm{l}\in\mathcal{E}^{(1,0)}}\partial_{\phi^i}\partial_{\phi^j}\|\rho_{\bm{l}}\|_\infty^2=\sum_{\bm{l}\in\mathcal{E}^{(1,0)}}\mathcal{K}^{(\bm{l})}_{kh}\varrho^k_{\bm{l}^{(i)}}\varrho^h_{\bm{l}^{(j)}} \, ,\\
    \Phi^{(1,0)}_k&=\sum_{\bm{l}\in\mathcal{E}^{(1,0)}}\partial_{\phi^k}\|\rho_{\bm{l}}\|^2_{\infty}|_{\phi=0} \, .
\end{align}
Notice that as $\mathcal{K}^{\bm{l}}$ is a symmetric positive definite matrix \cite{Grimm:2019ixq}, then $M^{(1,0)}$ as the sum of Gram matrices will be symmetric and \emph{semi}-positive definite, corresponding to the mass matrix (up to global saxionic factors) for the axions when only terms up to $\mathcal{E}^{(1,0)}$ are considered, so that $\Pi^{(0,1)}$ will be an attractor. The eigenspaces associated to the null eigenvalues of $M^{(1,0)}$ will correspond to flat directions of these terms of the potential.

With respect to this, we can have the following three cases:
\begin{itemize}
    \item The system has a \emph{unique solution} $\Pi^{(1,0)}=\{\phi_0\}$. The $n$ hyperplanes intersect in a single point. This is equivalent to the normal vectors to each of the $n$ hyperplanes being linearly independent, or equivalently $\rank M^{(1,0)}=n$. As we will see later, this point $\phi_0$ is asymptotically a global minimum for the all the axions, which are then stabilized (whether the stabilization point is $\phi_0$ or a point close to it will be later discussed in Section \ref{sec:how fast axions}).
    \item The system has \emph{infinitely many solutions}, with the $n$ hyperplanes intersecting at a locus $\Pi^{(1,0)}$ with dimension higher than 0. This can occur either because some of the hyperplanes are given by trivial optimization conditions (i.e. $\sum_{\bm{l}\in\mathcal{E}^{(1,0)}}\partial_{\phi^k}\|\rho_{\bm{l}}\|^2_{\infty}\equiv 0$ for some $k$) and/or because the normal vectors of the hyperplanes are linearly dependent but they do intersect. This is equivalent to $\rank M^{(1,0)}=\rank (M^{(1,0)}|\Phi^{(1,0)})<\hat{ n}$. The set of solutions corresponds to a flat direction at $\mathcal{E}^{(1,0)}$ level, which can then be stabilized to a single stabilization point $\phi_0$ by subleading terms $\mathcal{E}^{(i,j)}$.
    \item The system has \emph{no solutions}. This occurs when at least two of the hyperplanes are parallel and non-intersecting. This is equivalent to $\rank M^{(1,0)}<\rank (M^{(1,0)}|\Phi^{(1,0)})$. This means that two competing terms of the potential are trying to stabilize some axions to different finite values. This does not mean that the stabilization does not occur, having two competing terms dragging the axions to different finite values will en up in a combined action that attracts them to an intermediate (and thus finite) position. 
\end{itemize}

From this discussion we see that, at leading order in the asymptotic expansion for the potential, some axions get stabilized at finite values while others might correspond to flat directions. Let us stress that this result comes from the fact that the terms stabilizing the axions in potentials obtained through Asymptotic Hodge Theory are at most linear in them as discussed before. The presence of higher powers of the axions would involve not only intersections of hyperplanes but of generally curved hypersurfaces. Were these curved hypersurfaces to approach each other asymptotically (thus intersecting at infinity) the axions would not be stabilized at finite values but sent to infinity by the gradient flow. Luckily this is not the case.

We have thus narrowed our possibilities for $\mathcal{E}^{(1,0)}$ to only two cases. If the $\mathcal{E}^{(1,0)}$ minimum conditions have a single solution $\phi_0$, then we are done. As $M^{(1,0)}$ is definite positive (it is invertible, so it does not have flat directions) at all points, then the same can be said about the mass matrix for the axions, $M^{\phi}$, thus proving that $\phi_0$ is an stable global minimum. The contributions to $M^{\phi}$ by subleading terms of other $\mathcal{E}^{(i,j)}$ terms will be negligible as $\lambda\rightarrow\infty$.

Now for the second case, in which the saxions are located along a flat direction, we consider the following. Defining $\Pi^{(i,j)}$ in an analogous way as $\Pi^{(1,0)}$ in \eqref{Pi10}, $\Pi^{(1,i)}$ for $i=0,...,N_1$ will consist in affine submanifolds of $\R^{n}$, as $\mathcal{E}^{(1,i)}\subseteq\mathcal{E}^{(1)}$. We can progressively intersect 
\begin{equation}
    \tilde{\Pi}^{(1,k)}=\bigcap_{\substack{i=1\\\Pi^{(1,i)}\cap\tilde{\Pi}^{(1,i-1)}\neq\emptyset}}^k\Pi^{(1,i)}\qquad\text{for }k=1,...,N_{1} \, .
\end{equation}
Notice that nothing wrong happens if some $\Pi^{(1,k)}$ is disjoint with $(\cap_{i=0}^{k-1}\Pi^{(1,i)})$, as for $\lambda\rightarrow\infty$ the former term will be subleading with respect to the later ones. As $\dim\Pi^{(1,0)}\leq \dim\tilde{\Pi}^{(1,2)}\leq...\leq \dim\tilde{\Pi}^{(1,N_1)}$ we will progressively reduce the dimension of the flat attractor for the saxions. Those axions constrained to a single value $\phi^j_0$ will be stabilized by the potential, as we recall that the mass matrix will be a positive linear combination of semi-definite positive matrices, and possible higher order terms destabilizing $\phi^j$ will be subleading for $\lambda\rightarrow \infty$.

Now, if $\dim \tilde{\Pi}^{(1,N_1)}>0$ some axions will still be non-stabilized and we will have to turn to $\mathcal{E}^{(2)}$ for this. In principle given $\bm{l}\in\mathcal{E}^{(2)}$,  $\partial_{k}\|\rho_{\bm{l}}\|^2_{\infty}=-\mathcal{K}^{(\bm{l})}_{ij}\varrho^{i}_{\bm{l}}\varrho^j_{\bm{l}^{(k)}}$ would be cubic on the axions, as $\bm{l}^{(k)}\in\mathcal{E}^{(1)}$ for all $\bm{l}\in\mathcal{E}^{(2)}$ and $k=1,...,n$. Then the $\{\sum_{\bm{l}\in\mathcal{E}^{(2,i)}}\partial_{\phi^k}\|\rho_{\bm{l}}\|^2_{\infty}=0\}$ sets are cubic hypersurfaces of codimension 1, with $\Pi^{(2,i)}$ their intersection. We can then proceed as before, successively intersecting and obtaining 
\begin{equation}
    \tilde{\Pi}^{(2,k)}=\tilde{\Pi}^{(1,N_1)}\cap\left(\bigcap_{\substack{i=1\\\Pi^{(1,i)}\cap\tilde{\Pi}^{(2,i-1)}\cap\neq\emptyset}}^k\Pi^{(1,i)}\right)\qquad\text{for }k=1,...,N_{2},
\end{equation}
with $\dim \tilde{\Pi}^{(2,k)}\leq \tilde{\Pi}^{(2,k-1)}$. Again, those  $\Pi^{(2,i)}$ which do not intersect with $\tilde{\Pi}^{(2,k-1)}$ do not pose a problem for the stabilization problem, as they will become irrelevant for $\lambda\rightarrow \infty$. We can rinse and repeat, progressively obtaining $\tilde{\Pi}^{(i,j)}$ by successive intersections until all the axions are stabilized or we reach $\tilde\Pi^{N,N_N}$, with dimension higher than 0 and corresponding to a flat submanifold. As for the stability of these points, notice that now that we are not necessarily dealing with linear $\vec{\varrho}_{\bm{l}}$ we have
\begin{equation}
    \partial_{\phi^i}\partial_{\phi_j}\|\rho_{\bm{l}}\|^2_{\infty}=2\mathcal{K}^{(\bm{l})}_{hk}\left(\varrho^h_{\bm{l}^{(i)}}\varrho^k_{\bm{l}^{(j)}}+\varrho^h_{\bm{l}}\varrho^k_{\bm{l}^{(i,j)}}\right) \, .
\end{equation}
As previously discussed, while the first term corresponds to a Gram matrix, the second one might not at first correspond with a positive definite matrix. However, when all potential terms are taken into consideration,
\begin{equation}
     M^{\phi}_{ij}\sim 2\sum_{i=1}^N\sum_{j=1}^{N_i}\lambda^{(\beta^k\Delta l_k)_{(i,j)}}\sum_{\bm{l}\in\mathcal{E}^{(i,j)}}\mathcal{K}^{(\bm{l})}_{hk}\left(\varrho^h_{\bm{l}^{(i)}}\varrho^k_{\bm{l}^{(j)}}+\varrho^h_{\bm{l}}\varrho^k_{\bm{l}^{(i,j)}}\right),
\end{equation}
the terms dominating in $\lambda$ for each $M^{\phi}_{ij}$ component will be those with $\vec{\varrho}_{\bm{l}}$ linear, for which only the Gram matrix term appears. The rest of contributions are totally subleading for $\lambda\rightarrow\infty$. We thus conclude that the axion mass matrix $M^{\phi}$ is semi-positive definite for large enough $\lambda$, and thus the stabilized axions will be in a minimum of the potential or a flat direction.\\

Back to said not stabilized axions,  belonging to the flat $\tilde\Pi^{N,N_N}$, as explained in Appendix \ref{APP H damp ax}, even if they have some initial velocity $\phi^i(0)'\neq 0$, the Hubble damping of the expanding Universe will stop them from reaching arbitrarily high values, thus remaining in the growth sector regime in which we were working.

Note that in some cases the axionic functions $\|\rho_{\bm{l}}\|^2_\infty$ become null when stabilizing (as happens when a single function stabilizes some axion). This must be taken into account as then it is possible that when for the saxionic trajectory $\beta^i{\Delta l}^{\rm dom}_i=0$ there are no subleading terms sending the saxions to infinity along the attractor manifold.\\

 \subsection{Where and how fast do axions stabilize?\label{sec:how fast axions}}
 Once we have shown that axions appearing in the scalar potential are attracted towards a minimum, we need to establish the point where said axions stabilize. While in most cases said point corresponds with the potential minimum, in some instances the gradient flow dynamics will damp the axions to a point close but different than the minimum. Furthermore, one could wonder how fast this stabilization happens, or in other words, if it could be possible to have axions arbitrarily far from their stabilization point for arbitrarily high saxion values. As we will see now, axion stabilization is pretty fast, though it takes an infinite time to be completed.

 We will thus start with a saxion trajectory given by $\hat{\beta}$ (recall that the dominating terms of the potential setting this are independent from the axions), such that the most dominating term of the potential in which some $\phi^i$ appears can generically be written as $\tilde{\alpha}\lambda^{\Gamma^{(i)}}\sum_{h}(f_0^{(h)}\phi^i-f_1^{(h)})^2$, with $\tilde{\alpha}$ accounting for the $\alpha^i$ factors and $\{f_1^{(h)},f_2^{(h)}\}_h$ generic fluxes that might depend on quantized fluxes and previously stabilized axions. We allow for several terms of the same order stabilizing the axion, with the minimum being located at $\phi_0^i=\frac{\sum_hf_1^{(h)}f_0^{(h)}}{\sum_h(f_0^{(h)})^2}$. Parameterizing then $\phi^i$ with the geodesic distance traveled by the saxions, $D$, the equation of motion to first order will read
 \begin{equation}
     \dot{\phi^i}=-A^{(i)}e^{\omega^{(i)}D}(\phi^i-\phi^i_0),\qquad\text{ with}\left\{
     \begin{array}{rl}
          A^{(i)}&=\frac{4\tilde{\alpha}\sum_h(f_0^{(h)})^2}{d_iV_0\gamma_{\vec{f}}}>0\\
          \omega^{(i)}&=\Gamma^{(i)}+2\hat{\beta}^i+\gamma_{\vec{f}}
     \end{array}
     \right.
 \end{equation}
 where we have used that $\mathcal{F}(D)=\|\nabla V\|^{-1}=(\gamma_{\vec{f}}V(D))^{-1}=\gamma_{\vec{f}}^{-1}V_0^{-1}e^{\gamma D}$. The solution to the above ODE is found to be
 \begin{equation}
     \phi^i(D)=\left\{
     \begin{array}{lr}
          \phi_0^i+[\phi^i(0)-\phi_0^i]\exp\left\{\frac{A^{(i)}}{\omega^{(i)}}\left(1-e^{\omega^{(i)} D}\right)\right\}&\text{ if }\omega^{(i)}\neq 0\\
          \phi_0^i+[\phi^i(0)-\phi_0^i]e^{-A^{(i)} D}&\text{ if }\omega^{(i)}= 0
     \end{array}
     \right.
 \end{equation}
 The first case corresponds to a doubly exponential convergence, while the second one is simply exponential. We conclude then that for large $D$ values the saxions will be arbitrarily close to their stabilization points.
 It is immediate that for $\omega^{(i)}\geq 0$ then $\phi^i\rightarrow \phi^i_0$, while for $\omega^{(i)}<0$ $\phi^i\rightarrow\tilde{\phi}^i_0=\phi^i_0+[\phi^i(0)-\phi_0^i]e^{\frac{A^{(i)}}{\omega^{(i)}}}$.
 
 Now, regarding the possible values of $\omega^{(i)}$, we must note that as the $\bm{\Delta l}$ dominating the $\phi^i$ motion corresponds to a subleading term, then $\Gamma^{(i)}\leq -\gamma_{\vec{f}}-2\hat{\beta}^i$, so that $\omega^{(i)}\leq 0$. The $\omega^{(i)}=0$ is obtained for those saxions that are only one step from the end of the partially ordered chains (this will always be the case at least for one axion), while $\omega^{(i)}<0$ is given to those for which more steps appear\footnote{As in this case axion stabilization does make the associated potential term vanish, there is nothing radically different with the axion stabilizing to $\tilde{\phi}^i_0\neq \phi^i_0$.}. As an example, the former occurs for $b$ while the later for $c$ in \eqref{eq:ex bc}.
 
 We can now compute the distance \eqref{eq:field-metric} between the axions $\phi(D)$ at a certain saxionic distance $D$ and their ultimate stabilization point $\phi_0$ (we here replace $\phi_0$ by $\tilde{\phi}_0$ when it is the case). Noticing that the geodesics for fixed saxions are segments joining $\phi(D)$ and $\phi(0)$, it is not difficult to compute that
 \begin{equation}
     {d_{\rm{Eucl}}}(\phi(D),\phi_0)=\sqrt{\sum_{i=1}^{n}[\phi_0^i-\phi^i(D)]^2\frac{\Delta d_i}{2s^i(0)^2}e^{-2\hat{\beta}^iD}} \, .
 \end{equation}
 Now, as 
 \begin{equation}
     [\phi_0^i-\phi^i(D)]^2\sim\left\{
     \begin{array}{ll}
         e^{-2A^{(i)}D} &\text{if }\omega^{(i)}=0\\
         e^{2\omega^{(i)}D}&\text{if }\omega^{(i)}<0
     \end{array}
     \right.
 \end{equation}
 we conclude that ${d_{\rm{Eucl}}}(\phi(D),\phi_0)\sim e^{-\theta D}$, with $\theta=\min\left(\{A^{(i)}+\hat{\beta^i}\}_{\omega^{(i)}=0}\cup\{|\omega^{(i)}|+\hat{\beta^i}\}_{\omega^{(i)}<0}\right)>0$, and thus our trajectories are arbitrarily saxionic for high enough $D$. It is not difficult to show that indeed $\Omega \sim e^{-\theta D}$, with $\Omega$ the non-geodesity factor as defined under \eqref{eq:eV}.
 
 As it is evident, the axionic function that stabilized $\phi^i$ does not become null at any finite distance. One can then wonder if this stabilization makes the associated term irrelevant by this exponential suppression or if it is actually more dominant than other subleading terms. One can see that the term stabilizing $\phi^i$ is of order $\mathcal{O}(\exp\{-(2A^{(i)}+\hat{\beta}^j\Delta l_j)D\}$ for certain $\bm{\Delta l}\in\mathcal{E}$. The following subleading (in the saxionic sense) term will be at most of order $\mathcal{O}(\exp\{-(\hat{\beta}^j\Delta l_j+1)D\})$, and thus the ratio between the stabilizing term and the following saxionic term is of order $\mathcal{O}(\exp\{-(2A^{(i)}-1)D\})$. As one expects $V_0\ll 1$ in the asymptotic region, then $A^{(i)}\gg 1$, so that the stabilizing term is effectively suppressed in this stabilization and will not contribute to the dS coefficient.

\bibliographystyle{JHEP}
\bibliography{bibliography}

\end{document}